\documentclass[twocolumn,showkeys,superscriptaddress,longbibliography,floatfix]{revtex4-2}
\usepackage{graphicx, subfigure}
\usepackage{amssymb, amsmath,amssymb,amsfonts}
\usepackage{amsthm,mathrsfs,amsopn}
\usepackage{dcolumn}
\usepackage{bm}
\usepackage{color}
\usepackage[utf8]{inputenc}
\usepackage{mathtools}
\usepackage{float} 

\theoremstyle{plain} 

\definecolor{brickred}{rgb}{0.7, 0.25, 0.33}
\definecolor{applegreen}{rgb}{0.55, 0.71, 0.0}
\definecolor{darkblue}{rgb}{0.0, 0.0, 0.55}
\newcommand{\mar}[1]{\textcolor{black}{#1}}
\newcommand{\jen}[1]{\textcolor{black}{#1}}
\newcommand{\teo}[1]{\textcolor{black}{#1}}

\begin{document}

\title{Turing patterns on adaptive networks}

\author{Marie Dorchain}
\affiliation{Department of Mathematics and naXys, Namur Institute for Complex Systems,
University of Namur, Namur, Belgium}

\author{S. Nirmala Jenifer}
\affiliation{Department of Mathematics and naXys, Namur Institute for Complex Systems,
University of Namur, Namur, Belgium}
\affiliation{Department of Physics, Bharathidasan University, Tiruchirappalli 620024, Tamil Nadu, India}

\author{Timoteo Carletti}
\affiliation{Department of Mathematics and naXys, Namur Institute for Complex Systems,
University of Namur, Namur, Belgium}

\begin{abstract}
We are surrounded by spatio-temporal patterns resulting from the interaction of the numerous basic units constituting natural or human-made systems. In presence of diffusive-like coupling, Turing theory has been largely applied to explain the formation of such self-organized motifs \teo{on} networked systems, where reactions occur in the nodes and the available links are used for species to diffuse. In many relevant applications, those links \teo{do not have static weights}, as very often assumed, but evolve in time, and more importantly, they {\em adapt} their weights \teo{as functions of} the states of the nodes. In this work, we make one step forward, and we provide a general theory to prove the validity of Turing idea in the case of adaptive symmetric networks with positive weights. The conditions for the emergence of Turing instability rely on the spectral property of the Laplacian matrix and the model parameters, thus strengthening the interplay between dynamics and network topology. A rich variety of patterns are presented by using two prototype models of nonlinear dynamical systems, the Brusselator and the FitzHugh-Nagumo model. Because many empirical networks adapt to changes in the system states, our results pave the way for a thorough understanding of self-organization in real-world systems.
\end{abstract}

\maketitle

\section{Introduction}
\label{sec:intro}

We are surrounded by ordered patterns, i.e., regular structures spontaneously resulting from the many microscopic agents nonlinearly interacting with each other~\cite{PrigogineNicolis1967,Nicolis1977}. The emergence of coherent spatio-temporal motifs and self-organized phenomena seem thus ubiquitous in natural and human-made systems. One of the most widespread mechanisms invoked to explain those behaviors, was proposed by Alan Turing in the framework of morphogenesis~\cite{turing1990chemical}. In a nutshell, Turing considered a two-species reaction-diffusion system exhibiting a spatial homogeneous equilibrium and proved that, under suitable conditions, any arbitrarily small heterogeneous perturbation can trigger a diffusion-driven instability that pushes the system towards a new stable, possibly inhomogeneous, solution, i.e., a pattern. Successively, Gierer and Meinhardt showed that the presence of one activator and one inhibitor species is required to have Turing instability~\cite{GiererMeinhardt}. Beyond the initially proposed framework, the Turing mechanism has been invoked to explain the emergence of patterns in several and diverse domains, such as, neuroscience~\cite{bressloff2002geometric}, genetics~\cite{kondo2009}, economics~\cite{helbing2009pattern}, public goods games~\cite{Wakano2009} and even robotics~\cite{slavkov2018morphogenesis}. Because of its simplicity and mathematical elegance, nowadays the mechanism known as Turing instability, represents a pillar of self-organized structures in complex systems~\cite{pastorsatorrasvespignani2010}.

Those applications involve continuous supports for the reaction-diffusion systems (the interested reader is invited to consult the recent review about Turing pattern on continuous supports~\cite{KGMK2021}), however there are many empirical relevant cases where the underlying space has an intrinsically discrete structure (see~\cite{muolo2024review} for a recent review of Turing patterns on discrete supports), the latter being modeled as a complex network where reactions occur in the nodes and diffusion across the links~\cite{AlbertBarabasi,Strogatz01Nature,BBV2008,newmanbook}. Turing theory has then been improved to fully exploit the knowledge about complex networks by Nakao and Mikhailov~\cite{nakao2010turing}. By applying the meta population hypothesis~\cite{May} nodes are assumed to be large enough to host a well-stirred population and the reactions among the constituting units to be correctly described by an ordinary differential equation (ODE); then species diffuse across the nodes of the network by exploiting the available links. One can prove that the onset of Turing instability ultimately relies on the property of the spectrum of the Laplacian matrix associated with the network, encoding Fickean diffusion of species. \teo{In conclusion Turing patterns on networks result from a diffusion driven instability that spontaneously induces a differentiation of the network nodes
into activator-rich and activator-poor groups, creating thus a heterogeneous solution~\cite{nakao2010turing}.} Starting from these ideas, Turing patterns have been studied on directed network~\cite{asllani2014theory}, non-normal networks~\cite{top_resilience,jtb,malbor_teo} and also on defective ones~\cite{dorchain2023pattern}, namely once the Laplacian matrix does not allow for an eigenbasis. More recently, Turing instability has been extended to higher-order structures such as simplicial complexes~\cite{giambagli2022diffusion,muolo2024three} and hypergraphs~\cite{carlettifanellinicoletti2020,muolo2023turing}, by also considering their directionality~\cite{DSMT2023}.

\begin{figure*}
    \centering
    \includegraphics[width=0.8\linewidth]{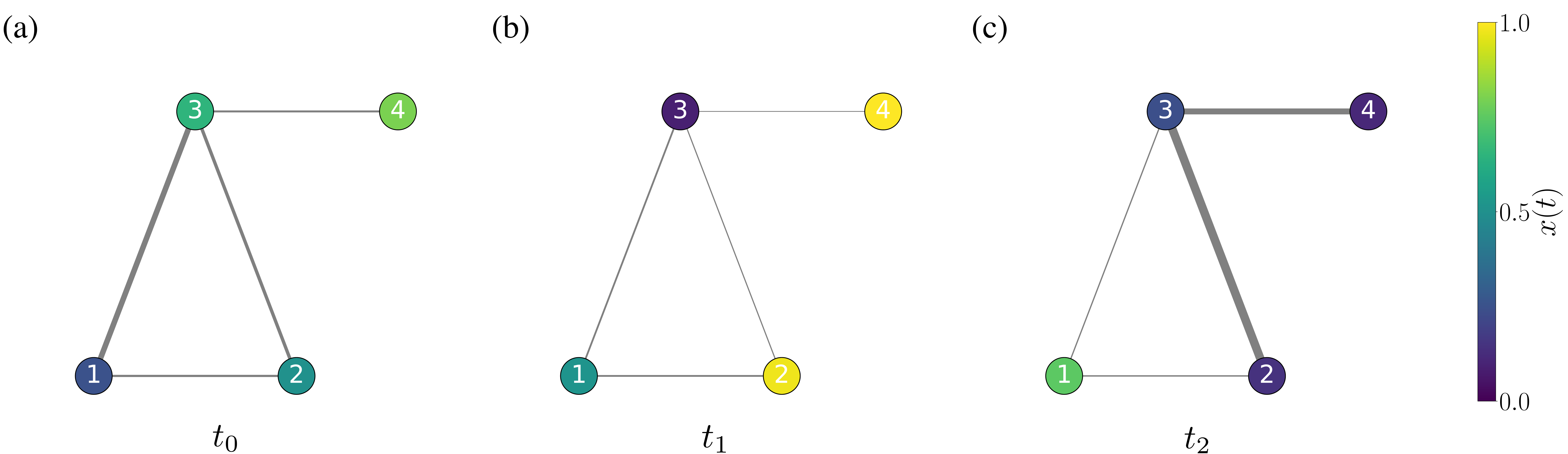}
    \caption{\jen{Cartoon representing an adaptive dynamical network. At $t=t_0$ (panel (a)), the nodes states, $x$, and the weights of the links are randomly assigned. At $t=t_1$ (panel (b)), the nodes states evolve so to adapt as a function of the sum of weights of the links they are connected to. In the presented example, the larger the weight, the larger the change in the node states, i.e., moving from dark blue to yellow as shown by the color bar. In response (panel (c) for $t=t_2$) the weights of the links change as a function of the difference between the states of the nodes they are connected to. In the present case, the smaller the difference, the stronger the connection, here represented by a thicker line. This realizes thus the feedback loop where the states affect the weights and vice versa.}}
    \label{fig:model-network}
\end{figure*}

In all the above cases, the network substrate is considered to be fixed, the number of nodes and the links weights do not change over time, or more precisely, their evolution is very slow with respect to natural time scale of the reaction terms that can thus be neglected. The opposite limit in which Turing patterns emerge in fast switching networks has been studied in~\cite{petit2017theory}, while in~\cite{carletti2022theory} the general theory of Turing instability on time-varying networks has been proposed, allowing thus to consider any rate for the reaction and the link dynamics. In those cases, however, the network dynamics is {\em exogenous}, namely links weight evolution is prescribed {\em a priori}~\cite{Holme2013,Holme2015,MR2016}. 
\mar{On the other hand, there are several phenomena in which the network structure evolves as a consequence of the node dynamics. Such a coevolution between node states and networks structure, e.g., link weights, creates a feedback loop, where the state of the nodes affects the connectivity of the network and the evolving link weights influence the node dynamics, it is thus an {\em endogenous} process. This class of systems is commonly referred to as adaptive or coevolutionary networks~\cite{GB2008,BGKKY2023}.}
\jen{A toy example is proposed in Fig.~\ref{fig:model-network}, here adaptation happens according to the ‘like-and-like’ process which strengthen connections between similar nodes~\cite{ito2001spontaneous}. We consider four nodes connected with four links; at time $t=t_0$ (panel (a)) we assign random values to node states and link weights. As time increases, they influence each other, node states evolve to new values as a function of the sum of weights of the links they are connected to. The larger the weight, the larger the change in the states, represented with a color code going from dark blue to yellow. In response, link weights change as a function of the difference between the states of the nodes they are connected to. The smaller the difference, the stronger the modification of the connection. In summary, node states affect the weights and vice versa, by creating a feedback loop.}
    
The prototype example of adaptive networks comes from brain dynamics, where synapses between neurons adjust their weight \teo{(synaptic efficacy)} based on some function of neuronal activity, \teo{e.g., relative difference of firing times}~\cite{MLFS1997,AN2000}. Epidemics or social systems are also relevant examples, where agents decide to create/destroy or strengthen/weaken connections~\cite{GDLDB2006,GB2008} based on the state of their neighbors, e.g., susceptible agents try to avoid infected ones, \teo{in turn stronger/weaker connections allow easier/more difficult transfer of opinion, information or virus}. There are also examples of chemical systems where reaction rates dynamically adapt according to the system variables~\cite{JK2001}. More generally, adaptive networks have attracted a lot of attention from scholars interested in synchronization~\cite{ZK2006,AA2009,TTSIetal2010,AA2011} and, to the best of our knowledge, a study about the emergence of Turing patterns on adaptive networks is lacking. \teo{We refer the interested reader to consult the reviews~\cite{GB2008,BGKKY2023} where they could find many more examples and applications of adaptive dynamical networks.}

Because of the relevance of Turing mechanism and the importance of adaptive networks, we propose in this work to fill the gap and provide a general theory of Turing instability on adaptive networks. We proposed two applications to emphasize the richness of the phenomena we can obtain, by considering the Brusselator~\cite{PrigogineNicolis1967} and the FitzHugh-Nagumo model~\cite{FitzHugh1961,Nagumo1962,Rinzel1973} as generic dynamical systems. The latter are defined on top of complex networks, whose edges are weighted and evolve in time as a function of the states of their end nodes. \mar{More precisely, our framework belongs to the class of adaptive weighted networks~\cite{BGKKY2023}, in which the underlying graph remains fixed while the link weights dynamically evolve according to the state of adjacent nodes.} Let us observe that weights evolution modifies the strength of the interaction among nodes and thus it impacts the diffusion across nodes, but there is not any deformation of the network. Similarly, nodes do not have a ``physical volume'' that changes in time, thus our model does not account for dilution or increase in concentration due to nodes shrinking or growth. We considered several examples of adaptive response functions, inspired by empirical observations and results available in the literature; in the following, we define the parabola-like and the cubic-like adaptive response, the former resembles the Hebbian response while the latter one is motivated by spike-timing dependent plasticity (STDP)~\cite{AA2009,AA2011,berner2019hierarchical}. Let us however stress that the framework is general enough to account for many other adaptive response functions. 

For the sake of simplicity and to emphasize the role of adaptation, we decided to work in the framework of Turing patterns emerging from a stationary stable equilibrium, let us however observe that a similar study, but with a more involved analysis, could be done in the case a Turing-like instability emerging from a limit cycle~\cite{CBF2015}. Our analysis relies on the study of the spectrum of the Laplacian matrix associated to the network and allows to determine conditions for the emergence of Turing patterns depending on models parameters and on the network topology as well. Let us observe that we do not impose any restriction on the symmetric network, except to have positive weights, and thus our results strengthen the interplay between dynamics of networks and the impact of the topology on the former. We have been able to show the existence of stationary, wave-like Turing patterns but also of ``bursty'' Turing patterns, namely the system tends to reach the homogeneous equilibrium, i.e., absence of patterns, but then it suddenly moves apart by creating a sort of spiking patterns, to eventually repeat the two phases indefinitely. For some adaptive response functions, we have shown that several link weights go to zero and thus the number of available links in the network is strongly reduced. In some cases, this process can break the network into small pieces; interestingly enough, those small modules are capable to sustain Turing patterns, otherwise impossible on such small groups of connected nodes. Stated differently, the links adaptation slowly brings the system to a pattern with nodes hosting different species densities, then some links are removed and the species remain trapped into this patchy configuration.

The work is organized as follows. In Section~\ref{sec:model} we present the general framework of Turing patterns on adaptive networks and we develop the theory ensuring the emergence of such heterogeneous solutions. Section~\ref{sec:numrel} is devoted to the presentation of some numerical results to corroborate the general theory and to emphasize the richness of the possible phenomena. Eventually in Section~\ref{sec:conclusions} we summarize our conclusions.

\section{The model}
\label{sec:model}

Let us consider a symmetric network made of $n$ nodes whose links are encoded by the $n\times n$ binary adjacency matrix $\mathbf{A}$, such that $a_{ij}=1$ if and only if there is a link connecting nodes $i$ and $j$. Moreover we assume links to be positively weighted, namely there exists a time-varying matrix $\mathbf{W}(t)\in\mathbb{R}_+^{n\times n}$ such that $w_{ij}(t)>0$ if and only if $a_{ij}=1$. Let us eventually define the weighted, time-varying, adjacency matrix $\mathcal{A}$, whose entries are given by $\mathcal{A}_{ij}=a_{ij}w_{ij}$.

Let $s_i=\sum_j\mathcal{A}_{ij}$ be the strength of node $i$ and let $\mathcal{L}$ be the Laplacian matrix of the weighted time-varying network
\begin{equation}
    \label{eq:Laplwt}
    \mathcal{L}_{ij}=\mathcal{A}_{ij}-\delta_{ij}s_i\, . 
\end{equation}
\teo{Let us emphasize that the matrices $\mathcal{L}$ and $\mathcal{A}$ are functions of the weights $w_{ij}$ and thus they vary because of the time dependence of the weights, however to lighten the notations we prefer not adopt the notation $\mathcal{L}(w(t))$ and $\mathcal{A}(w(t))$.} We can then define a reaction-diffusion system involving two species, $u$ and $v$, reacting on nodes and diffusing via a diffusive, time-dependent, coupling, more precisely
\begin{equation}
\label{eq:modeluv}
\begin{cases}
\displaystyle \frac{du_i}{dt} = f(u_i,v_i)+D_u\sum_j \mathcal{L}_{ij} u_j\\
\displaystyle \frac{dv_i}{dt} = g(u_i,v_i)+D_v\sum_j \mathcal{L}_{ij} v_j\quad \forall i=1,\dots, n\, ,
\end{cases}
\end{equation}
where $f$ and $g$ are two nonlinear functions describing the reaction occurring on each node, $u_i(t)$, resp. $v_i(t)$, the density of species $u$, resp. $v$, in the $i$-th node at time $t$; $D_u>0$ (resp. $D_v>0$) is the diffusion coefficient of species $u$ (resp. $v$). We moreover assume the existence of a stable homogeneous equilibrium, $u_i=u^*$ and $v_i=v^*$ for all $i\in\{1,\dots,n\}$ such that
\begin{equation}
\label{eq:fgzero}
f(u^*,v^*)=g(u^*,v^*)=0\, ,
\end{equation}
and the Jacobian matrix
\begin{equation}
\label{eq:J0}
\mathbf{J}_0=\left(
\begin{matrix}
 \partial_ u f & \partial_v f\\ \partial_ u g & \partial_v g
\end{matrix}
\right)\Big\rvert_{u^*,v^*}\, ,
\end{equation}
is stable, namely its spectrum contains only eigenvalues with negative real part.

\subsection{Adaptive Mechanisms}
\label{ssec:adaptmech}
\begin{figure}[ht]
\centering
\includegraphics[scale=0.25]{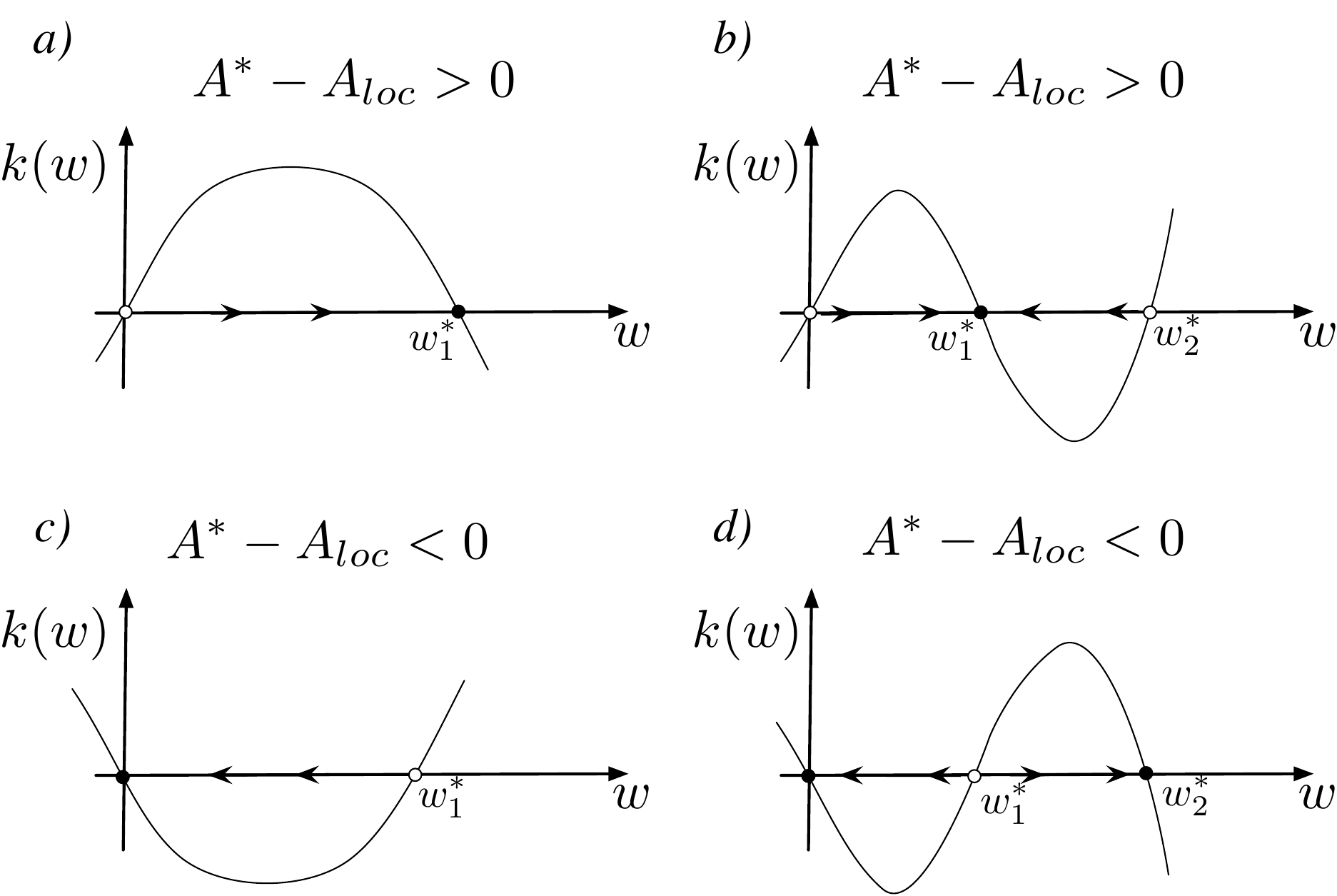}
\caption{Cartoon representing four generic behaviors for the nonlinear adaptive response function ruling the weights evolution. Panel a) quadratic-like with positive concavity, $w_1^*$ is a stable fixed point (black circle); panel c) quadratic-like with negative concavity, $w_1^*$ is an unstable fixed point (white circle). Panel b) cubic-like with signature ``$+\, -$'', $w_1^*$ is a stable fixed point (black circle); panel d) cubic-like with signature ``$-\, +$'', $w_1^*$ is an unstable fixed point (white circle). In each panel $A_{\mathit{loc}}$ is the local patterns amplitude and the arrows on the horizontal axis denote the time evolution of weights $w_{ij}(t)$ according to Eq.~\eqref{eq:modelw3a}, i.e., the arrows point in the direction of increasing weights.}
\label{fig:Mod3}
\end{figure}

The weights of the links can evolve by following different adaptive mechanisms; however in many applications the behavior of a given link is assumed to depend on the state of its end nodes, this is the approach we will hereby also adopt. In the realm of Turing patterns a quantity of high relevance is the patterns amplitude defined as the deviation of the system state with respect to the homogeneous equilibrium, namely
\begin{equation}
\label{eq:Amp}
A(\vec{u},\vec{v})=\frac{1}{2n}\sum \left[(u_i(t)-u^*)^2+(v_i(t)-v^*)^2\right]\, ,
\end{equation}
where $\vec{u}=(u_1,\dots,u_n)^\top$, resp. $\vec{v}=(v_1,\dots,v_n)^\top$, denotes the state vector. The latter is however a global quantity, to fit thus in the above described scheme where the evolution of $w_{ij}$ is driven by local information from nodes $i$ and $j$, we introduce a {\em local patterns amplitude}~\footnote{Let us observe that the dynamics obtained by using the global amplitude~\eqref{eq:Amp} is very similar to the one resulting from the local one~\eqref{eq:Aloc} as we show in Appendix~\ref{sec:globAmpli}}
\begin{widetext}
\begin{equation}
\label{eq:Aloc}
A_{\mathit{loc}} (u_i,u_j,v_i,v_j)= \frac{(u_i(t)-u^*)^2+(u_j(t)-u^*)^2+(v_i(t)-v^*)^2+(v_j(t)-v^*)^2}{4}\, .
\end{equation}
\end{widetext}
We can thus define the ODE ruling the links evolution by
\begin{equation}
\label{eq:modelw3a}
 \frac{dw_{ij}}{dt} = a_{ij} \left(A^*-A_{\mathit{loc}}(u_i,u_j,v_i,v_j)\right) k(w_{ij})\, ,
\end{equation}
where $k(w)$ is some nonlinear function. In the following, we will consider two main examples of adaptive response functions, $k(w)=w(w_1^*-w)$ and $k(w)=w(w_1^*-w)(w_2^*-w)$, $0<w_1<w_2$. In Fig.~\ref{fig:Mod3} we schematically display those functions, the stable equilibria are depicted by a black circle while the unstable ones by a white circle; the arrows on the horizontal axis denote the time evolution of the weights, $w_{ij}(t)$, given by~\eqref{eq:modelw3a}, in this way one can easily appreciate the qualitative behavior of the system. Let us observe that those function are inspired by the ones used in the literature dealing with Hebbian learning in neural systems, e.g., the parabola-like, and the change of synaptic weight due to neurons activity (STDP), e.g., the cubic-like. In Appendix~\ref{sec:anotheradapt} a third example of adaptive response function has been studied. Finally, in Eq.~\eqref{eq:modelw3a}, $A^*\geq 0$ is a control parameter that allows the adaptation to switch from Hebbian, or STDP, to anti-Hebbian, or anti-STDP, and to induce a sort of frustration in the dynamics, as we hereby explain. 

The rationale of the equation for $dw_{ij}/dt$ is the following: if the densities in nodes $i$ and $j$ are close enough to the homogeneous solution such that $A_{\mathit{loc}}  < A^*$, then the rate of change for the weights is given by the sign of $k(w)$, e.g., $w_{ij}$ will increase toward $w_1^*$ in the parabola-like case (or approaches $w_1^*$ from both sides in the cubic-like case), on the other hand, if $A_{\mathit{loc}}>A^*$, then $w_{ij}$ will move away from $w_1^*$, being the derivative negative. Observe that the former case, i.e., the parabola-like, recalls the Hebbian learning rule used in neuroscience, or the anti-Hebbian one; namely for a pair of neurons with similar phases, the plasticity function returns a large positive value, hence the weight between them will be strengthened; on the other hand, weights between neurons with well separated phases will be weakened. The latter case, i.e., the cubic-like behavior, can be associated with the spike-timing dependent plasticity (STDP)~\cite{AA2009,AA2011}.

\subsection{Turing instability for adaptive systems}
\label{ssec:Tinstadapt}

The reaction-diffusion system defined on an adaptive network is thus described by the set of equations
\begin{equation}
\label{eq:modeluvw}
\begin{cases}
\displaystyle \frac{du_i}{dt} = f(u_i,v_i)+D_u\sum_\ell \mathcal{L}_{i\ell} u_\ell\\
\displaystyle \frac{dv_i}{dt} = g(u_i,v_i)+D_v\sum_\ell \mathcal{L}_{i\ell} v_\ell\\
\displaystyle  \frac{dw_{ij}}{dt} = a_{ij} \left(A^*-A_{\mathit{loc}}(u_i,u_j,v_i,v_j)\right) k(w_{ij})\, ,
\end{cases}
\end{equation}
for all $i$ and $j$ in $\{1,\dots,n\}$. Let us observe that $A_{\mathit{loc}}(u^*,u^*,v^*,v^*)=0$ and $\sum_\ell \mathcal{L}_{i\ell} (w(t))=0$ for all $t$, hence $u_i=u^*$, $v_i=v^*$ and $w_{ij} =w^*_1$ for all $i$ and $j$ is a stable equilibrium of~\eqref{eq:modeluvw}. Let us stress that such homogeneous equilibrium always exists because of the property of the Laplacian matrix and the existence of the equilibrium~\eqref{eq:fgzero}. To prove the existence of a Turing instability, we must show that the latter equilibrium becomes unstable in the presence of heterogeneous perturbation and suitable diffusion coefficients $D_u$ and $D_v$. To prove this claim, we perform a linear stability analysis of~\eqref{eq:modeluvw} close to this equilibrium; we hence introduce small perturbations, $\delta u_i=u_i-u^*$, $\delta v_i=v_i-v^*$ and $\delta w_{ij}=w_{ij}-w_1^*$, and by retaining only linear terms we get from~\eqref{eq:modeluvw}
\begin{equation}
\label{eq:model3deltauv2}
\begin{cases}
\displaystyle \frac{d\delta u_i}{dt} = \partial_u f \delta u_i+\partial_v f \delta v_i+D_u w_1^*\sum_j L_{ij}\delta u_j\\
\displaystyle \frac{d\delta v_i}{dt} = \partial_u g \delta u_i+\partial_v g \delta v_i+D_v w_1^* \sum_j L_{ij}\delta v_j\\
\displaystyle \frac{d\delta w_{ij}}{dt} = a_{ij}A^*\partial_{w}k\delta w_{ij}\, ,
\end{cases}
\end{equation}
where the derivatives of $f$ and $g$ are evaluated at $(u^*,v^*)$ while the derivatives of $k$ at $w_1^*$, and we introduced the Laplacian matrix of the underlying unweighted static network
\begin{equation}
\label{eq:Lap}
L_{ij} =a_{ij}-\delta_{ij}\sum_\ell a_{i\ell}\, .
\end{equation}
{\color{black}To prove how to obtain ~\eqref{eq:model3deltauv2} from~\eqref{eq:modeluvw} let us consider the definition of $\mathcal{L}(w)$ evaluated on $w_{ij}=w_1^*+\delta w_{ij}$:
\begin{eqnarray*}
 \mathcal{L}_{ij}&=&a_{ij}(w_1^*+\delta w_{ij})-\delta_{ij}\sum_\ell a_{i\ell}(w_1^*+\delta w_{i\ell})\\
 &=&a_{ij}w_1^*-w_1^*\delta_{ij}\sum_\ell a_{i\ell}+\delta w_{ij} a_{ij}-\delta_{ij}\sum_\ell a_{i\ell}\delta w_{i\ell}\\
 &=&w_1^*L_{ij} +\delta w_{ij} a_{ij}-\delta_{ij}\sum_\ell a_{i\ell}\delta w_{i\ell}, .
\end{eqnarray*}
Inserting the latter into Eq.~\eqref{eq:modeluvw}, where $u_i=u^*+\delta u_i$, we eventually get $\sum_j\mathcal{L}_{ij}\delta u_j =  w_1^*\sum_j {L}_{ij} \delta u_j$, by neglecting terms in $\delta u_j \delta w_{ij}$. A similar result can be clearly obtained for $\sum_j\mathcal{L}_{ij}\delta v_j$.}

Let us emphasize that the obtained linearized system~\eqref{eq:model3deltauv2} is governed by constant coefficients, since the Laplacian matrix reduces to the one of the underlying static network evaluated at equilibrium weights and also the Jacobian of the linear part has been evaluated on the equilibrium $(u^*,v^*)$. As a consequence, the linearized dynamics is autonomous, which allows us to use standard spectral analysis to infer about the stability or lack thereof.
Let us observe that $A_{\mathit{loc}}$ has a second order zero at the homogeneous equilibrium hence it does not contribute to the first order expansion. By introducing the stack vectors  $\delta \vec{x}=(\delta u_1,\delta v_1,\dots,\delta u_n,\delta v_n,)^\top$ and $\delta \vec{w}=(\delta w_{11},\dots,\delta w_{1n},\delta w_{21},\dots,\delta w_{2n},\dots,\delta w_{n1},\dots,\delta w_{nn})^\top$, the latter system of equations can be written in compact matrix form
\begin{equation}
\label{eq:model3deltauvMat2}
\frac{d}{dt}\left(
\begin{matrix}
 \delta \vec{x}\\\delta\vec{w}
\end{matrix}\right)=\left(
\begin{matrix}
 \mathbf{S} & \mathbf{O}\\\mathbf{O}^\top & A^*\mathbf{A}'\partial_w k
\end{matrix}
\right)\left(
\begin{matrix}
 \delta \vec{x}\\\delta\vec{w}
\end{matrix}\right)\equiv \mathbf{M}\left(
\begin{matrix}
 \delta \vec{x}\\\delta\vec{w}
\end{matrix}\right)\,,
\end{equation}
$\mathbf{M}$ is thus a $(2n+n^2)\times (2n+n^2)$ block matrix, $\mathbf{A}'=\mathrm{diag}(a_{11},a_{12},\dots,a_{1n},\dots, a_{nn})$ is the $n^2$ diagonal matrix whose elements are the entries of the adjacency matrix $\mathbf{A}$ ``unrolled'', $\mathbf{O}$ is a rectangular matrix of size $2n\times n^2$ whose entries are all $0$. The matrix $\mathbf{S}$ is given by
\begin{equation}
\label{eq:matrixS}
\mathbf{S}=\mathbf{I}_n\otimes \mathbf{J}_0+w_1^*L\otimes \left(
\begin{smallmatrix}
 D_u & 0\\0 & D_v
\end{smallmatrix}\right)\, .
\end{equation}
The stability of system~\eqref{eq:model3deltauvMat2} is determined by the spectrum of the matrix $\mathbf{M}$ and because of its blocks shape the latter is the spectrum of $\mathbf{S}$ together with the eigenvalues $A^* \partial_w k(w_1^*)$ and $0$, the former has multiplicity the number of links present in the unweighted static network, i.e., $E=\sum_{ij}a_{ij}/2$, while the latter has multiplicity $n^2-2E$. 

Let us thus study the spectrum of $\mathbf{S}$. To make a step forward let us introduce the eigenbasis of the Laplacian matrix $\mathbf{L}$, namely $\mathbf{L}\vec{\phi}^{(\alpha)} = \Lambda^{(\alpha)}\vec{\phi}^{(\alpha)}$, $\alpha=1,\dots, n$. Because of the property of the Laplacian matrix we have  $\Lambda^{(1)}=0$, $\vec{\phi}^{(1)}\sim (1,\dots,1)^\top$, $\Lambda^{(\alpha)}<0$ for all $\alpha >1$ and $\vec{\phi}^{(\alpha)}\cdot \vec{\phi}^{(\beta)}=\delta_{\alpha\beta}$. Let us eventually define the matrix $\mathbf{V}$ given by
\begin{equation*}
 \mathbf{V}=\left(
\begin{matrix}
 \Phi\otimes \mathbf{I}_2 & \mathbf{O}\\\mathbf{O}^\top & \mathbf{I}_{n^2}
\end{matrix}\right)\, ,
\end{equation*}
where $\Phi$ is the $n\times n$ matrix whose columns are the eigenvectors $\vec{\phi}^{(\alpha)}$, namely $\Phi^\top\mathcal{L}\Phi=\mathrm{diag}(0,\Lambda^{(2)},\dots,\Lambda^{(n)})\equiv \mathbf{\Lambda}$. One can thus obtain
\begin{equation*}
 \mathbf{V}^\top \mathbf{M}\mathbf{V} = \left(
\begin{matrix}
 \mathbf{I}_n\otimes \mathbf{J}_0+w_1^*\mathbf{\Lambda}\otimes \left(
\begin{smallmatrix}
 D_u&0\\0& D_v
\end{smallmatrix}\right) & \mathbf{O}\\\mathbf{O}^\top & \mathbf{C}''
\end{matrix}\right)\, ,
\end{equation*}
where $\mathbf{C}''$ is the $n^2\times n^2$ diagonal matrix whose elements are $A^* \partial_w k(w_1^*)$ repeated $E$ times and $0$ repeated $n^2-2E$ times. The spectrum of $\mathbf{S}$ is obtained by solving the $n$ characteristic problems
\begin{equation}
\label{eq:reldispeq}
\det \left[ \mathbf{J}_0+w_1^*\Lambda^{(\alpha)}\left(
\begin{smallmatrix}
 D_u&0\\0& D_v
\end{smallmatrix}\right)-\lambda \mathbf{I}_2\right]=0\quad \alpha=1,\dots,n\, ,
\end{equation}
from which one can determine the dispersion relation, $\rho(\Lambda^{(\alpha)})=\max \Re\lambda_{\pm}(\Lambda^{(\alpha)})$ where
\begin{widetext}
\begin{equation*}
 \lambda_{\pm}(\Lambda^{(\alpha)}) = \frac{\mathrm{tr}\mathbf{J}_0+w_1^*\Lambda^{(\alpha)}(D_u+D_v)\pm\sqrt{\left(\mathrm{tr}\mathbf{J}_0+w_1^*\Lambda^{(\alpha)}(D_u+D_v)\right)^2-4 \Delta}}{2}\, ,
\end{equation*}
with $\Delta = \left[\det \mathbf{J}_0+w_1^*\Lambda^{(\alpha)}\left(D_u \partial_v g+D_v\partial_u f\right)+(w_1^*)^2D_uD_v(\Lambda^{(\alpha)})^2\right]$\, .
\end{widetext}
 Let us recall that the equilibrium $(u^*,v^*)$ is stable, hence $\mathrm{tr}\mathbf{J}_0<0$, moreover $w_1^*>0$ and $\Lambda^{(\alpha)}\leq 0$, from which we can conclude that 
\begin{equation*}
 \mathrm{tr}\mathbf{J}_0+w_1^*\Lambda^{(\alpha)}(D_u+D_v)<0\, .
\end{equation*}
To study the sign of the above roots, hence of the dispersion relation, we replace in Eq.~\eqref{eq:reldispeq}, $\Lambda^{(\alpha)}$ with a continuous non-positive variable, say $x$, and thus consider the function $x\mapsto\rho(x)$, that we still name dispersion relation. Then $\rho(x) >0$ if $\Delta <0$, namely
\begin{equation}
\label{eq:TIcond}
\begin{cases}
\displaystyle \det \mathbf{J}_0 - \frac{(D_u\partial_v g+D_v\partial_u f)^2}{4D_uD_v}<0\\
\displaystyle D_u\partial_v g+D_v\partial_u f>0 \, ,
\end{cases}
\end{equation}
Let us remember that the homogeneous equilibrium $(u^*,v^*)$ should be stable, hence $\mathrm{tr}\mathbf{J}_0=\partial_u f +\partial_v g <0$ and $\det \mathbf{J}_0=\partial_u f \partial_v g-\partial_u g \partial_v f>0$. The former condition implies $\partial_u f \partial_v g<0$, namely we are facing to an inhibitor, $v$ and $\partial_v g<0$, and an activator, $u$ and $\partial_u f>0$. The latter result together with the second equation in~\eqref{eq:TIcond} allow to conclude that
\begin{equation*}
    D_v\partial_u f > -D_u \partial_v g \Rightarrow \frac{D_v}{D_u} > -\frac{\partial_v g}{\partial_u f}>1\, ,
\end{equation*}
where the last implication follows from $\partial_u f +\partial_v g <0$. In conclusion Turing instability can emerge once we are dealing with an inhibitor and an activator and the former diffuses faster than the second. We have thus found the same conditions determined by Gierer and Meinhardt~\cite{GiererMeinhardt}.

Let us now briefly describe the system behavior. By construction $\partial_w k(w_1^*)<0$ and thus links would converge to $w_1^*$, however let us assume condition~\eqref{eq:TIcond} to hold true, then the equilibrium $(u^*,v^*)$ is unstable, hence $A_{\mathit{loc}}$ increases and if it will overcome $A^*$ then the equilibrium $w_1^*$ will also become unstable (the adaptive response function will change its sign) and $w_{ij}(t)$ move away from $w_1^*$. If (sufficiently) many weights become very small then diffusion will be slowed down and patterns stop to emerge, $u_i$ and $v_i$ go back to $u^*$ and $v^*$. But this means that $A_{\mathit{loc}}$ could decrease below $A^*$ and thus $w_1^*$ turns out to be stable again,  links weights will increase and allow the concentrations $u_i$ and $v_i$ to depart from the equilibrium in a (possibly) never ending cycle. 

In conclusion we have found conditions for the onset of Turing instability~\eqref{eq:TIcond} similar to the ones we could determine for a static network~\cite{nakao2010turing}, however the network adaptation enters into the play via the parameter $A^*$, that could give rise to stationary patterns or oscillatory ones. In Appendix~\ref{sec:anotheradapt} we consider another adaptive mechanism and study the emergence of Turing instability in the FitzHugh-Nagumo model.

\section{Numerical results}
\label{sec:numrel}

The aim of this section is to corroborate the presented theory with some dedicated numerical simulations. The simulations hereby presented  have been done by using the standard Runge-Kutta 4th order method. For the sake of definitiveness, we decided to use the Brusselator model~\cite{PrigogineNicolis1967}, a prototype dynamical system exhibiting Turing patterns, largely used in the literature for this purpose~\cite{muolo2024review}. The general system~\eqref{eq:modeluvw} is then replaced by \teo{the following one, written by using non-dimensional variables and parameters}
\begin{eqnarray}
\label{eq:modeluvwBrus}
\begin{dcases}
 \frac{du_i}{dt} = 1 - (b+1)u_i + c u^2_{i}v_i +D_u\sum_j \mathcal{L}_{ij} u_j \\ 
 \frac{dv_i}{dt} = bu_i - cu^2_{i}v_{i}+D_v\sum_j \mathcal{L}_{ij} v_j \\
   \frac{dw_{ij}}{dt} = a_{ij} \left(A^*-A_{\mathit{loc}}(u_i,u_j,v_i,v_j)\right) k(w_{ij})\, ,
\end{dcases}
\end{eqnarray}
where $b>1$ and $c>0$ are model parameters and the equilibrium of the reaction part is given by $u^*=1$ and $v^*=b/c$. $D_u>0$, resp. $D_v>0$, is the diffusion coefficient of species $u$, resp. $v$. Only the link weights $w_{ij}$ evolve in time, while the network topology remains fixed, \teo{namely the number of nodes, of links and their position do not change. Let us stress that the Laplace matrix $\mathcal{L}$ is a function of the weights, hence the nodes dynamics is influenced by the latter (see first and second equations~\eqref{eq:modeluvwBrus}), on the other hand (see third equation~\eqref{eq:modeluvwBrus}) the weights evolution depends on the node variables. This feedback loop is the very definition of adaptive network}. 

In the following we will be interested in the adaptive dynamics, in particular the functional form of the adaptive response and the threshold parameter $A^*$, for this reason we fixed the remaining model parameters to some generic values, ensuring the emergence of Turing patterns as prescribed by Eqs.~\eqref{eq:TIcond}, more precisely we set $b = 4.0$, $c = 6.5$, $D_u = 0.07$ and $D_v = 0.7$. We also fix the size of the network to $n=20$ nodes.

Let us start our analysis by considering the parabola-like adaptive response (see panels a) and c) in Fig.~\ref{fig:Mod3}), more precisely we set
\begin{equation}
\label{eq:adaptrespparab}
k(w)=w(w_1^*-w) \text{ with $w_1^*=1/2$}\, .
\end{equation}
Following the Turing scheme, initial conditions for the reacting species $u$, resp. $v$, are drawn from the uniform distribution $U[-\delta,\delta]$ centered in $u^*$, resp. $v^*$; the initial weights are distributed uniformly at random in the interval $(w_1^*-\delta,w_1^*)$~\footnote{Let us observe that on purpose we do not initialize the weights beyond $w_1^*$, indeed in the case the latter is unstable, i.e., when $A^*<A_{loc}$, then those weights could grow unbounded.}, in both cases we set $\delta=0.01$. The underlying unweighted network is assumed to be all-to-all, namely $a_{ij}=1$ for all $i\neq j$, and $a_{ii}=0$.

\begin{figure}[ht]
\centering
\includegraphics[scale=0.18]{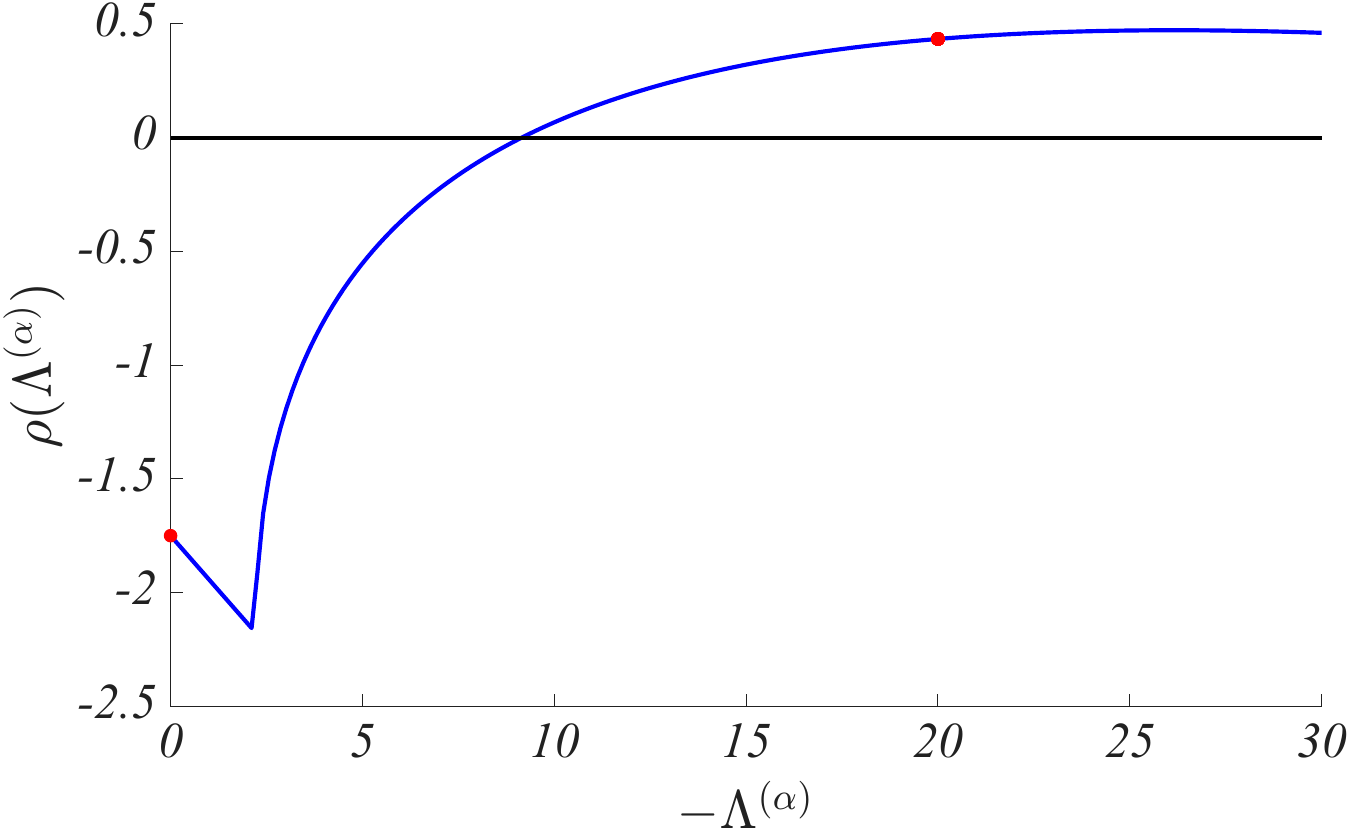}\includegraphics[scale=0.18]{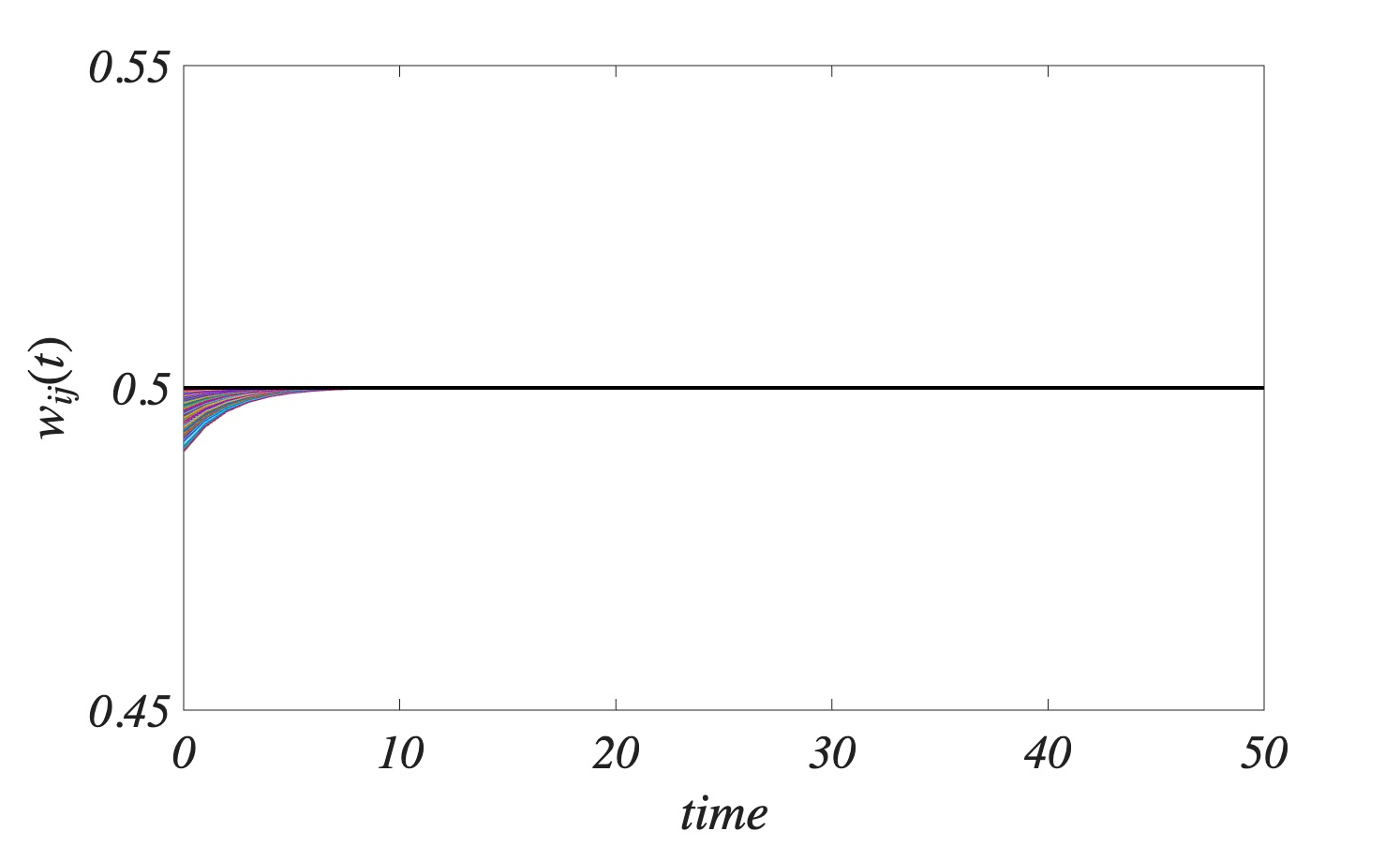}\\
\includegraphics[scale=0.18]{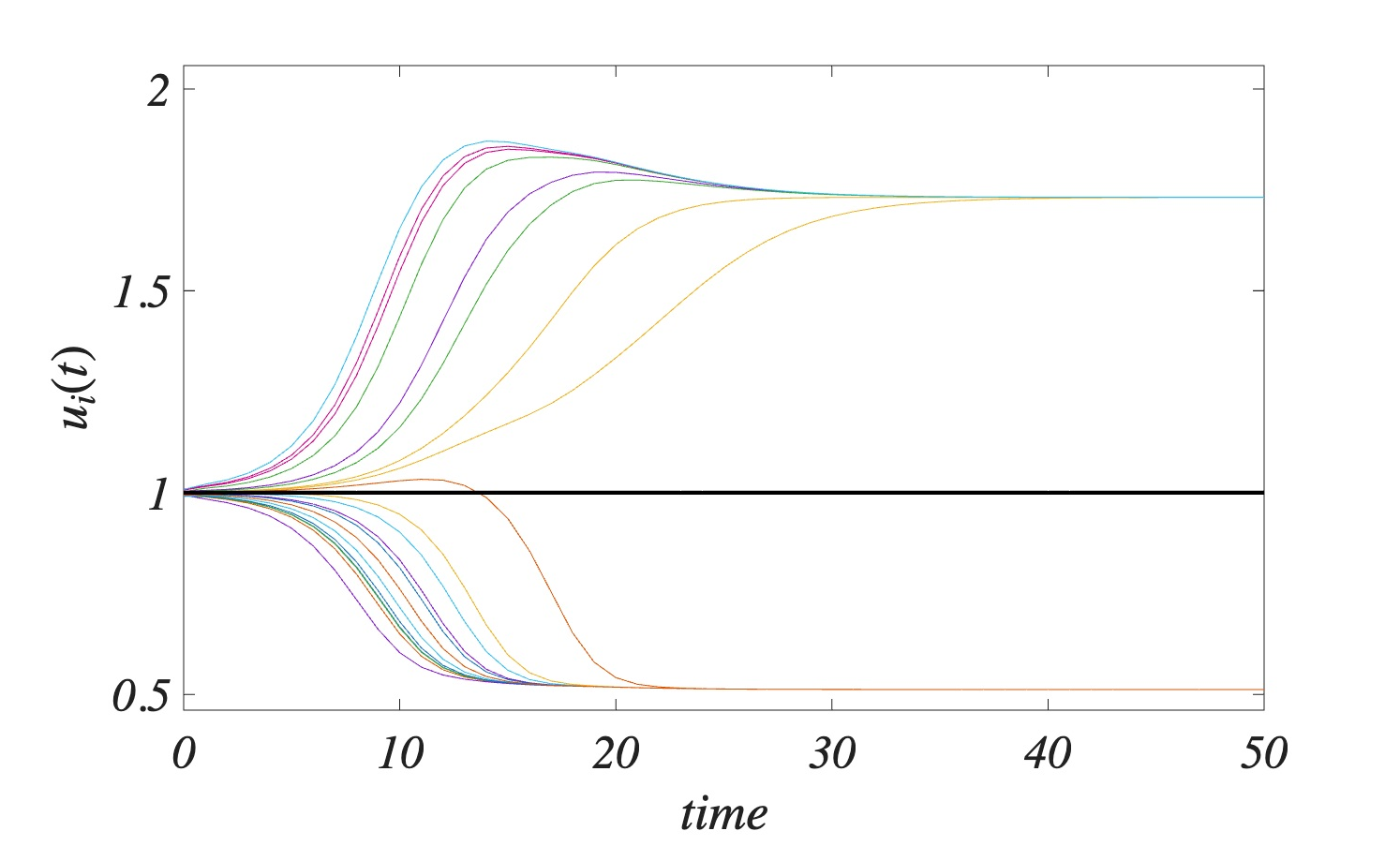}\includegraphics[scale=0.18]{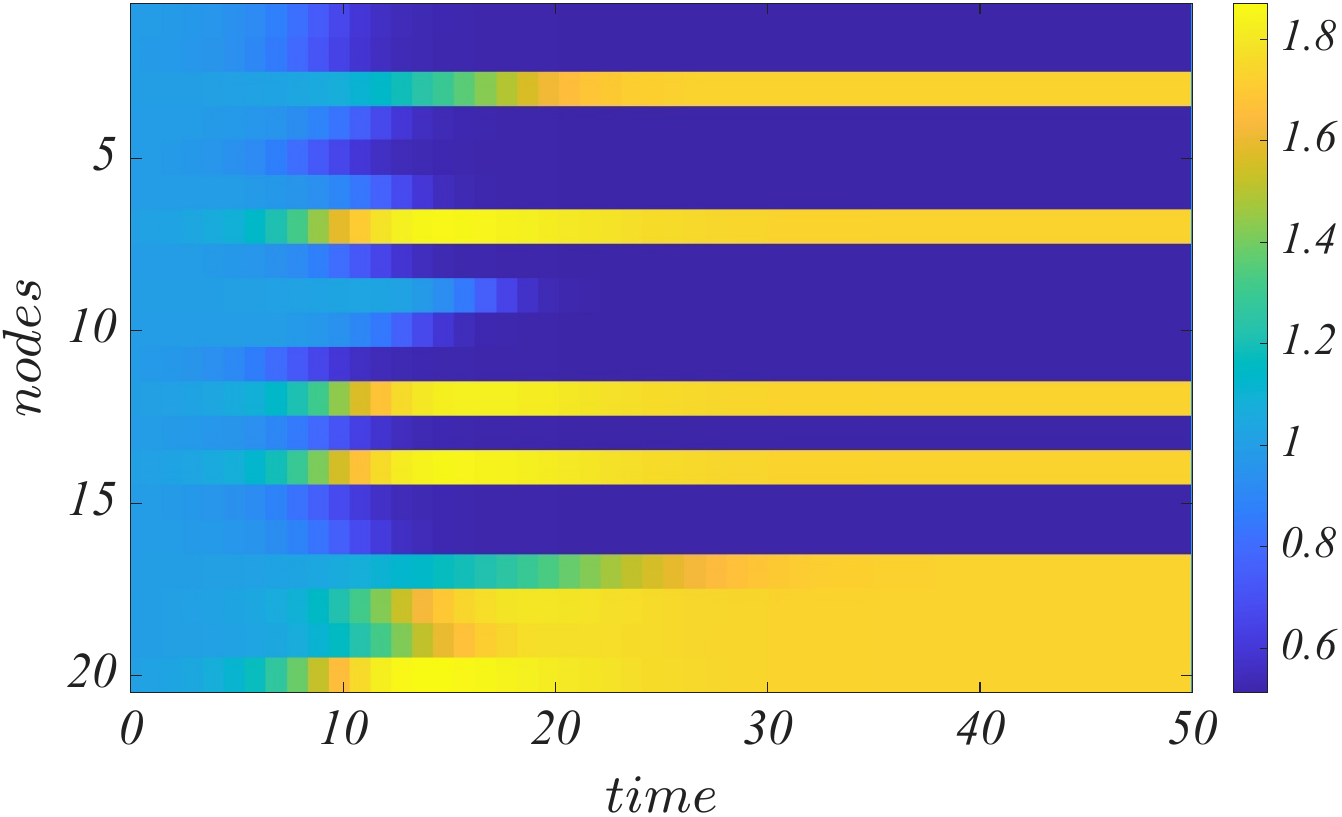}\\
\includegraphics[scale=0.18]{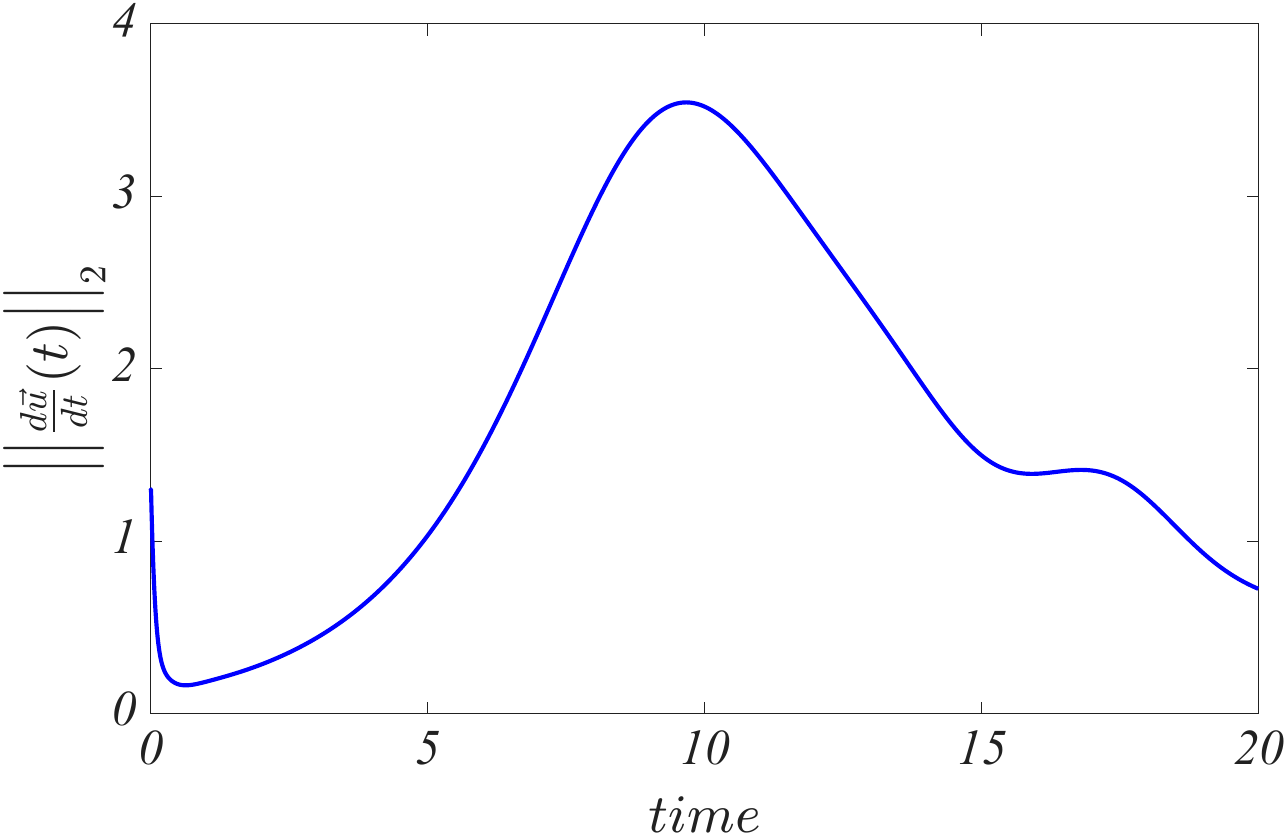}
\caption{Turing patterns with the adaptive response $k(w)=w(1/2-w)$ and $A^*=1.0$. Top left panel:  dispersion relation of the initial network, where red dots represent the dispersion relation evaluated on the discrete Laplacian spectrum, $\rho(\Lambda^{(\alpha)})$, while the blue curve stands for $\rho(x)$ and it has been drawn for ease of visualization. Top right panel: time evolution of the weights $w_{ij}(t)$. \teo{Middle} left panel: time evolution of species $u$, $u_i(t)$, \teo{Middle} right panel: space-time view of the time evolution of species $u$, the horizontal axis denotes time while the vertical one, space, i.e., nodes indexes, then the point with coordinates $(t,i)$ is colored according to the value of $u_i(t)$ (see colorbar). \teo{Bottom panel: time evolution of the (discrete) $L^2$ norm of $\frac{d\vec{u}}{dt}$}. \teo{See Supplementary Movie 1 to appreciate the dynamics of the adaptive network~\cite{SuppMovie1}.}}
\label{fig:Astarlargeparab}
\end{figure}

Let us consider first two extreme cases $A^*\gg 1$ and $A^*=0$. In the former, the inequality $A^*-A_{loc}>0$ is always satisfied, $w_1^*$ is thus an attractor and $w_{ij}(t)\rightarrow w_1^*$ as time increases, the network converges thus to a complete weighted network where all weights have value $w_1^*$. Because by assumption this network supports Turing patterns for the choice of the  parameters, the adaptive case will \teo{do the same; a numerical confirmation of this claim is shown in Fig.~\ref{fig:Astarlargeparab} where we display the positive dispersion relation (red dot in the top left panel), the time evolution of weights converging to the value $w_1^*=1/2$ (top right panel). The dynamics of species $u$ in each node (middle panels) and the $L^2$ norm of the velocity $d\vec{u}/dt$ (bottom panel)}
\begin{equation}
{\color{black}\Big|\Big| \frac{d\vec{u}}{dt}(t)\Big|\Big|_2=\sqrt{\sum_i \left(\frac{du_i}{dt}(t)\right)^2}}\, .
\end{equation}
\teo{Let us observe that the latter exhibits the typical signature of a Turing pattern, i.e., an initial decrease because of the diffusion process is followed by an exponential growth due to the Turing instability, namely the existence of unstable modes, to eventually decrease again once the pattern stabilizes.}

\begin{figure}[ht]
\centering
\includegraphics[scale=0.18]{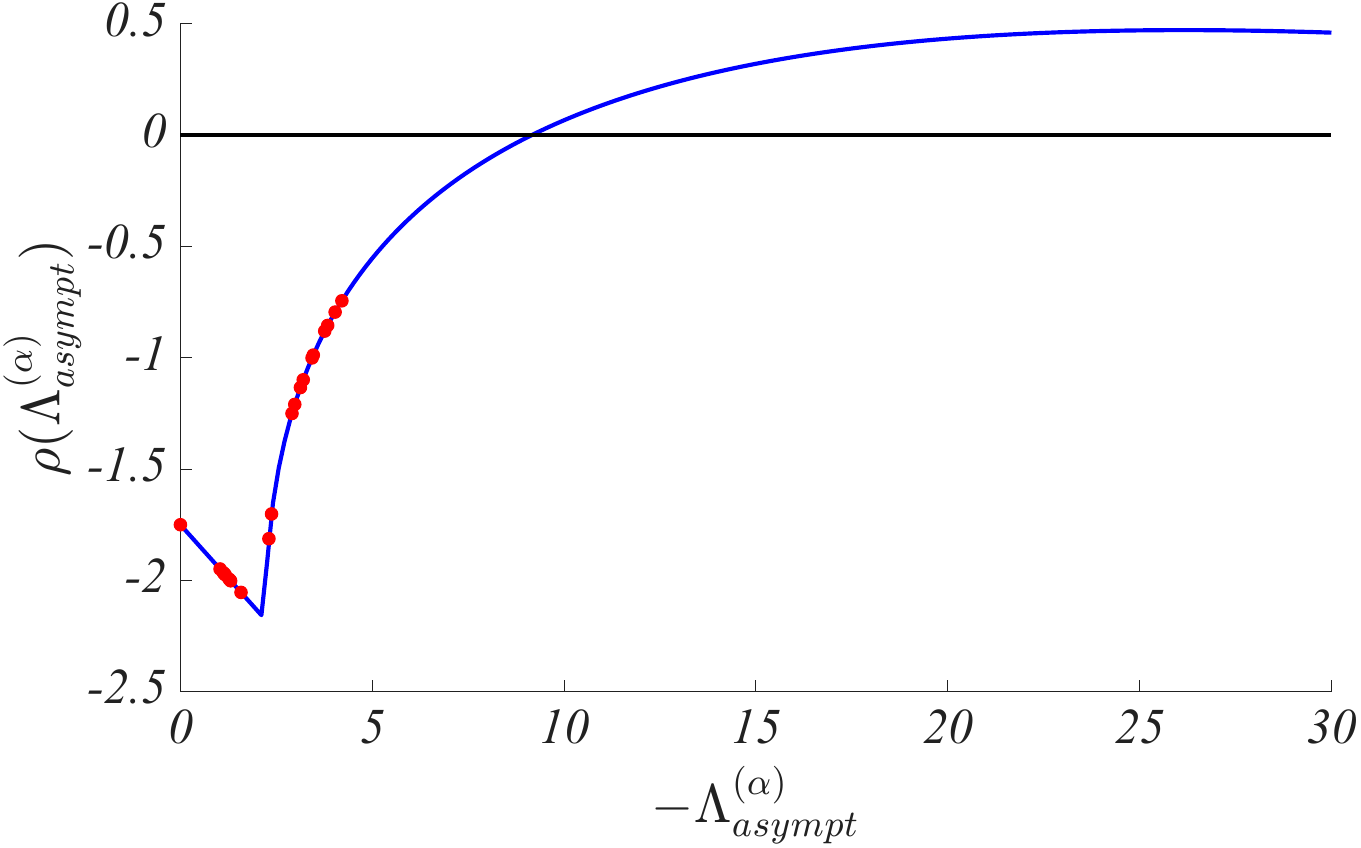}\includegraphics[scale=0.18]{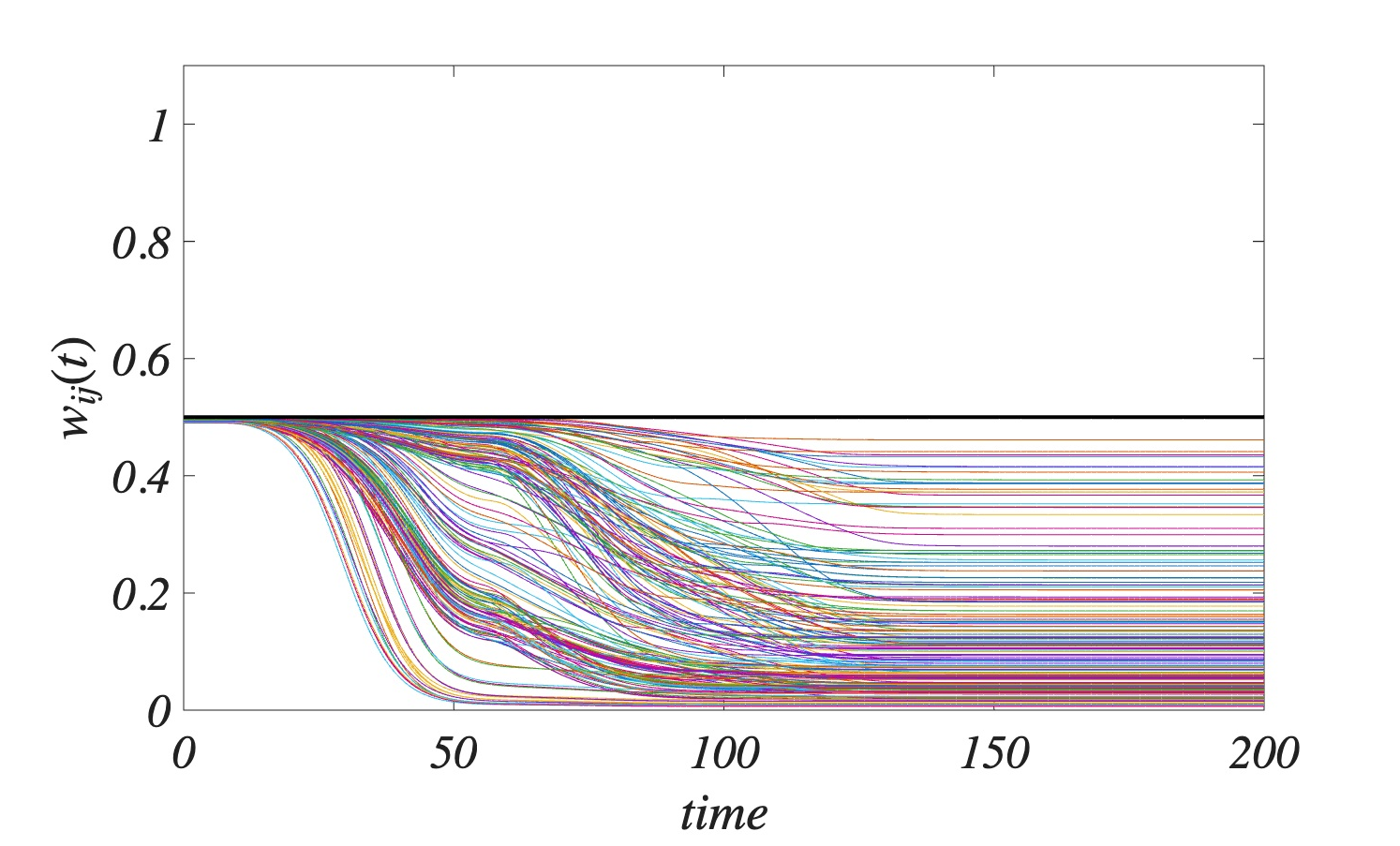}\\
\includegraphics[scale=0.18]{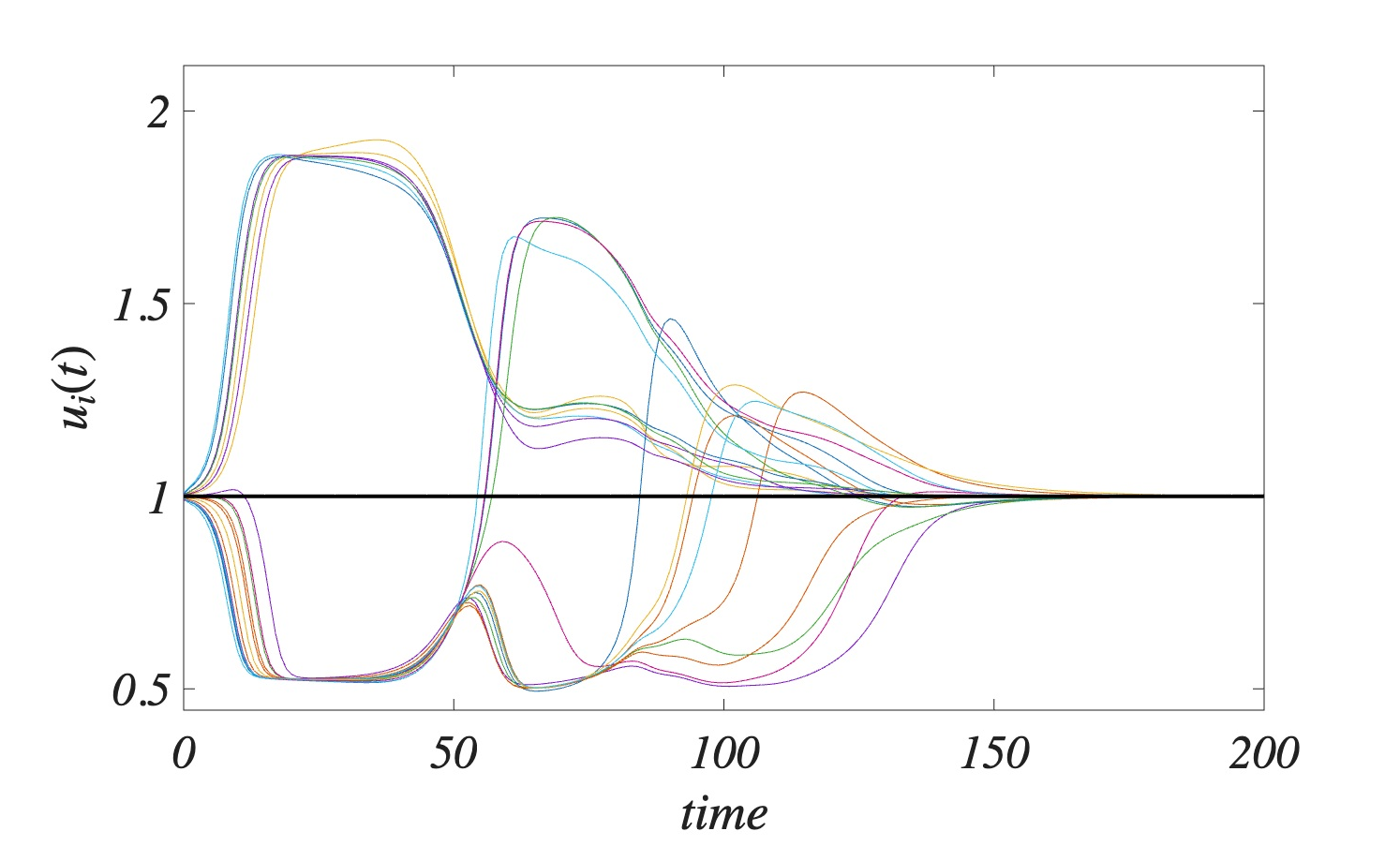}\includegraphics[scale=0.18]{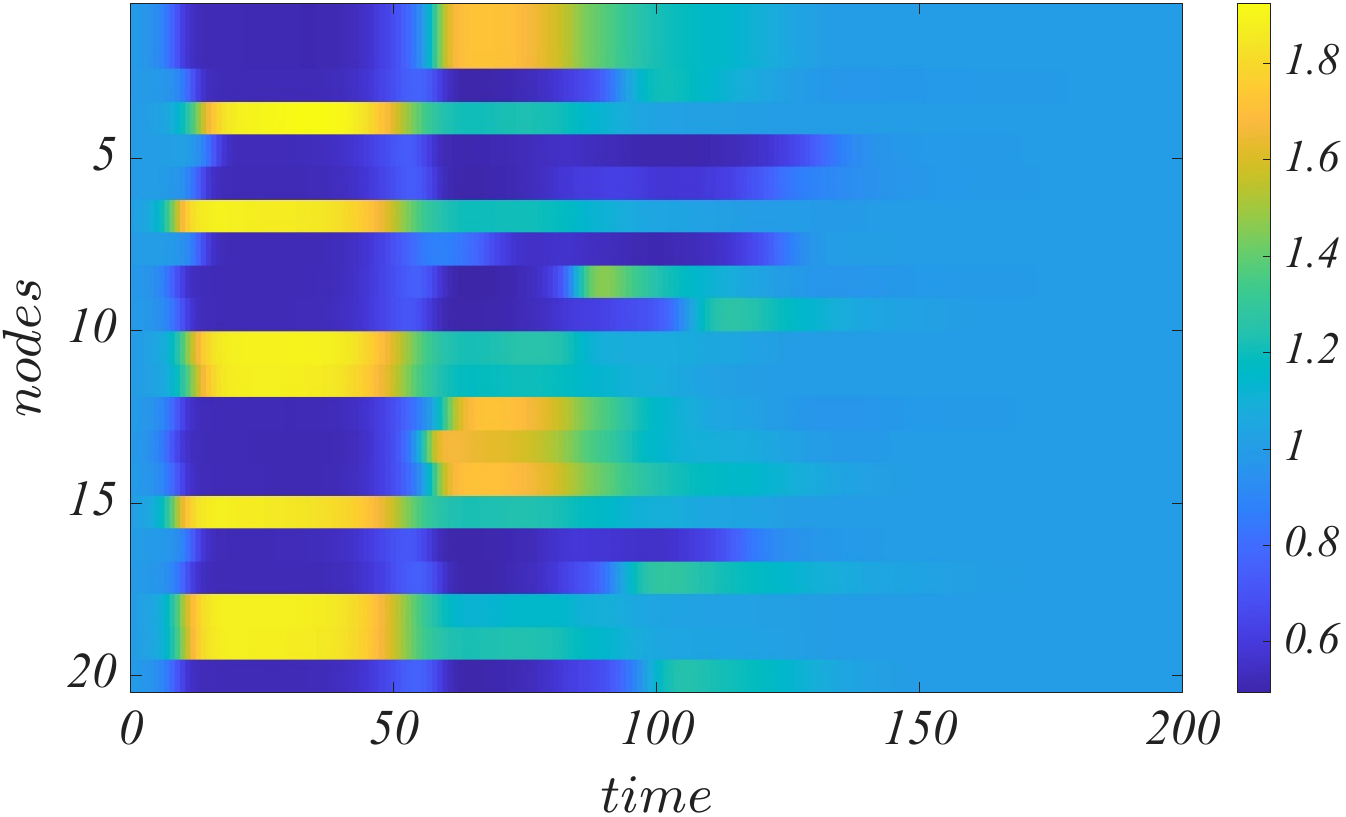}\\
\includegraphics[scale=0.18]{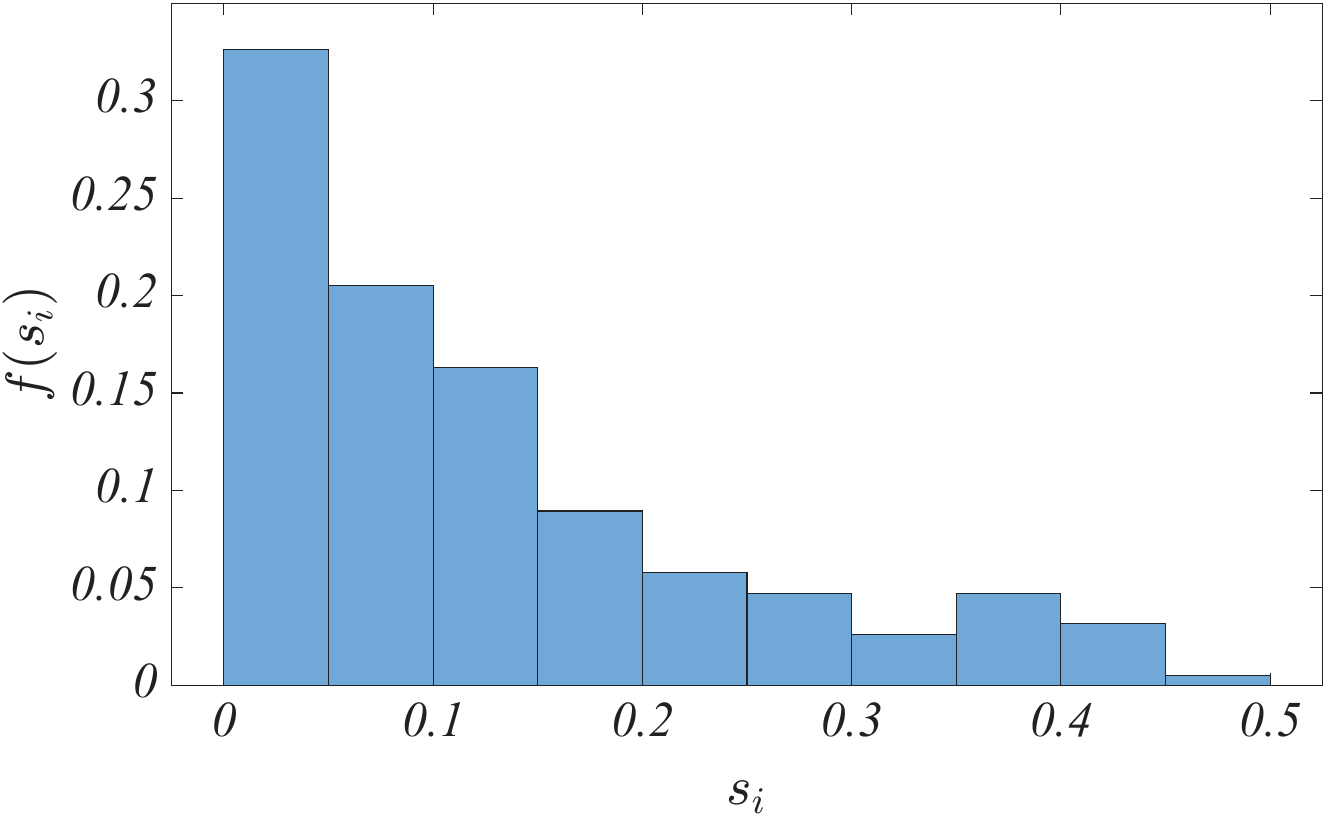}
\caption{Turing patterns with the adaptive response $k(w)=w(1/2-w)$ and $A^*=0.0$. Top left panel:  dispersion relation for the asymptotic network, the red dots denote the dispersion relation computed on the Laplacian eigenvalues, $\rho(\Lambda^{(\alpha)})$, while the blue curve stands for $\rho(x)$ and it has been drawn for ease of visualization. Top right panel: time evolution of the weights $w_{ij}(t)$. Bottom left panel: time evolution of species $u$, $u_i(t)$, bottom right: space-time view of the time evolution of species $u$, the horizontal axis denotes time while the vertical one, space, i.e., nodes indexes, then the point with coordinates $(t,i)$ is colored according to the value of $u_i(t)$ (see colorbar). Lower panel: frequency distribution of asymptotic links strength, $s_i(t)$ for $t\gg 1$. \teo{See Supplementary Movie 2 to appreciate the dynamics of the adaptive network~\cite{SuppMovie2}.}}
\label{fig:Astarsmallparab}
\end{figure}

In the second case, $A^*=0$, we trivially have $A^*-A_{loc}<0$, thus $w_1^*$ is a repeller and $w_{ij}(t)$ moves toward zero. Recall that at time $t=0$ the system satisfies the conditions for the emergence of Turing patterns, hence $u$ and $v$ move away from $u^*$ and $v^*$ while $w_{ij}(t)$ decrease, i.e., move away from $w_1^*$. Hence as time flows, the spectrum of the time-varying Laplacian matrix shrinks up to a value for which $\rho(\Lambda^{(\alpha)})<0$ for all $\alpha$, namely Turing patterns cannot develop anymore, $u_i(t)$ and $v_i(t)$ move back toward $u^*$ and $v^*$, thus the time evolution of $w_{ij}(t)$ slow down and the weights stop adapting, because $A_{loc}\rightarrow 0$. Eventually, we thus find a complete weighted network with heterogeneous weights not supporting Turing patterns (see Fig.~\ref{fig:Astarsmallparab}). We can thus conclude that in this case adaptation prevents the emergence of Turing patterns, that results to be transient.

\begin{figure*}[ht]
\centering
\includegraphics[scale=0.25]{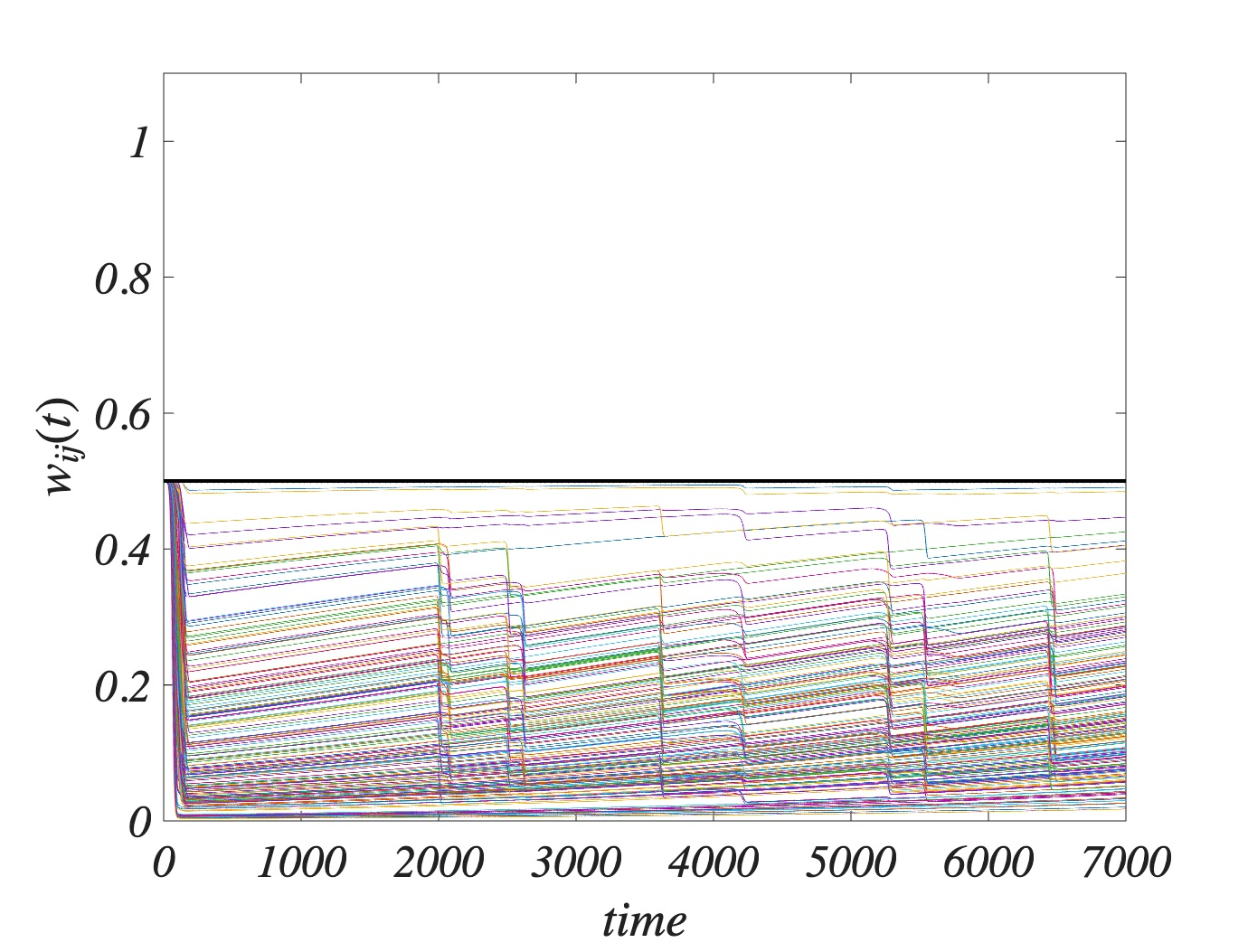}\includegraphics[scale=0.25]{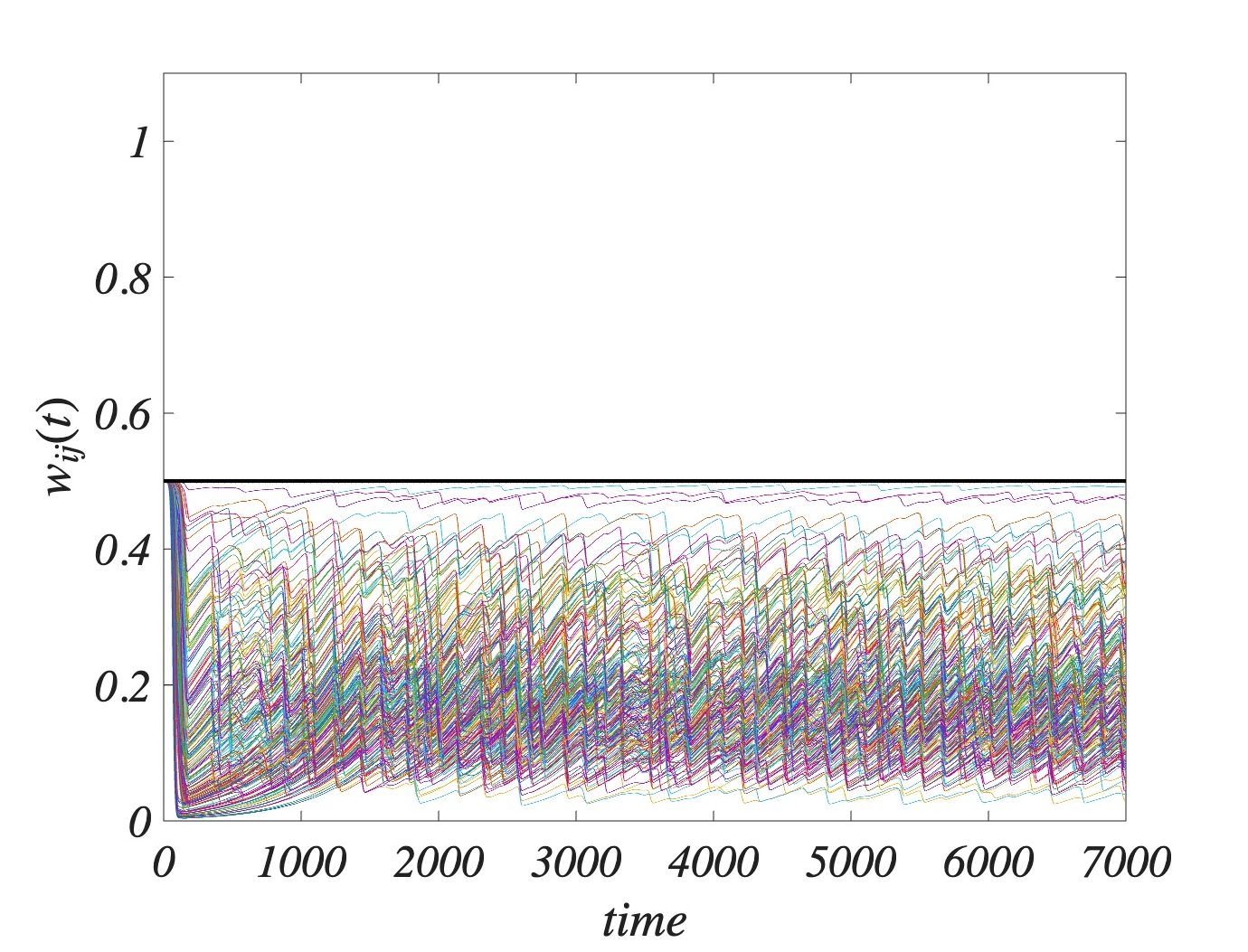}\includegraphics[scale=0.25]{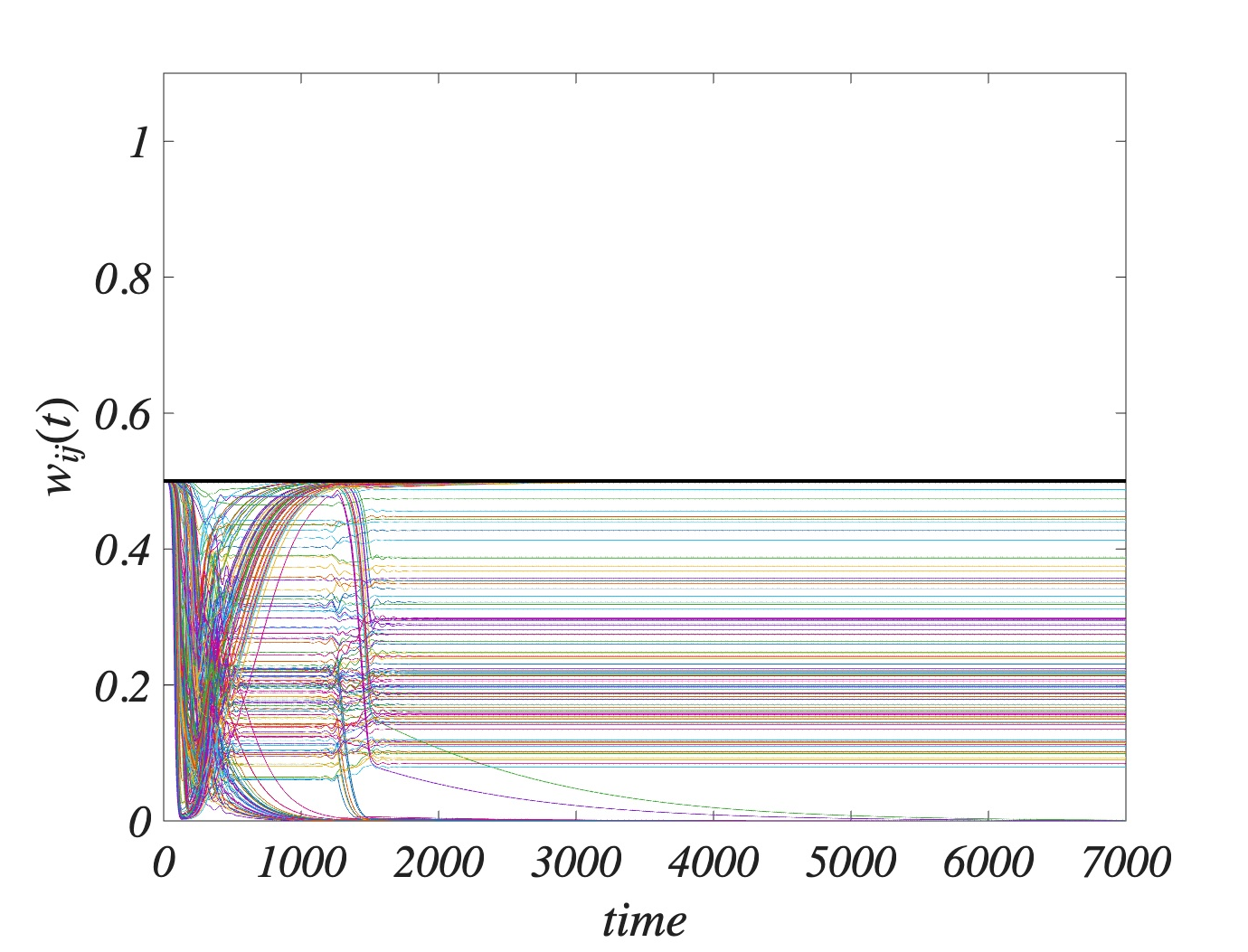}\\
\includegraphics[scale=0.25]{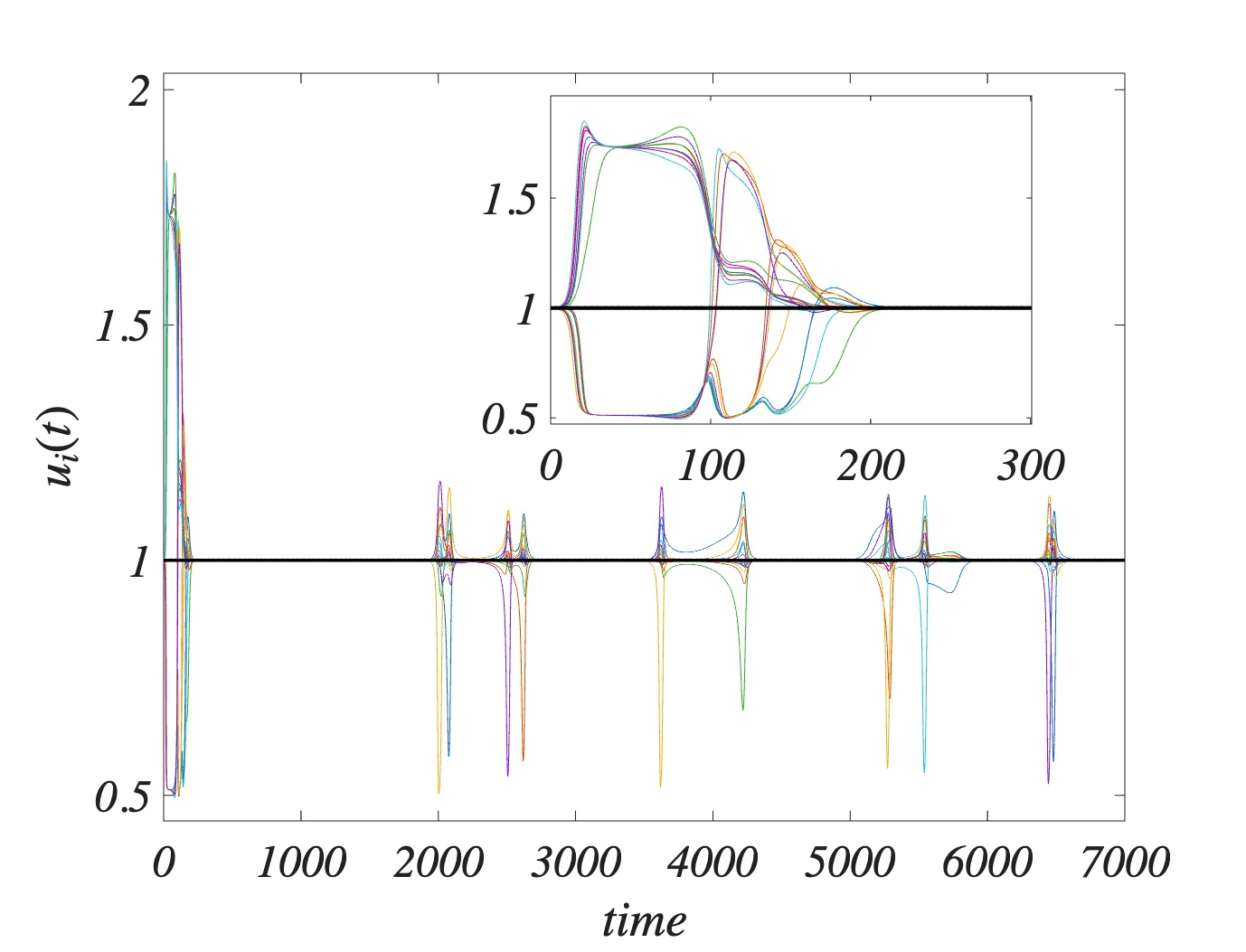}\includegraphics[scale=0.25]{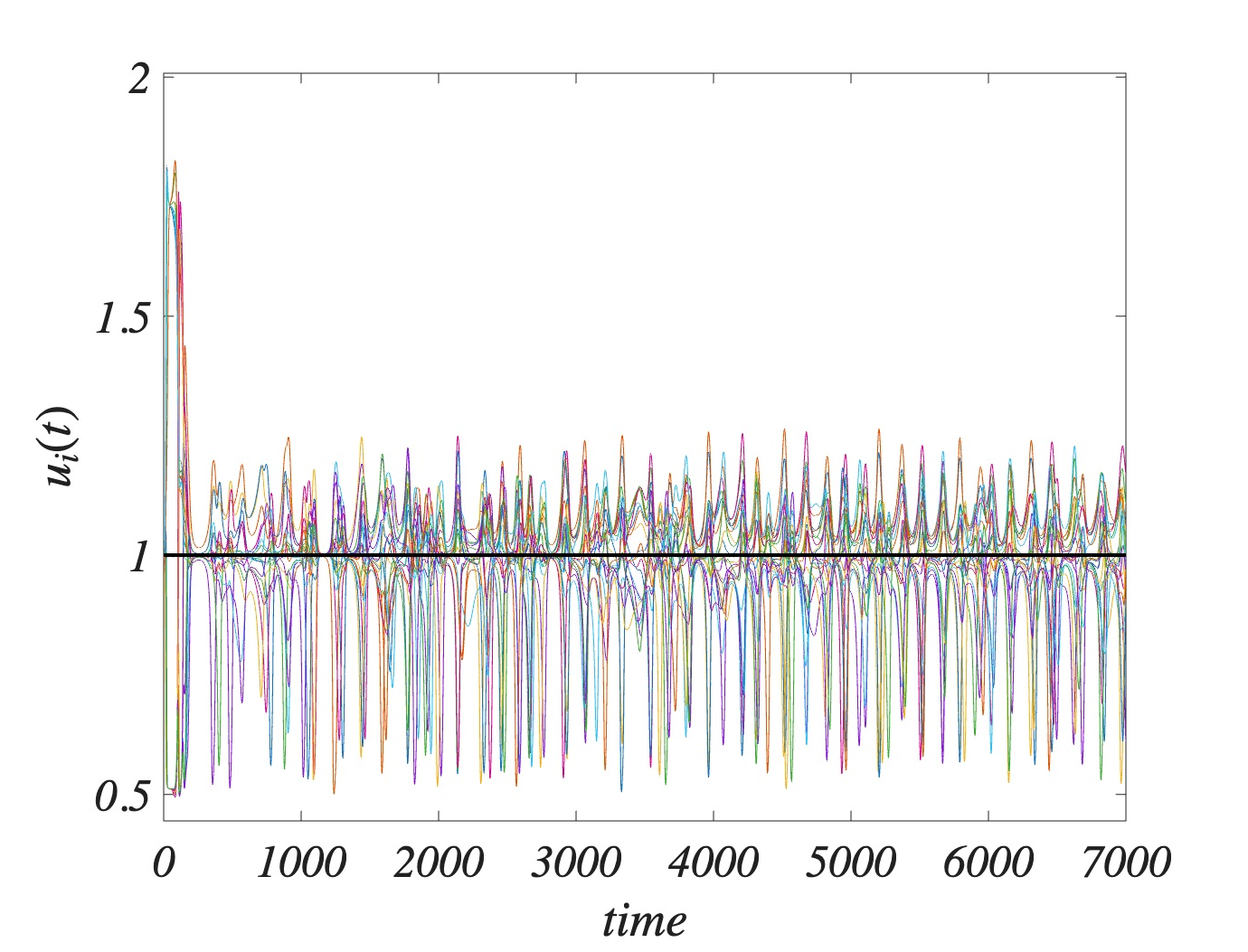}\includegraphics[scale=0.25]{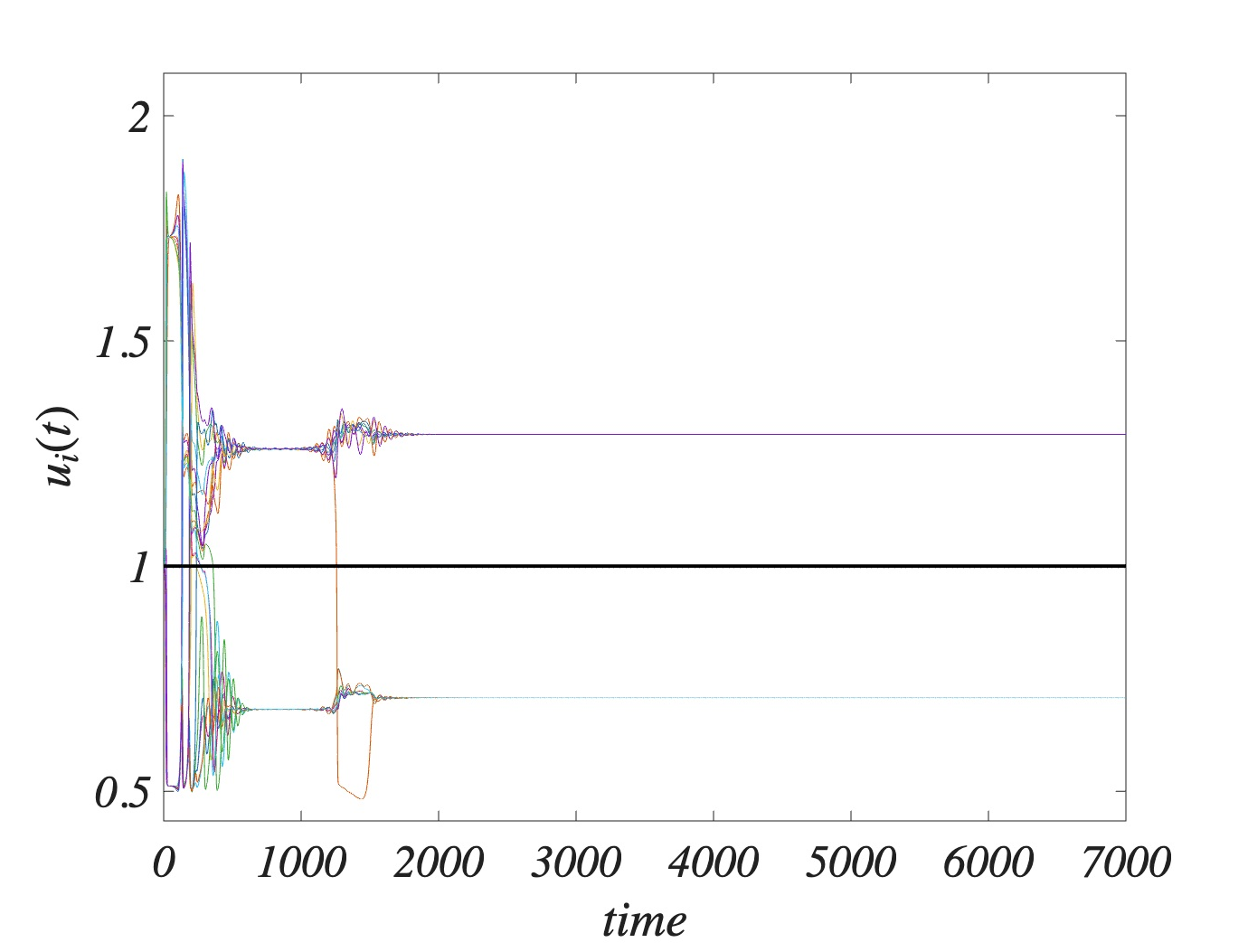}\\
\includegraphics[scale=0.25]{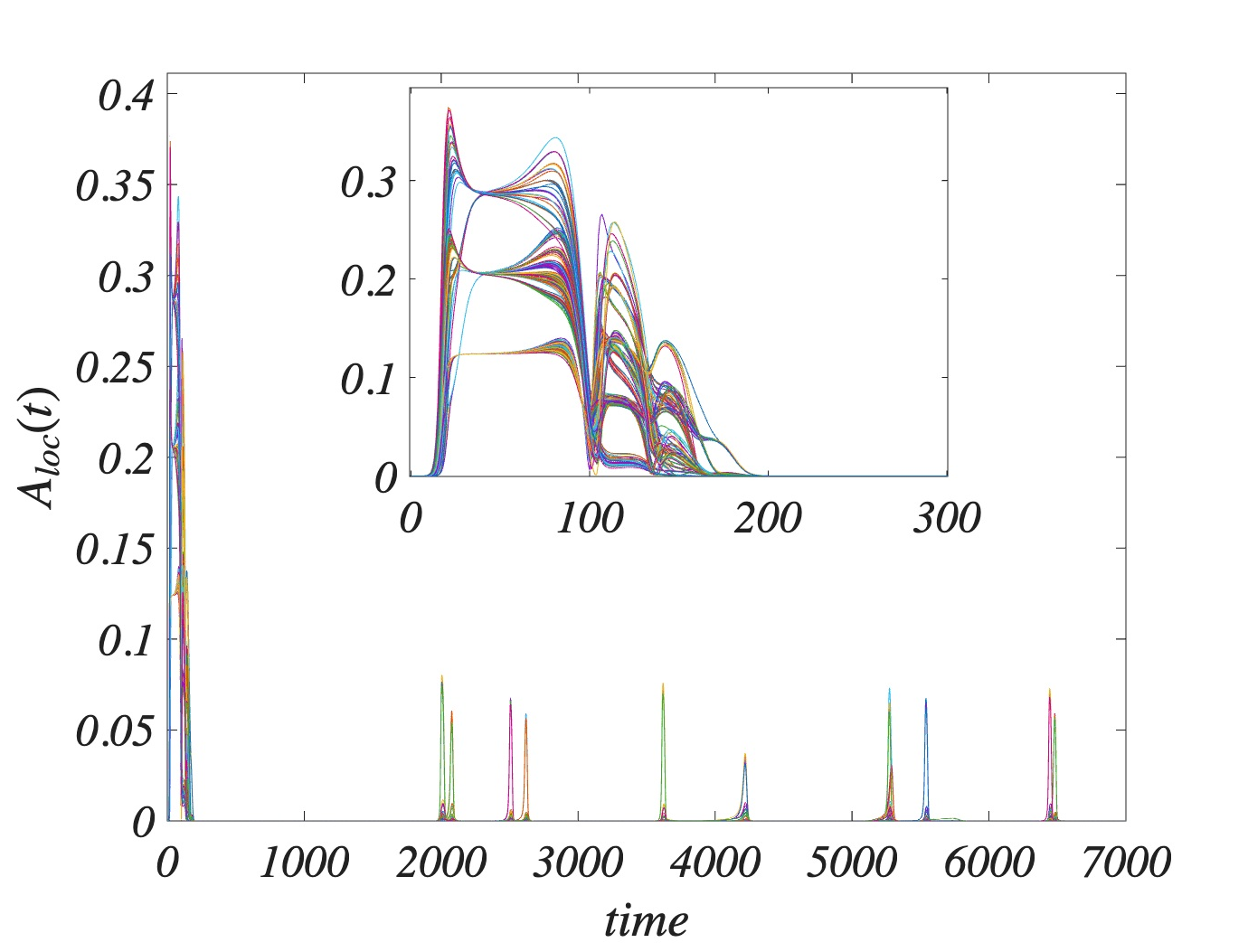}\includegraphics[scale=0.25]{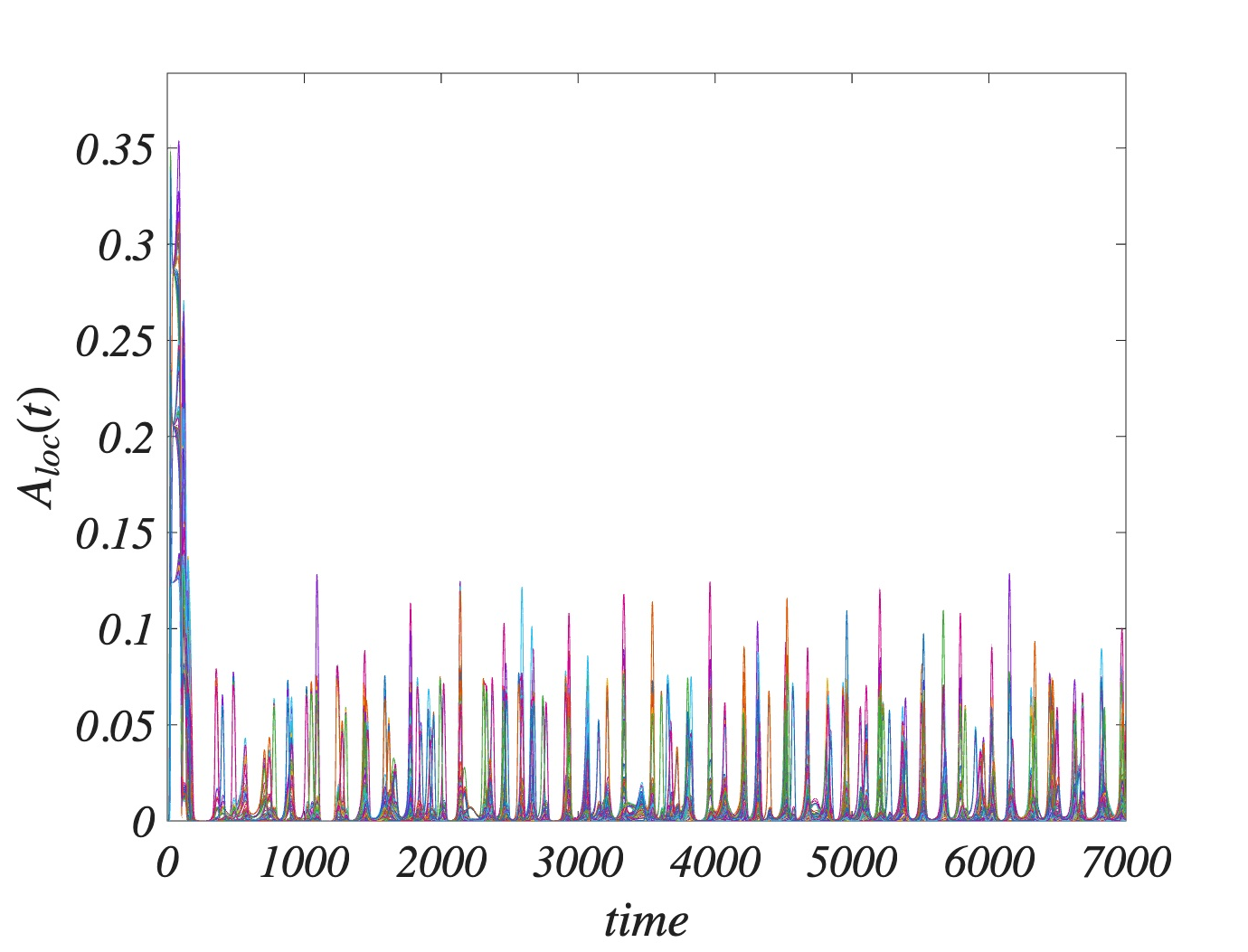}\includegraphics[scale=0.25]{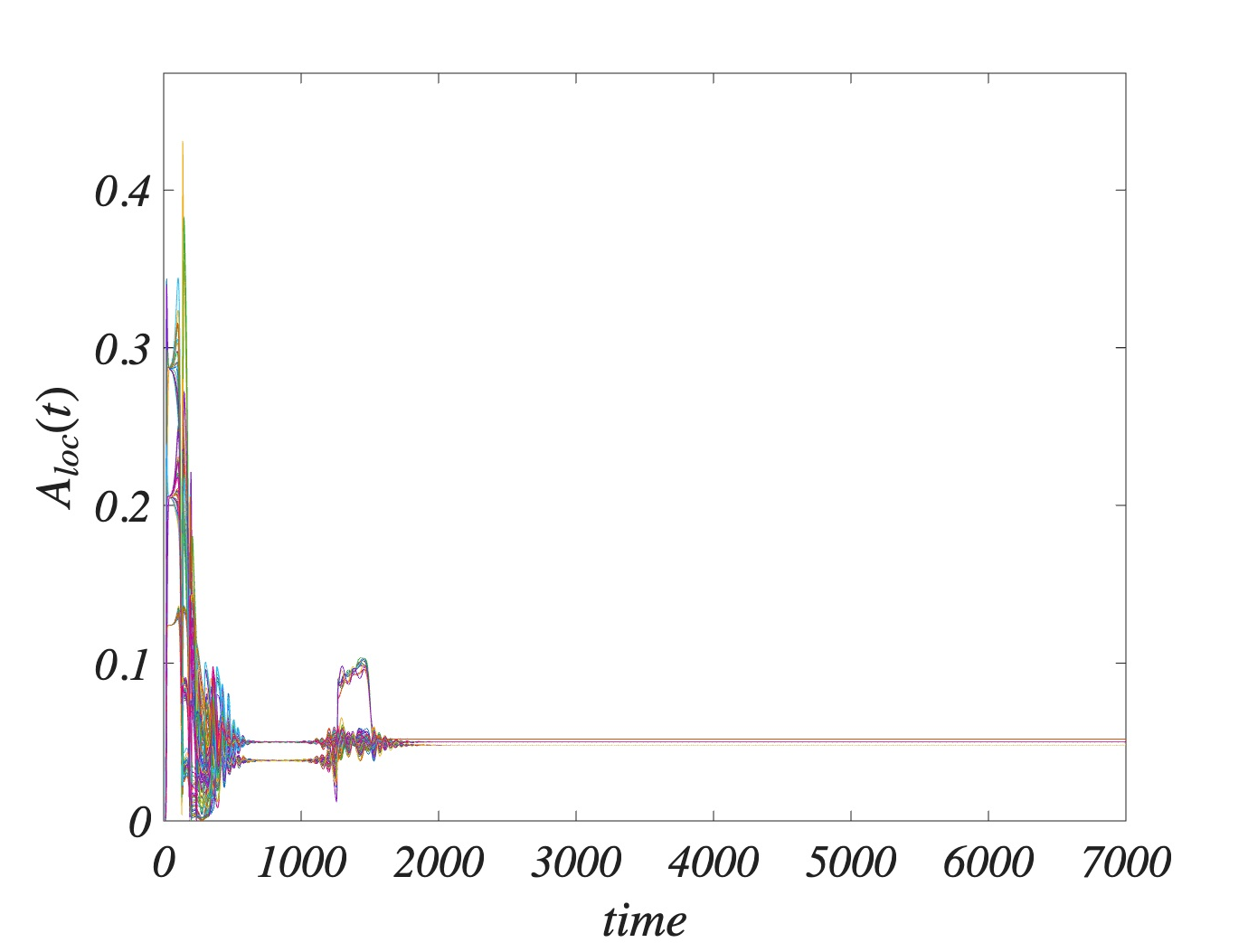}
\caption{The impact of $A^*$ in the parabola-like adaptive response. Top row: $w_{ij}(t)$, middle row: $u_i(t)$, lower row $A_{loc}(t)$. Left column $A^*=5.0\,10^{-4}$, middle column $A^*=5.0\,10^{-3}$, right column $A^*=5.0\,10^{-2}$. \teo{See Supplementary Movie 3, 4 and 5 to appreciate the dynamics of the adaptive network~\cite{SuppMovie3,SuppMovie4,SuppMovie5}.}}
\label{fig:Astarvarparab}
\end{figure*}

In the remaining case, $A^*>0$, the system exhibits a ``weird'' behavior with the possibility of stationary, oscillating or bursting patterns, depending on the initial conditions and the value of $A^*$. Indeed as previously explained the system is facing two antagonistic forces, from one side $A_{loc}$ tends to increase because $u_i$ and $v_i$ move away from the equilibrium, while $w_{ij}$ approach $w_1^*$; on the other hand, a too large $A_{loc}$ pushes $w_{ij}$ far from $w_1^*$ that in turn revert the direction of $u_i$ and $v_i$. The key factor is thus the value of $A^*$ at which the dynamics change. Because of our choice of initial conditions $A_{loc}(0)\sim \delta^2$, this will hence set a ``natural'' scale for $A^*$. The results in Fig.~\ref{fig:Astarvarparab} show the behavior of $w_{ij}(t)$ (top row), $u_i(t)$ (middle row) and $A_{loc}(t)$ (bottom row) for three values of the threshold parameter $A^*$. In the left column we set $A^*=5.0\,10^{-4}$, this can be considered to be a ``small'' value being of the order of $5 A_{loc}(0)$; we can observe (see inset) that at short time $u_i(t)$ deviate from $u^*$ and $A_{loc}(t)$ increase, while $w_{ij}(t)$ move away from $w_1^*$ and thus they decrease to $0$. At some point (data not shown) the spectrum of the time-dependent Laplacian matrix could not any longer support Turing patterns and thus $u_i(t)$ tip and return toward $u^*$ and so $A_{loc}(t)$ decrease, and once it will reach $A^*$ this will induce a change of behavior for $w_{ij}(t)$ that slowly increase. This phenomenon repeats several times in an irregular and bursty behavior. 

The middle column of Fig.~\ref{fig:Astarvarparab} refers to an intermediate value $A^*=5.0\,10^{-3}$, of the order of $50 A_{loc}(0)$; a behavior similar to the one presented for $A^*=5.0\,10^{-4}$ is shown but on a faster time scale, indeed being $A^*$ larger, $A_{loc}(t)$ takes less time to reach it. Finally, on the right column we present the case of a large value $A^*=5.0\,10^{-2}$, of the order of $500 A_{loc}(0)$, here the intermittent behavior cannot be sustained and, after a transient, $w_{ij}(t)$ settles to constant values, i.e., the network reaches an all-to-all configuration with heterogeneous weights capable to support stationary Turing patterns. A similar behavior is observed by using the global amplitude $A(\vec{u},\vec{v})$ defined in~\eqref{eq:Amp}. For the sake of completeness we report in Appendix~\ref{sec:globAmpli} the numerical results obtained by using the global amplitude; Fig.~\ref{fig:4withAuv} is thus the equivalent of Fig.~\ref{fig:Astarvarparab} and we can observe that it exhibits the same qualitative dependence on $A^*$, in particular, a faster convergence as $A^*$ increases.

After the parabola-like case, let us now consider the cubic-like adaptive response (see panels b) and d) in Fig.~\ref{fig:Mod3}), more precisely we set
\begin{equation}
\label{eq:adaptrespcub}
k(w)=w(w_1^*-w)(w_2^*-w) \, , w_1^*=1/2 \text{ and }w_2^*=1\, .
\end{equation}
We consider initial conditions as in the previous case but now initial weights are distributed uniformly at random in the interval $(w_1^*-\delta,w_1^*+\delta)$~\footnote{Let us observe that by taking initial weights in $(w_1^*-\delta,w_1^*)$, i.e.,  as in the parabola-like case, we will obtain the same behavior as in the former case because $w_{ij}(t)$ will be bounded to lie in $(0,w_1^*)$ where the two adaptive responses have the same functional shape. So, to fully exploit the cubic-like case, it is mandatory to have initial weights below and above $w_1^*$.}, with again $\delta=0.01$. The underlying unweighted network is still assumed to be all-to-all, namely $a_{ij}=1$ for all $i\neq j$, and $a_{ii}=0$.

Once we consider a large value of the threshold we obtain the same dynamics as before (data not shown): $u_i(t)$ and $v_i(t)$ will move away from the equilibrium hence $A_{loc}(t)$ increase, however $A^*-A_{loc}(t)>0$ is always satisfied because of the choice of $A^*\gg 1$. This implies that $w_1^*$ is an attractor for the weights and the network converges to a weighted complete graph where all the links have value $w_1^*$; this network supports Turing patterns by assumption, and thus also the adaptive system does. 

\begin{figure*}[ht]
\centering
\includegraphics[scale=0.25]{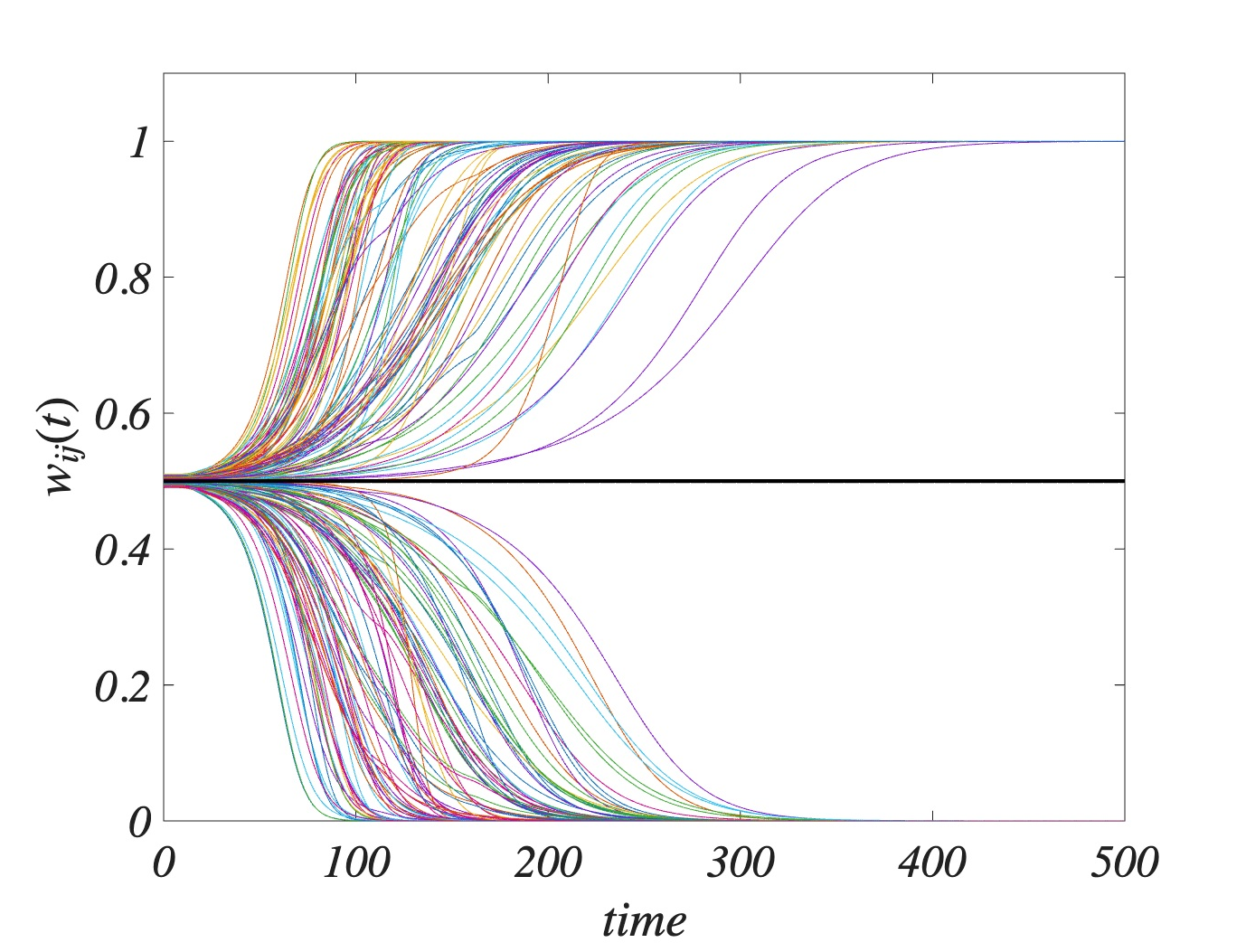}\includegraphics[scale=0.25]{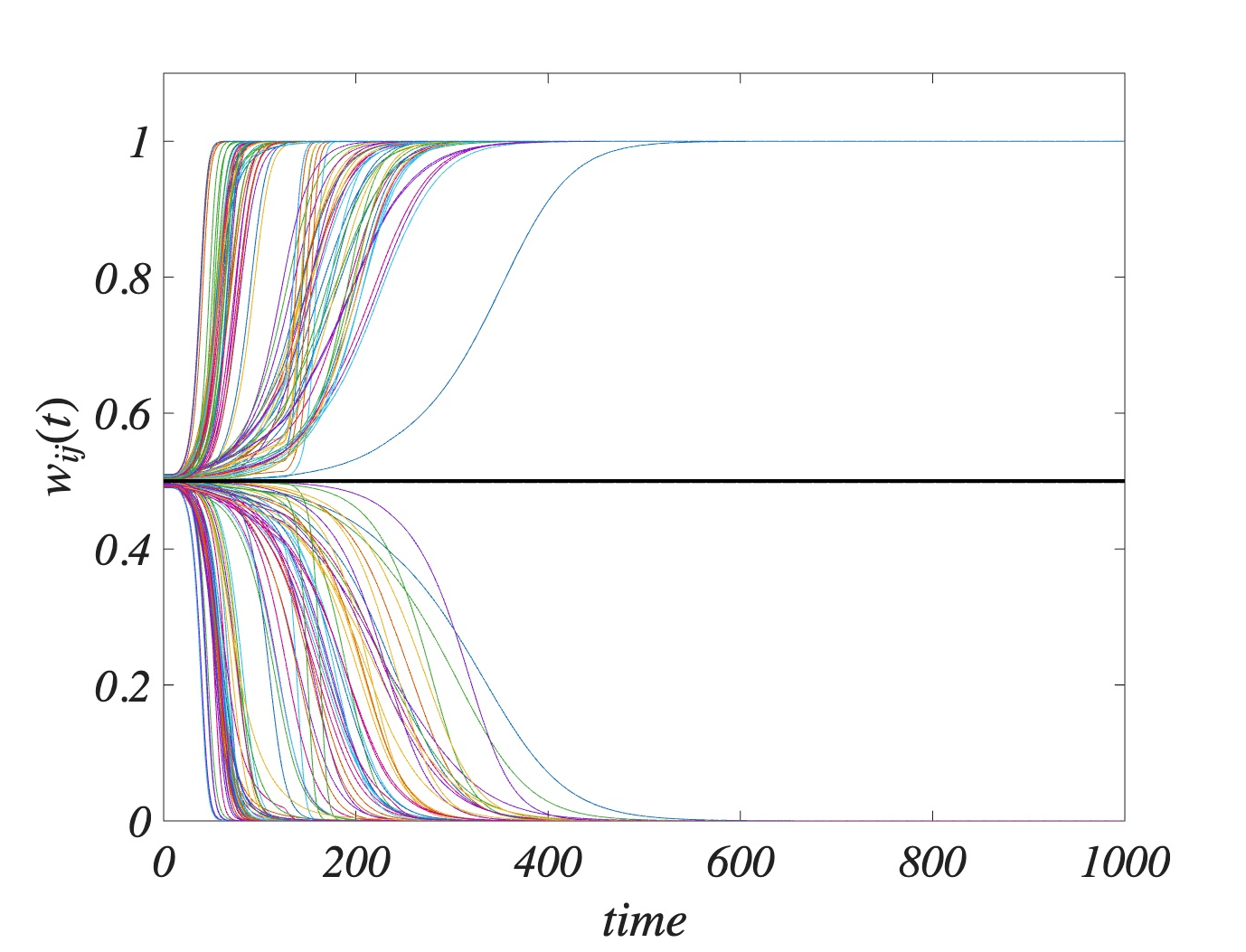}\includegraphics[scale=0.25]{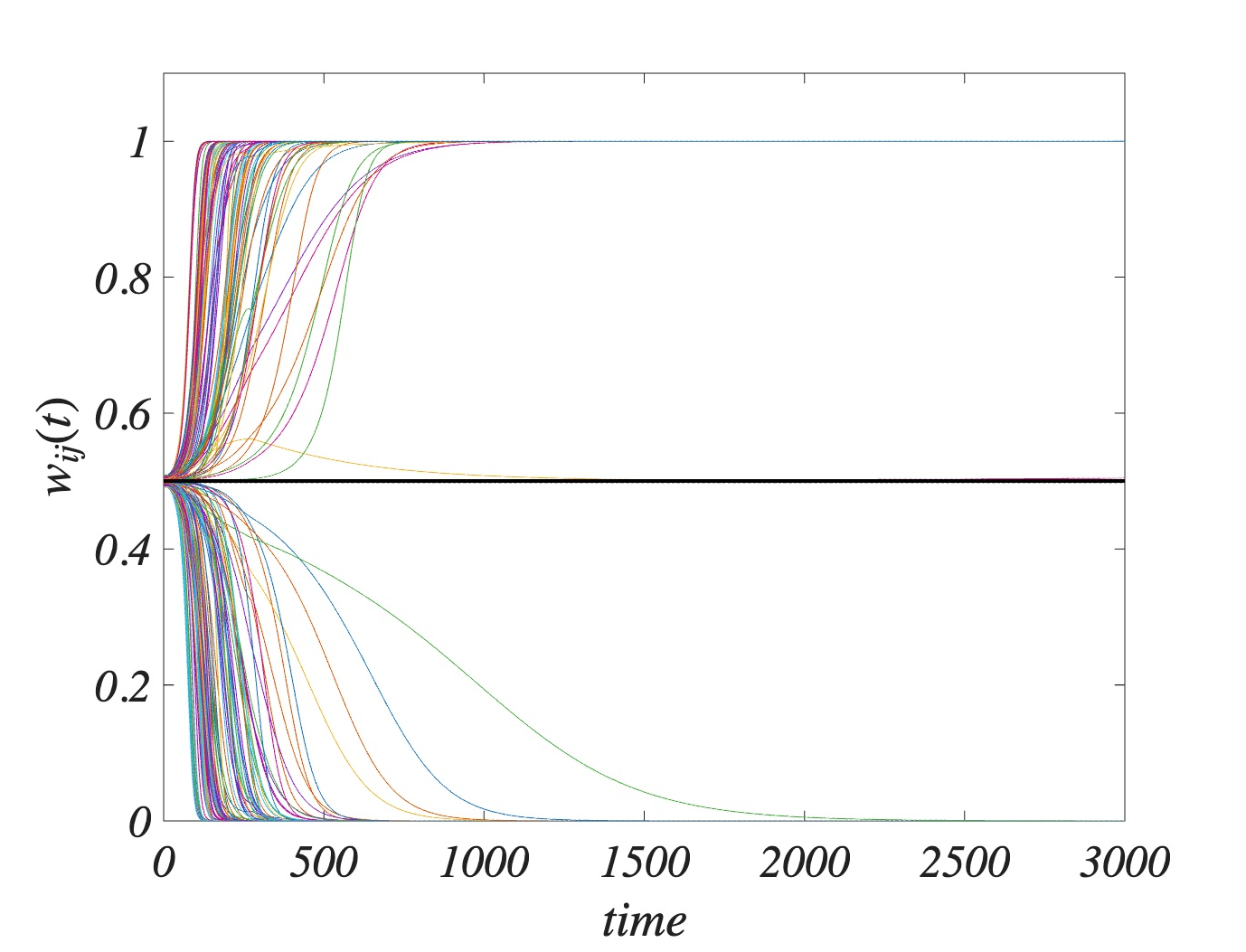}\\
\includegraphics[scale=0.25]{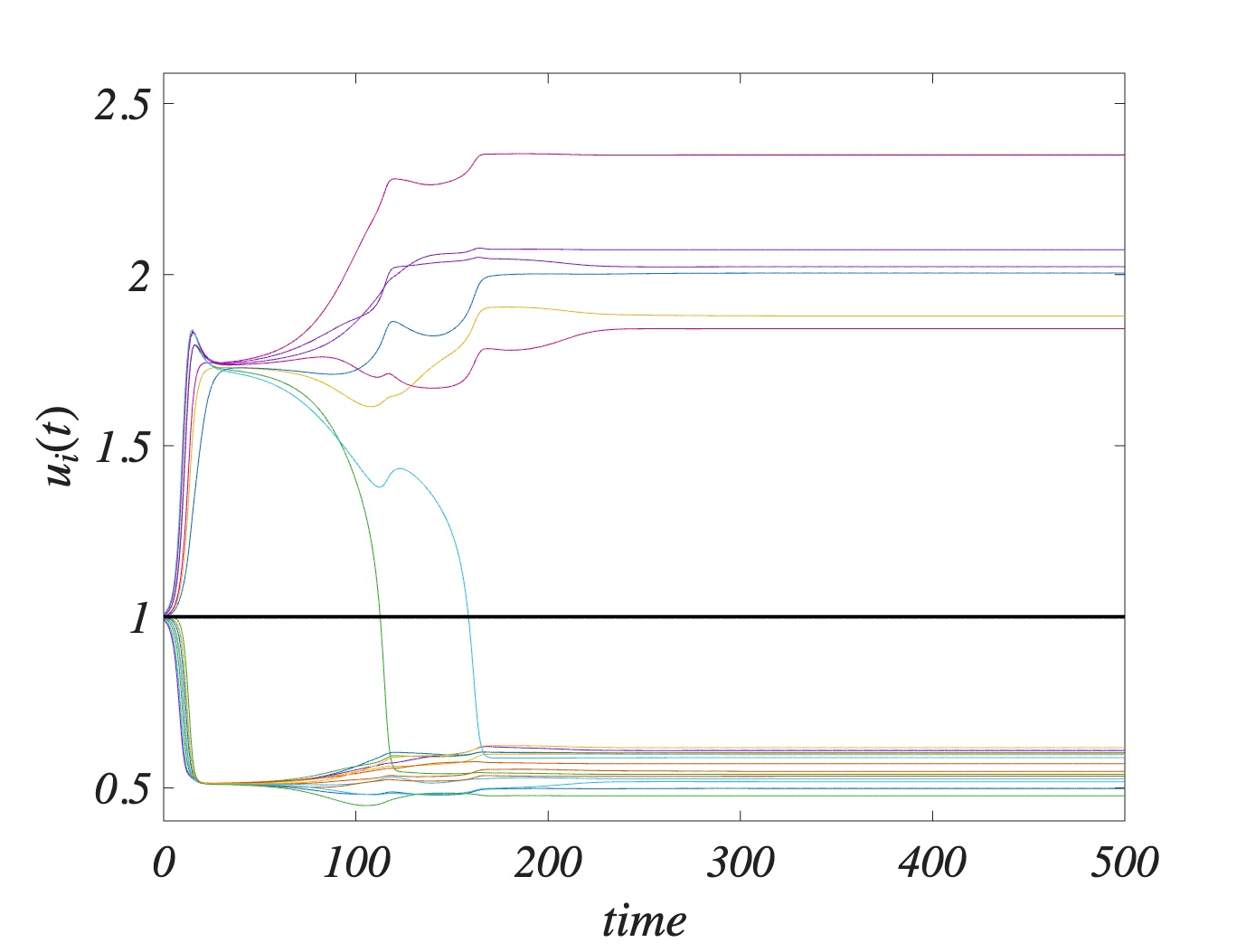}\includegraphics[scale=0.25]{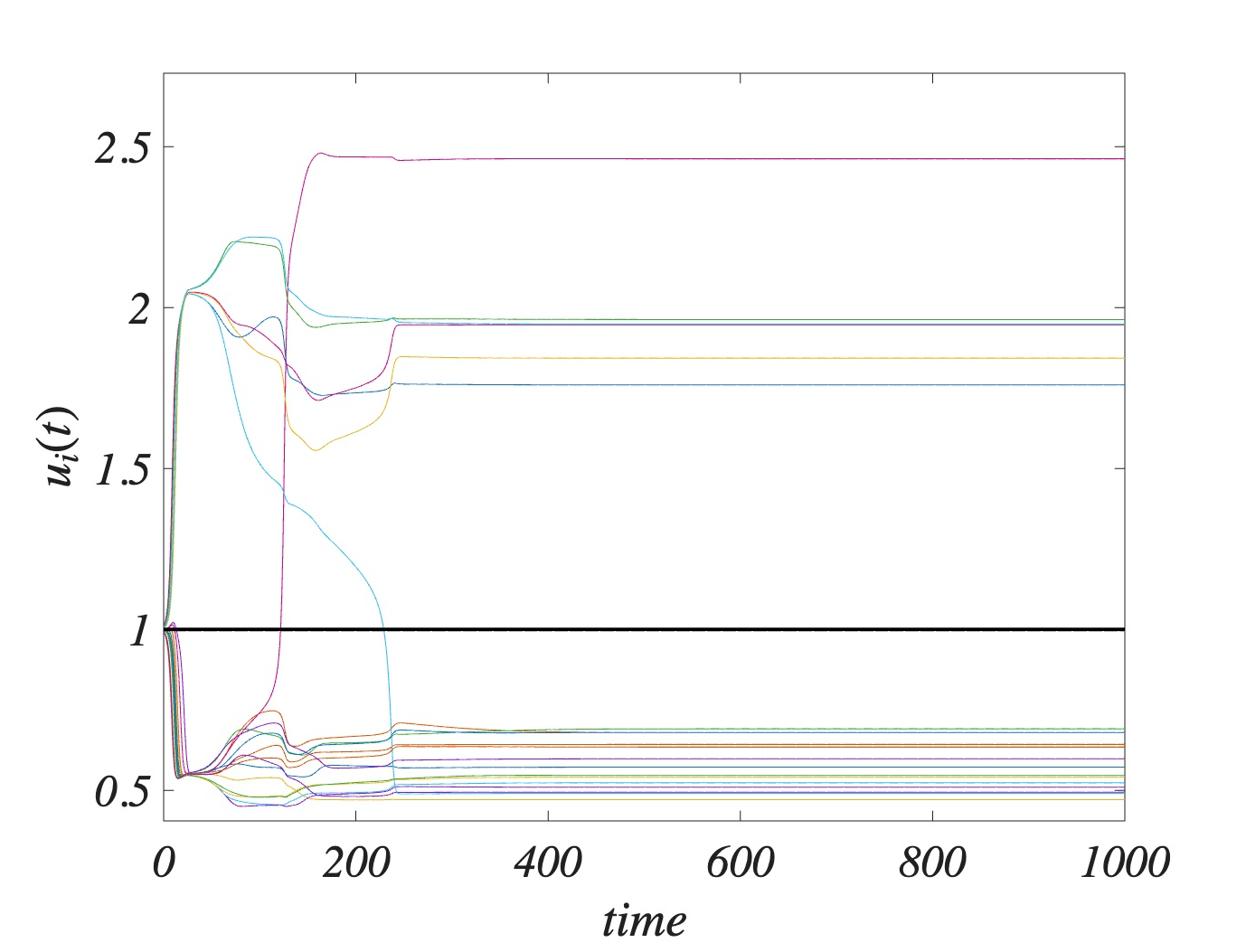}\includegraphics[scale=0.25]{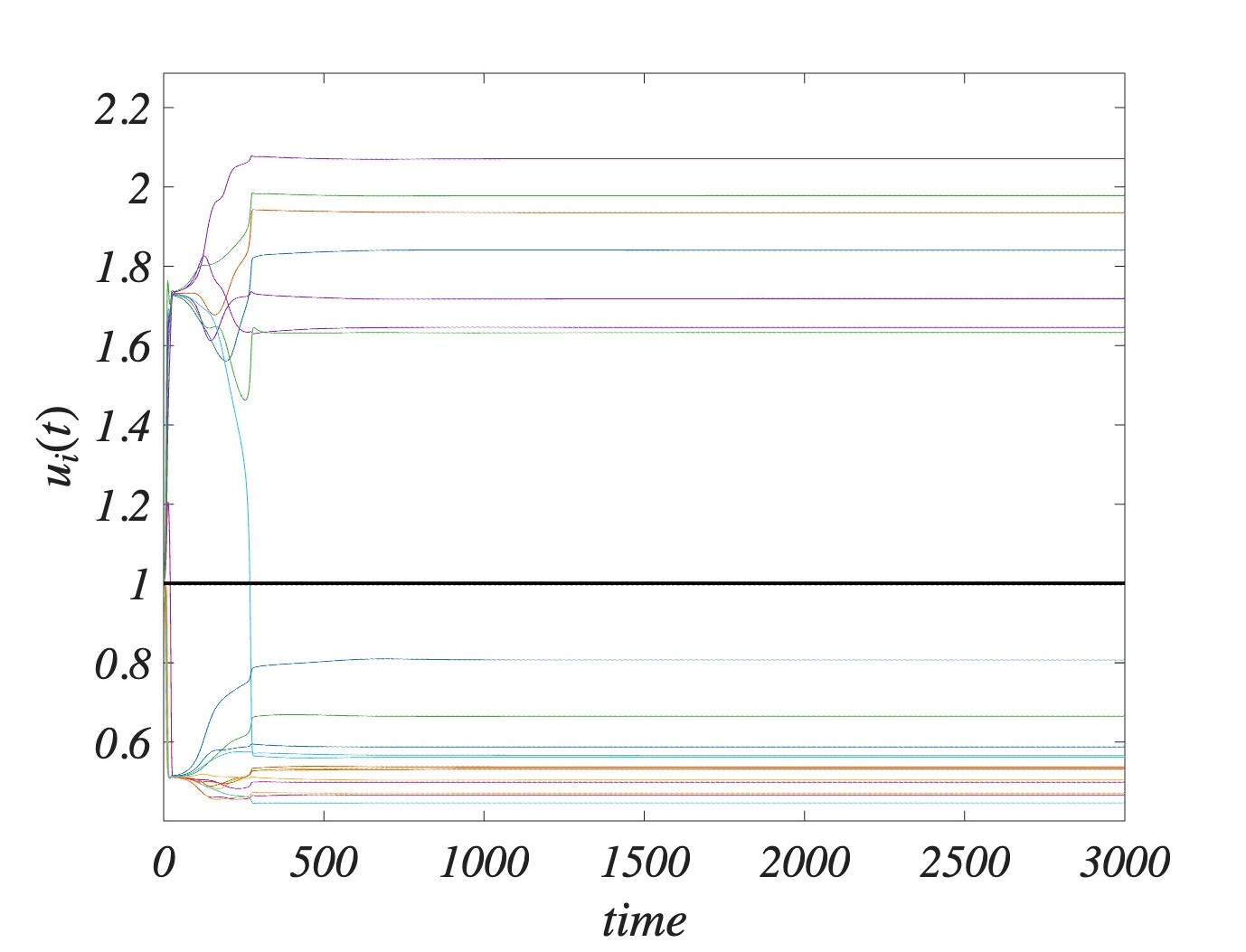}\\
\includegraphics[scale=0.25]{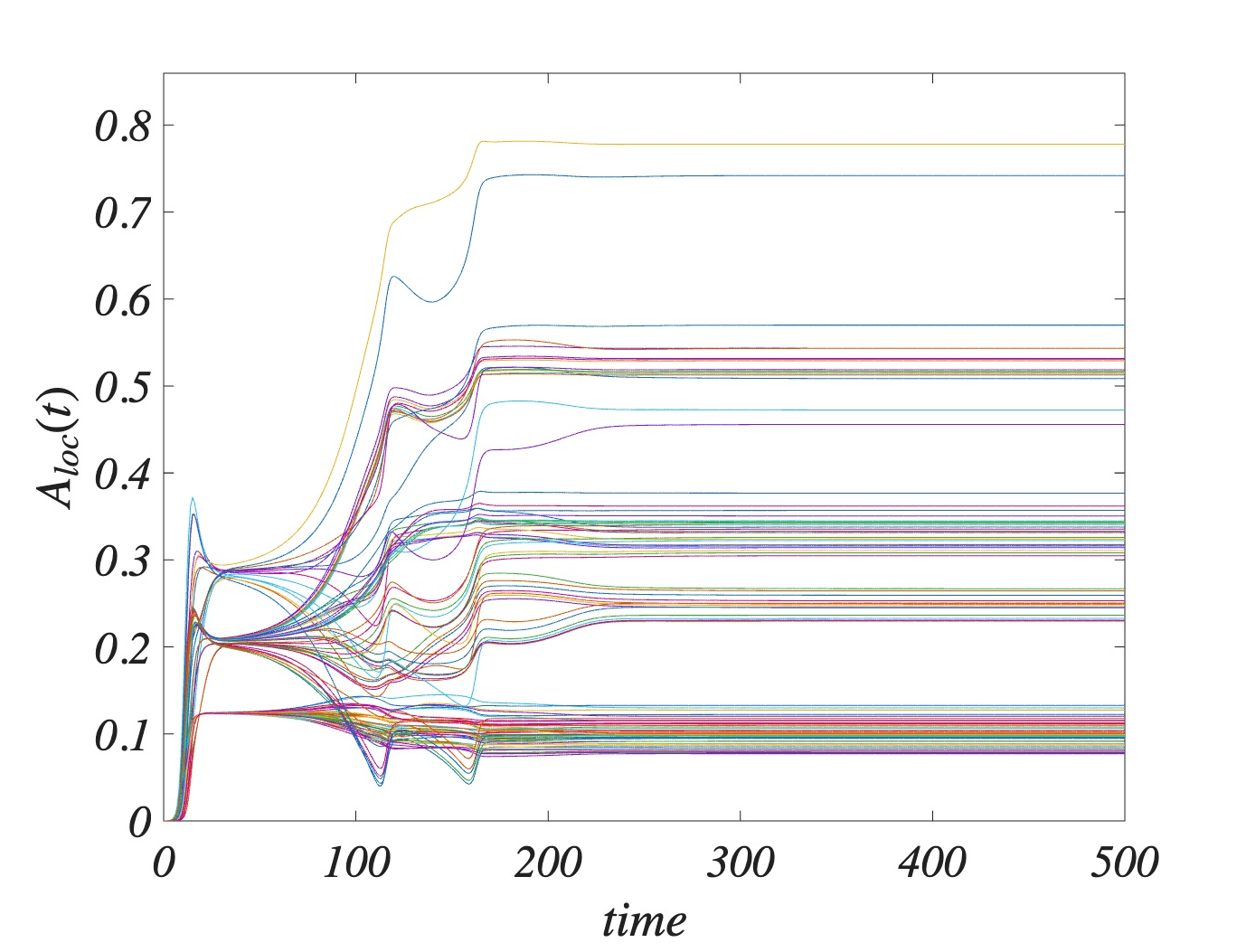}\includegraphics[scale=0.25]{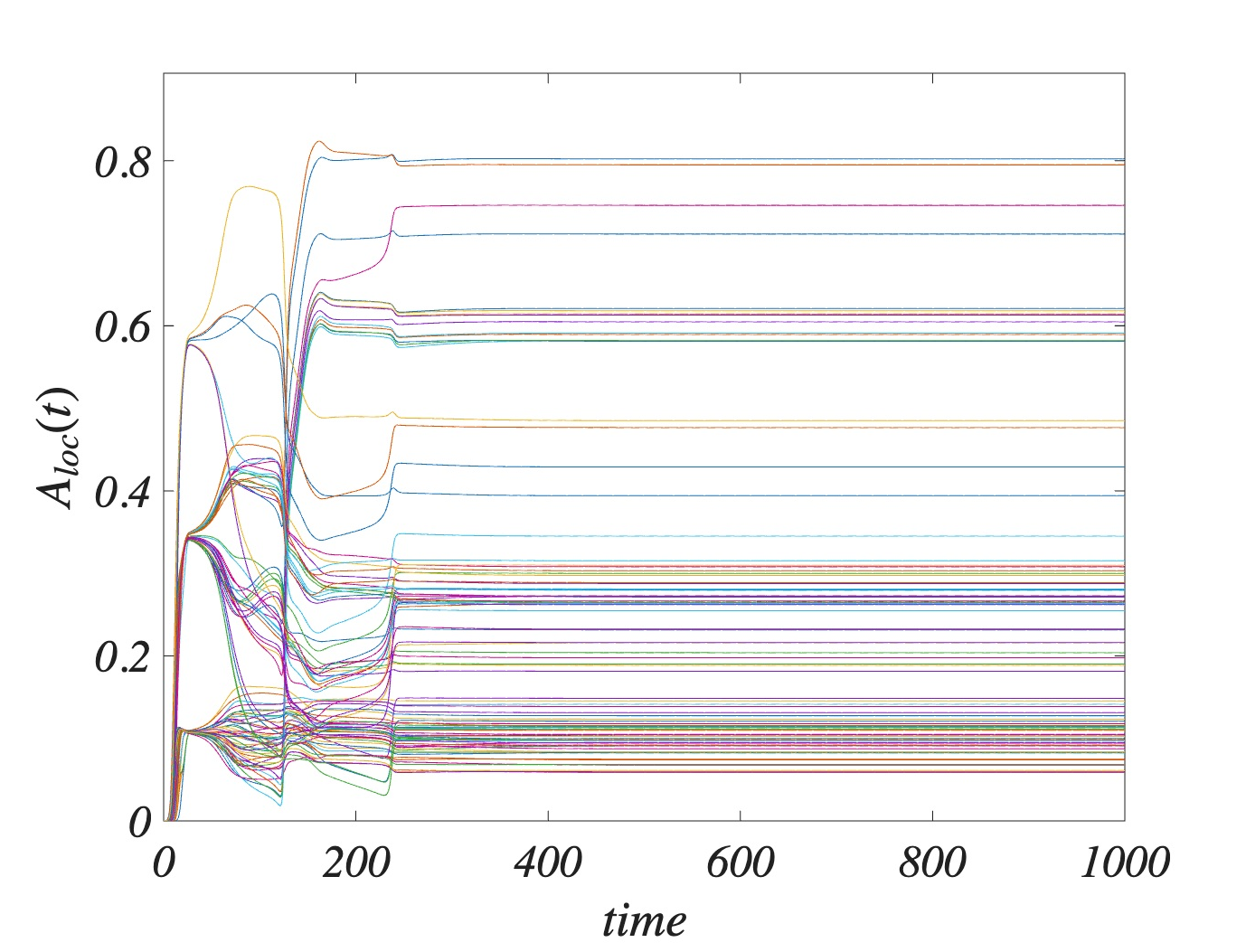}\includegraphics[scale=0.25]{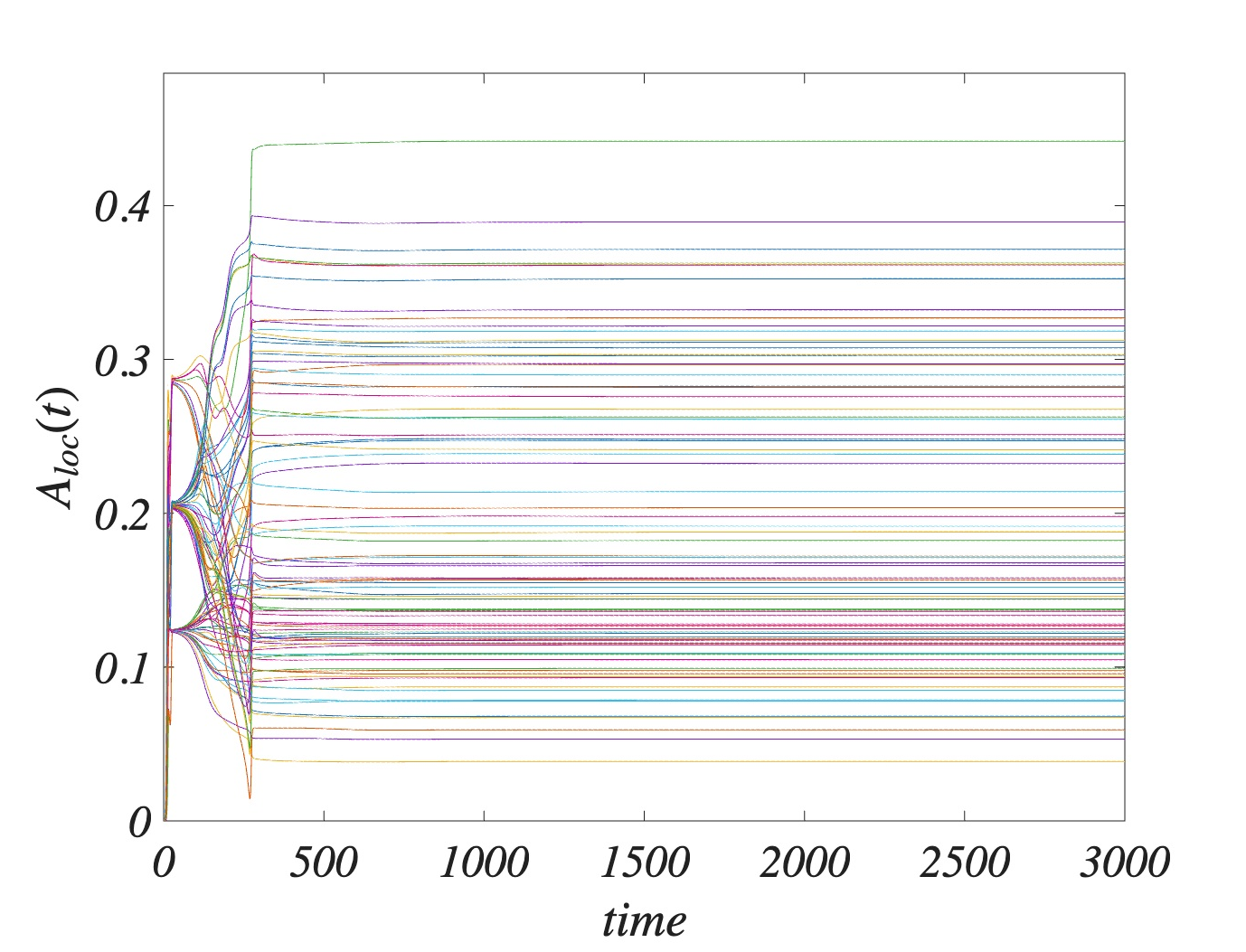}
\caption{The impact of $A^*$ in the cubic-like adaptive response. Top row: $w_{ij}(t)$, middle row: $u_i(t)$, lower row $A_{loc}(t)$. Left column $A^*=0$, middle column $A^*=5.0\,10^{-4}$, right column $A^*=5.0\,10^{-2}$.}
\label{fig:Astarvarcubic}
\end{figure*}

Conversely in the remaining cases, $A^*\geq 0$ but not too large, the cubic-like adaptive response coupled with the choice of initial conditions for the weights returns a different behavior: almost half of the links disappear (this is because of the symmetric way about $w_1^*$ initial conditions have been set), i.e., their weights become extremely small, and the resulting network supports Turing patterns (see Fig.~\ref{fig:Astarvarcubic}). Indeed, the growth of $A_{loc}(t)$ and the relatively small value of $A^*$ imply that sooner or later $A^*-A_{loc}(t)<0$, hence $w_1^*$ becomes unstable and the weights move away from it; roughly speaking the weights such that $w_{ij}(0)>w_1^*$ will converge to $w_2^*$ while those satisfying $w_{ij}(0)<w_1^*$ will converge to $0$. The resulting network will support Turing patterns. This behavior is shown in Fig.~\ref{fig:Astarvarcubic} where we display $w_{ij}(t)$ (top row), $u_i(t)$ (middle row) and $A_{loc}(t)$ (bottom row) for three values of the threshold parameter $A^*$: $A^*=0$ (left column),  $A^*=5.0\,10^{-4}$ (middle column) and $A^*=5.0\,10^{-2}$ (right column). 

\begin{figure}[ht]
\centering
\includegraphics[scale=0.25]{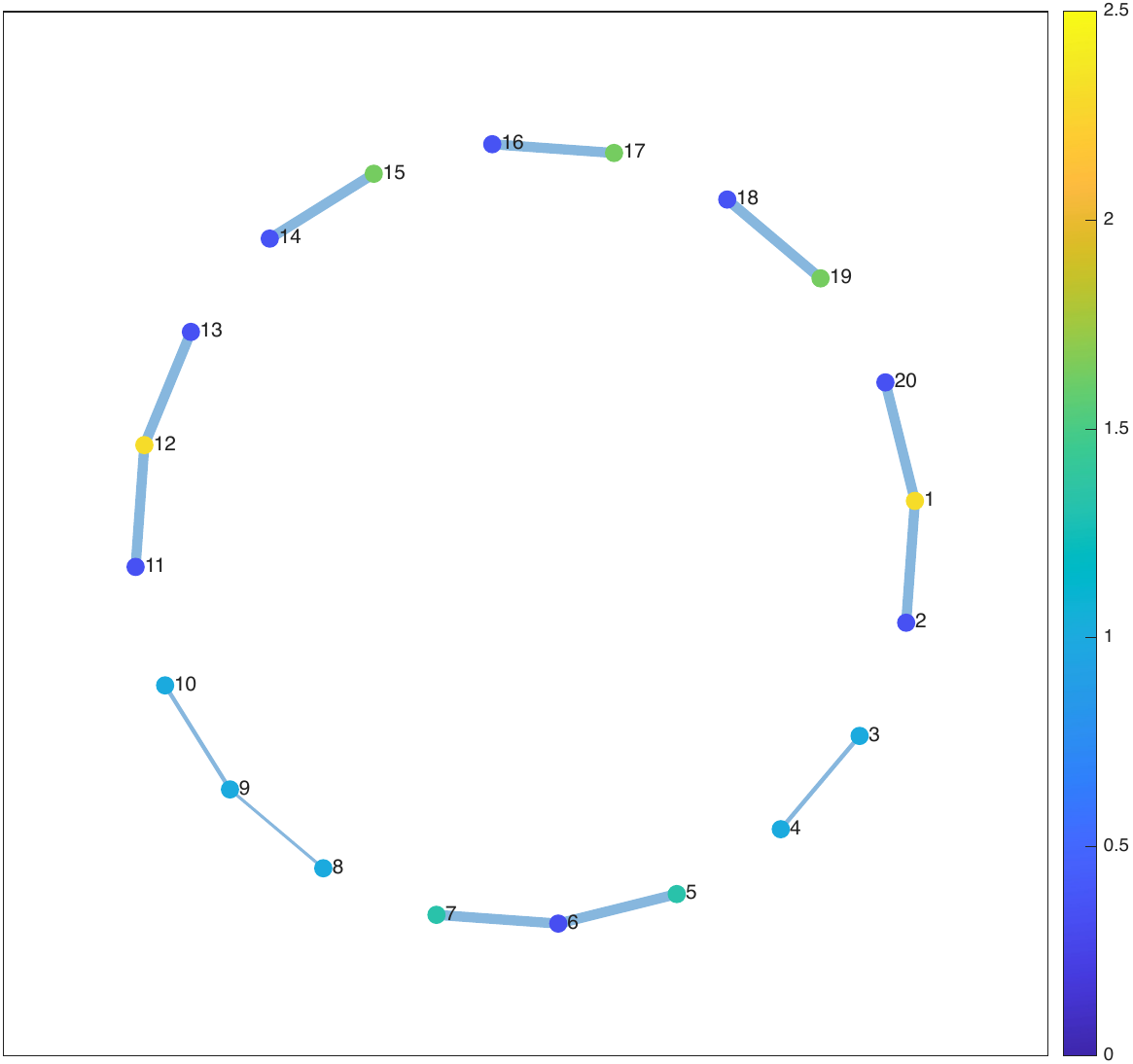}\\
\includegraphics[scale=0.25]{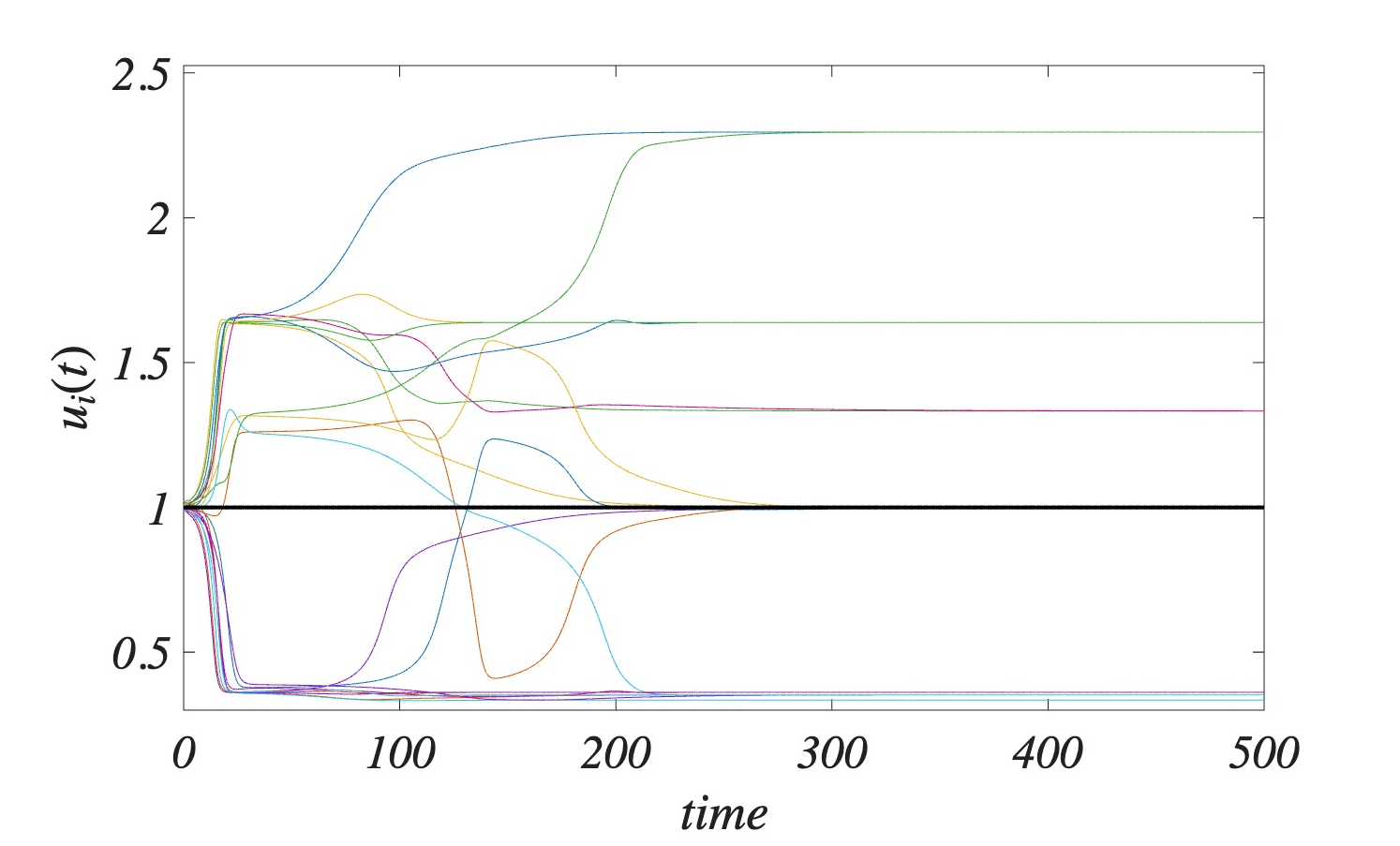}
\caption{Adaptation breaks the network and induce Turing patterns in small networks. We consider the cubic adaptive response, $k(w)=w(1/2-w)(1-w)$ and $A^*=0$, defined on a regular ring where each node has degree $2$ at time zero. Top panel: Starting from a ring configuration where each node has two neighbors,  some weights converge to zero (data not shown) and the ring breaks into eight subnetworks of sizes $2$ and $3$. The thickness of the links is proportional to the asymptotic stationary weight while the color of the nodes to the asymptotic stationary density of $u$ (see colorbar). Bottom panel: time evolution of species $u$ in each node, $u_i(t)$. The remaining parameters are $b=4$, $c=6.5$, $D_u=0.07$ and $D_v=1.7$. \teo{See Supplementary Movie 6 to appreciate the dynamics of the adaptive network~\cite{SuppMovie6}.}}
\label{fig:AdaptBreak}
\end{figure}

We have thus observed that adaptation can reduce the number of available edges, i.e., with positive weights; we can thus design cases where by removing sufficiently many edges the network breaks into small subnetworks and this challenges the emergence of Turing patterns, indeed the latter are very difficult to be observed in networks composed by few nodes. In particular we considered a $2$-regular ring, i.e., a $1$-dimensional ring where each one of the $n=20$ nodes has initial degree $2$, then we let the system to adapt with cubic response~\eqref{eq:modeluvwBrus} with $w_1^*=1/2$, $w_2^*=1$ and $A^*=0$. The results reported in Fig.~\ref{fig:AdaptBreak} clearly show that the initial ring is broken into $10$ smaller subnetworks composed by $2$ or $3$ nodes (top panel), interestingly enough several of them support Turing patterns otherwise impossible in so small groups of nodes (see groups $3-4$ and $8-9-10$). Those patterns are created by the slow adaptation of the weights: nodes start to differentiate - accumulate or dilute the density of $u$ and $v$ - once the links are still active, but then the weights vanish and species cannot move anymore, they are thus trapped into the patterns.

Let us conclude this section by showing that adaptation can induce the emergence of Turing patterns otherwise impossible in the case of static networks. We consider thus a network whose initial weights return a negative dispersion relation, hence the node densities, $u_i(t)$ and $v_i(t)$, will converge to the homogeneous equilibrium, $u^*$ and $v^*$, if they are initialized close enough to the latter. If the weights cannot evolve, the system will thus exhibit the same density in each node. On the other hand, if we let the weights to dynamically adapt to the evolution of the density, then one can ``force'' them to move to values associated to positive dispersion relation, thus ensuring the emergence of Turing patterns.

For the sake of definitiveness, we will show this phenomenon by considering the cubic adaptive response, let us however observe that a similar result can be shown to hold true also in the case of quadratic adaptive response (see Appendix~\ref{sec:pattadaptquad}). In Fig.~\ref{fig:AdaptTPcub} we report the results obtained by using a network composed by $n=9$ nodes, the model parameters used are the same as the previous example, but $b= 4.0$ and $c = 6.8$, and weights are initialized in $(w_1^* - \delta , w_1^*+\delta)$, with $w_1^*=1/2$ and $\delta=0.1$. The dispersion relation computed by using the initial weights, and assuming they do not evolve, is negative (see top panel), however the cubic adaptive response, $k(w)=w(w_1^*-w)(w_2^*-w)$, with $w_2^*=1$ and $A^*=5.0\, 10^{-4}$, allows some of the weights to increase (see middle panel) and thus determine a positive dispersion relation from which patterns can emerge (see bottom panel).
\begin{figure}[ht]
\centering
\includegraphics[scale=0.25]{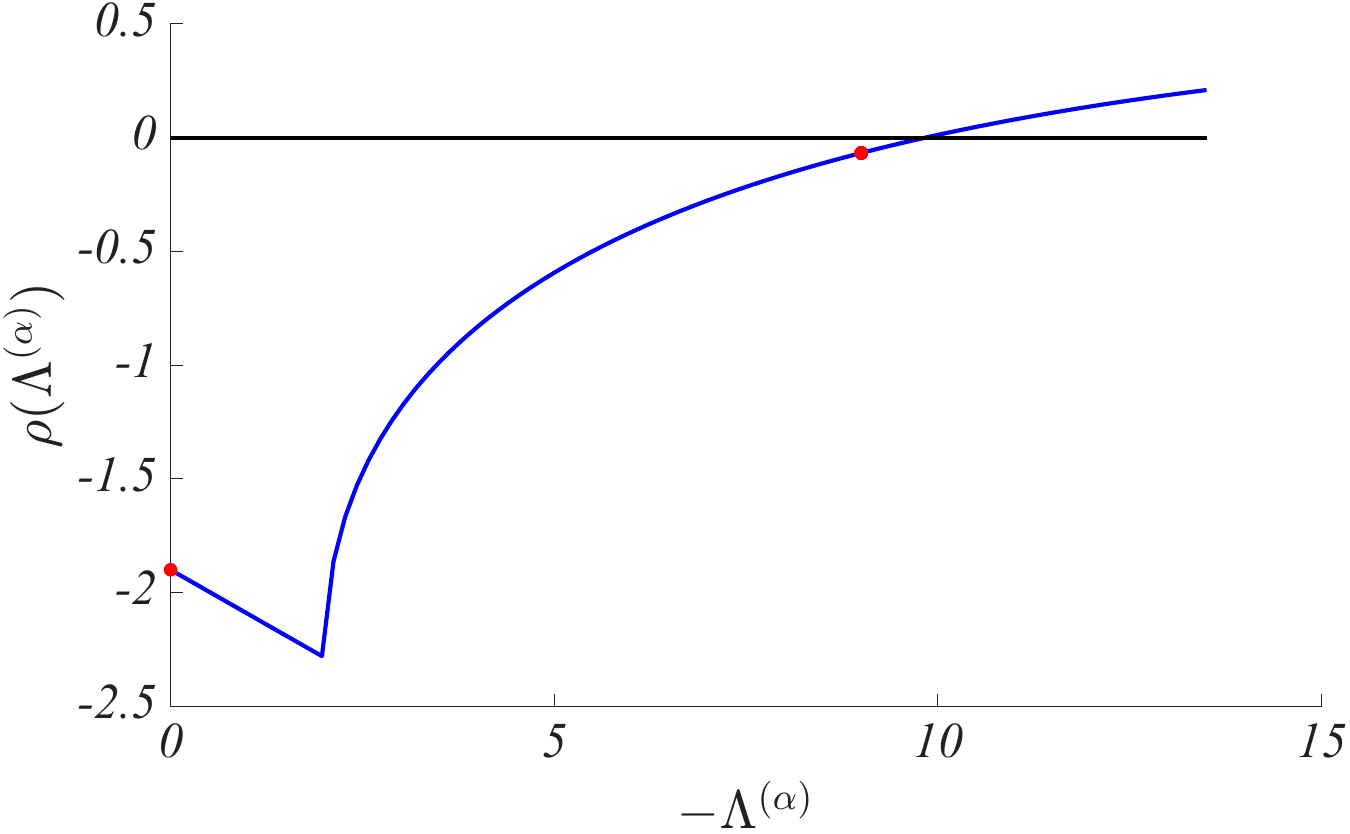}\\
\includegraphics[scale=0.25]{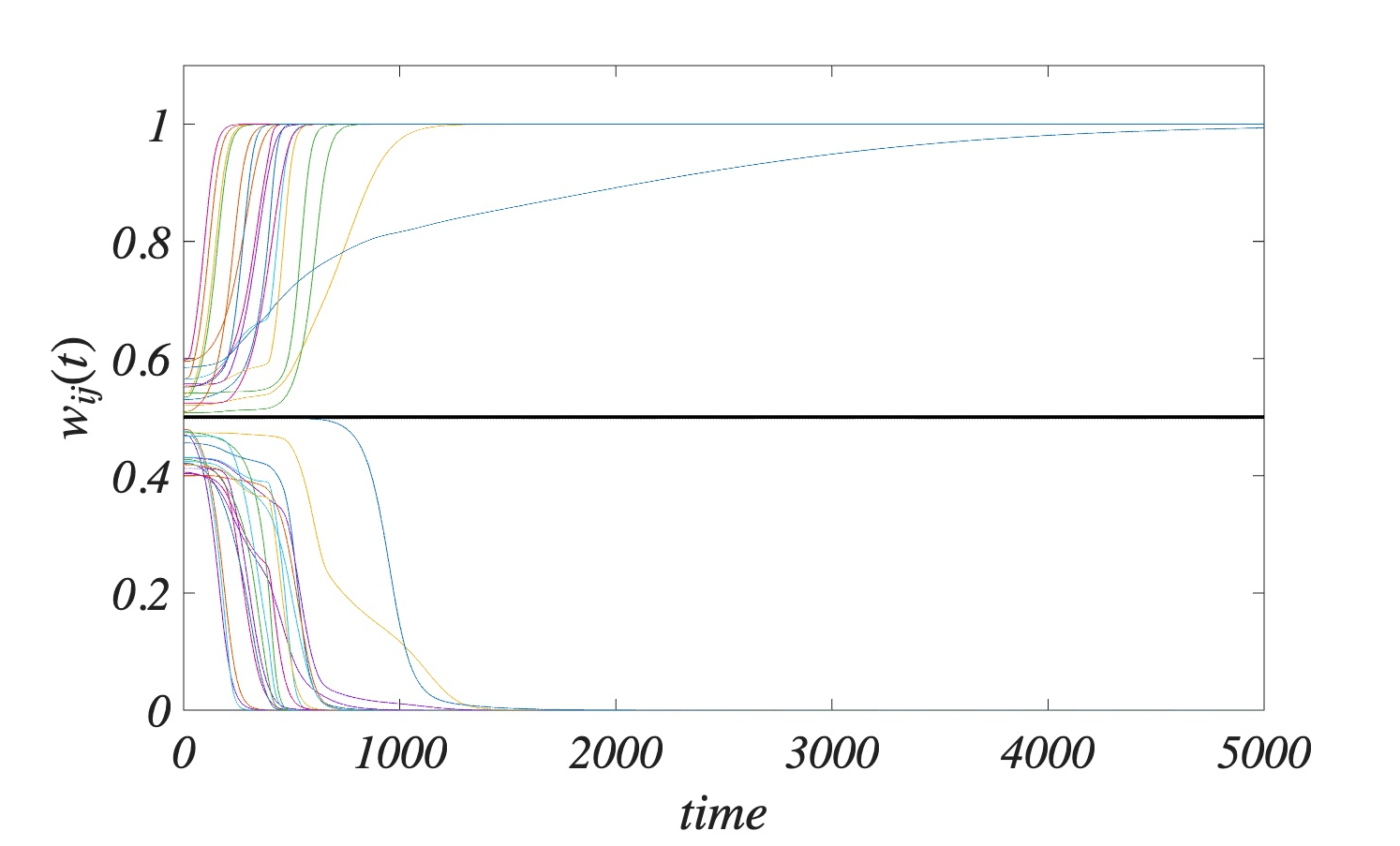}\\
\includegraphics[scale=0.25]{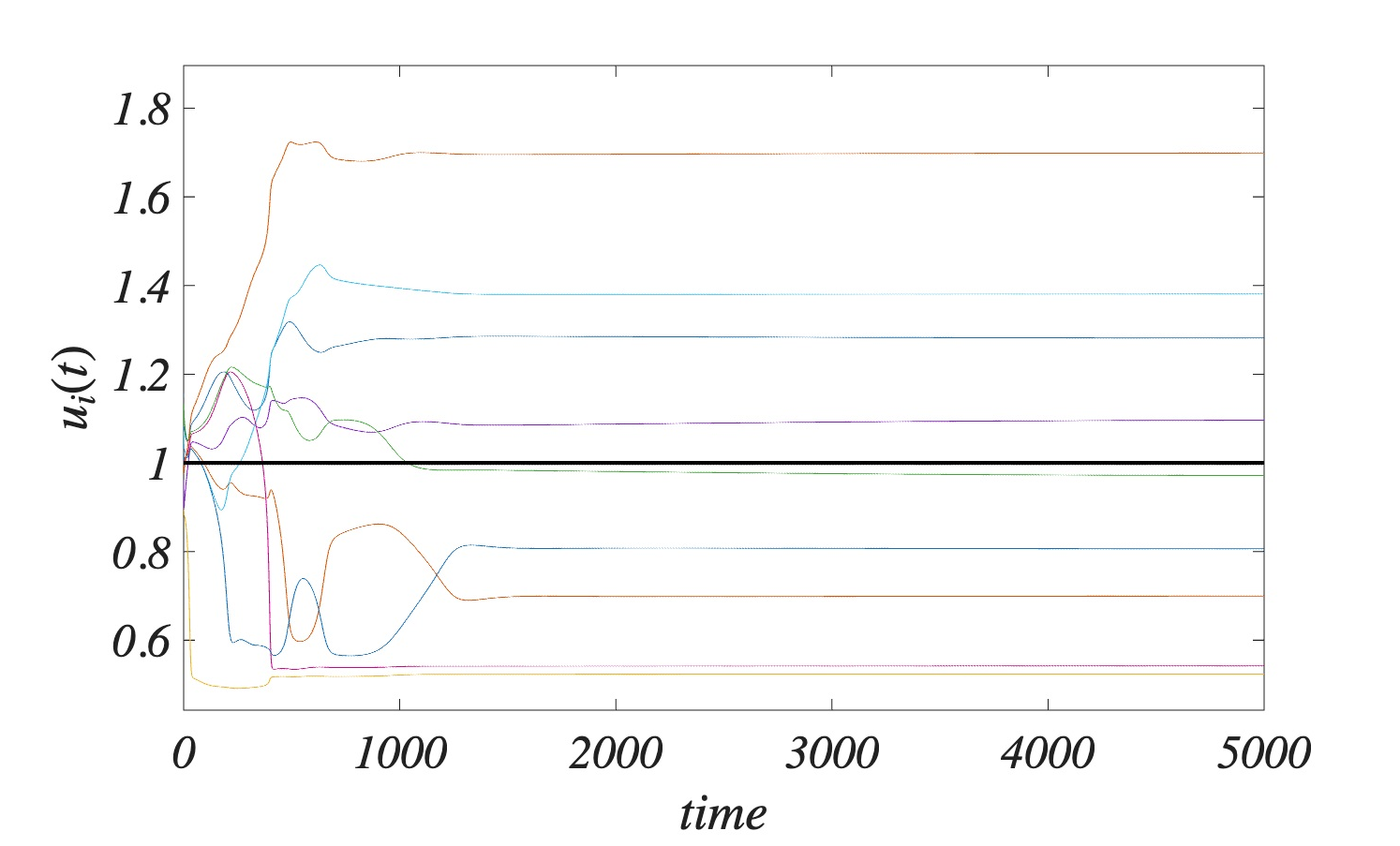}
\caption{Turing patterns induced by the cubic adaptive response, $k(w)=w(1/2-w)(1-w)$ and $A^*=5.0\, 10^{-4}$. Top panel: dispersion relation for the initial static network, the red dots denote the dispersion relation computed for the Laplacian eigenvalues,  $\rho(\Lambda^{(\alpha)})$, while the blue curve stands for $\rho(x)$ and it has been drawn for ease of visualization; middle panel: time evolution of the weights $w_{ij}(t)$. Bottom panel: time evolution of species $u$ in each node, $u_i(t)$.}
\label{fig:AdaptTPcub}
\end{figure}

\section{Conclusions}
\label{sec:conclusions}

In this work we proposed a theory of Turing patterns on adaptive networks, namely we allow links weights to dynamically evolve depending on the nodes states, \teo{the latter being in turn influenced by a nonlinear function of the weights}. We thus enlarge even more the applicability domain of Turing theory to explain the emergence of self-organized patterns. Let us stress that in the proposed framework, the dynamical weights are responsible for the strength of the interaction among nodes and thus they modify diffusion of species $u$ and $v$ across nodes; this modeling scheme is aligned with many other ones in the framework of adaptive dynamical systems defined on networks. Hence we do not take into account any ``physical'' deformation of the network, nor for links neither for nodes, thus our model does not account for dilution or increase in concentration due to nodes shrinking or growth. This is an interesting research direction that could be studied in a forthcoming work.

The proposed theory relies on the use of the Laplacian eigenvalues to compute the dispersion relation, allowing to determine the onset of patterns depending on the model parameters and network topology. The theory is general enough to be applied to any complex network; let us observe that in this work we limited ourselves to consider symmetric networks with positive weights, but the theory can be generalized to deal with the undirected case as well. Moreover it would be interesting to study the interplay between the adaptive mechanism and the network directionality; stated differently, one could study if weights evolution induces directionality in an otherwise initially undirected network, once Turing patterns emerge.

Already in the symmetric network framework the dynamics is very rich. We have shown the existence of parameters and adaptive response functions for which Turing patterns emerge, even once they could not in the counterpart of static network. We have also emphasized the existence of ``bursty'' Turing patterns, where the system alternates between a resting phase toward the homogeneous solution followed by a sudden divergence from the latter, thus creating short breathing patterns. Interestingly enough, we have proved that links adaptation can reduce the number of available links in the network by forcing their weights to vanish and thus break the network into small parts. Despite the latter being composed of few nodes, they can support Turing patterns because the dynamics slowly confine species densities to such a configuration, namely, far enough from the homogeneous equilibrium to be unable to fall back to it once isolated in small modules.

Let us observe that in this work we assumed, as very often done by other scholars in the framework of Turing patterns, that the diffusion coefficients do not depend on the spatial location but are intrinsic property of the diffusing species; one can relax this assumption by considering
\begin{equation*}
\begin{dcases}
\frac{d{u_i}}{dt}=f(u_i,v_i)+D_{u,i}\sum_\ell \mathcal{L}_{i\ell}u_\ell\\
\frac{d{v_i}}{dt}=g(u_i,v_i)+D_{v,i}\sum_\ell \mathcal{L}_{i\ell}v_\ell\, .
\end{dcases}
\end{equation*}
However this case can be analyzed by ``absorbing'' the diffusion coefficients into the Laplacian matrices, i.e., by defining
\begin{eqnarray*}
\hat{\mathcal{L}}^{(u)}_{ij}= D_{u,i} \mathcal{L}_{ij} \text{ and } \hat{\mathcal{L}}^{(v)}_{ij}= D_{v,i} \mathcal{L}_{ij}\, .
\end{eqnarray*}
Proceeding in this way, we are facing to asymmetric matrices, $\hat{\mathcal{L}}^{(u)}_{ij}\neq \hat{\mathcal{L}}^{(u)}_{ji}$ and $\hat{\mathcal{L}}^{(v)}_{ij}\neq \hat{\mathcal{L}}^{(v)}_{ji}$, whose spectrum can thus be complex and this property should be taken into account in deriving the condition for the instability. Moreover each species diffuses according to a different Laplacian matrix, i.e., $\hat{\mathcal{L}}^{(u)}_{ij}\neq \hat{\mathcal{L}}^{(v)}_{ij}$, namely the system is defined on a multigraph and not a simple graph, i.e., couples of nodes can be connected by two links, one for each involved species. Turing patterns have been already studied in the case of directed networks~\cite{asllani2014theory} and in the case of multigraphs~\cite{Asllani2016}, but, to the best of our knowledge, never by consider together those properties. This is thus an interesting research direction that deserves to be analyzed by including also adaptation in a forthcoming work.

In this work we have considered a deterministic setting, however Turing patterns have been studied in presence of intrinsic or demographic noise~\cite{fluctTuring,stochTuring} and new interesting phenomena have been brought to the fore, i.e., the presence of quasipatterns and an enlargement of the parameters region for which Turing patterns emerge. It would thus be interesting to extend the adaptive framework hereby considered as to consider the presence of noise.

Our work thus opens the way to several other applications of patterns emergence on adaptive complex networks, beyond the examples hereby proposed and may pave the path to better understanding the emergence of different types of instabilities and spatio-temporal patterns that occur in real-world non-equilibrium systems~\cite{kelso2012multistability}.\\

\noindent\textbf{Acknowledgments}. \teo{We would like to thank the anonymous Reviewer who suggested to consider the time evolution of the $L^2$ norm of the rate of change of species $u$ to demonstrate the presence of Turing patterns.}

\clearpage
\bibliography{reference}

\clearpage
\newpage
\onecolumngrid
\appendix

\renewcommand\theequation{{S-\arabic{equation}}}
\renewcommand\thetable{{S-\Roman{table}}}
\renewcommand\thefigure{{S-\arabic{figure}}}
\setcounter{equation}{0}
\setcounter{figure}{0}
\setcounter{section}{0}

\section{Patterns induced by adaptive quadratic case}
\label{sec:pattadaptquad}
For the sake of completeness we hereby present the results about Turing patterns induced by a parabola-like adaptive response function, to complement the ones presented in Section~\ref{sec:numrel} obtained by using a cubic-like adaptive function.

\begin{figure}[ht]
\centering
\includegraphics[scale=0.22]{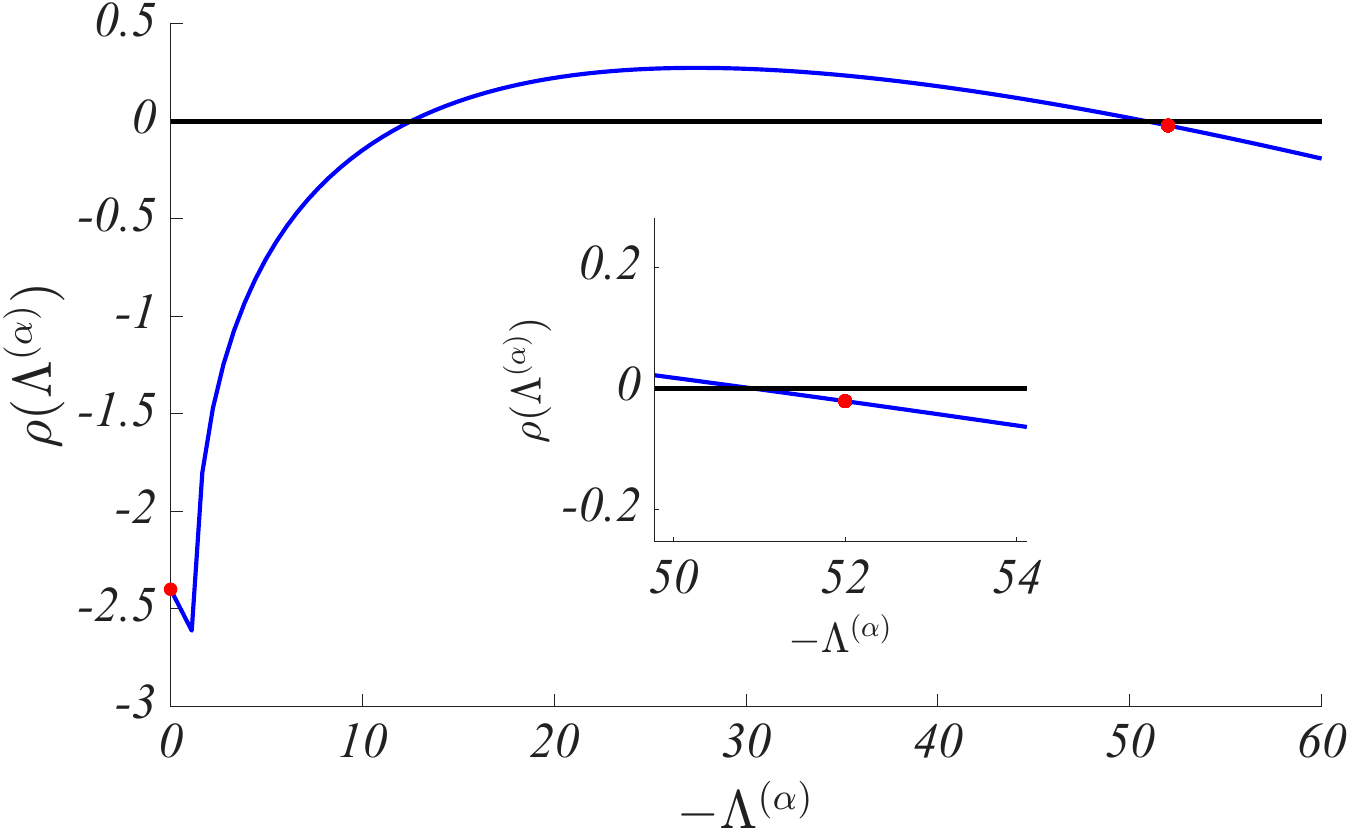}\quad
\includegraphics[scale=0.22]{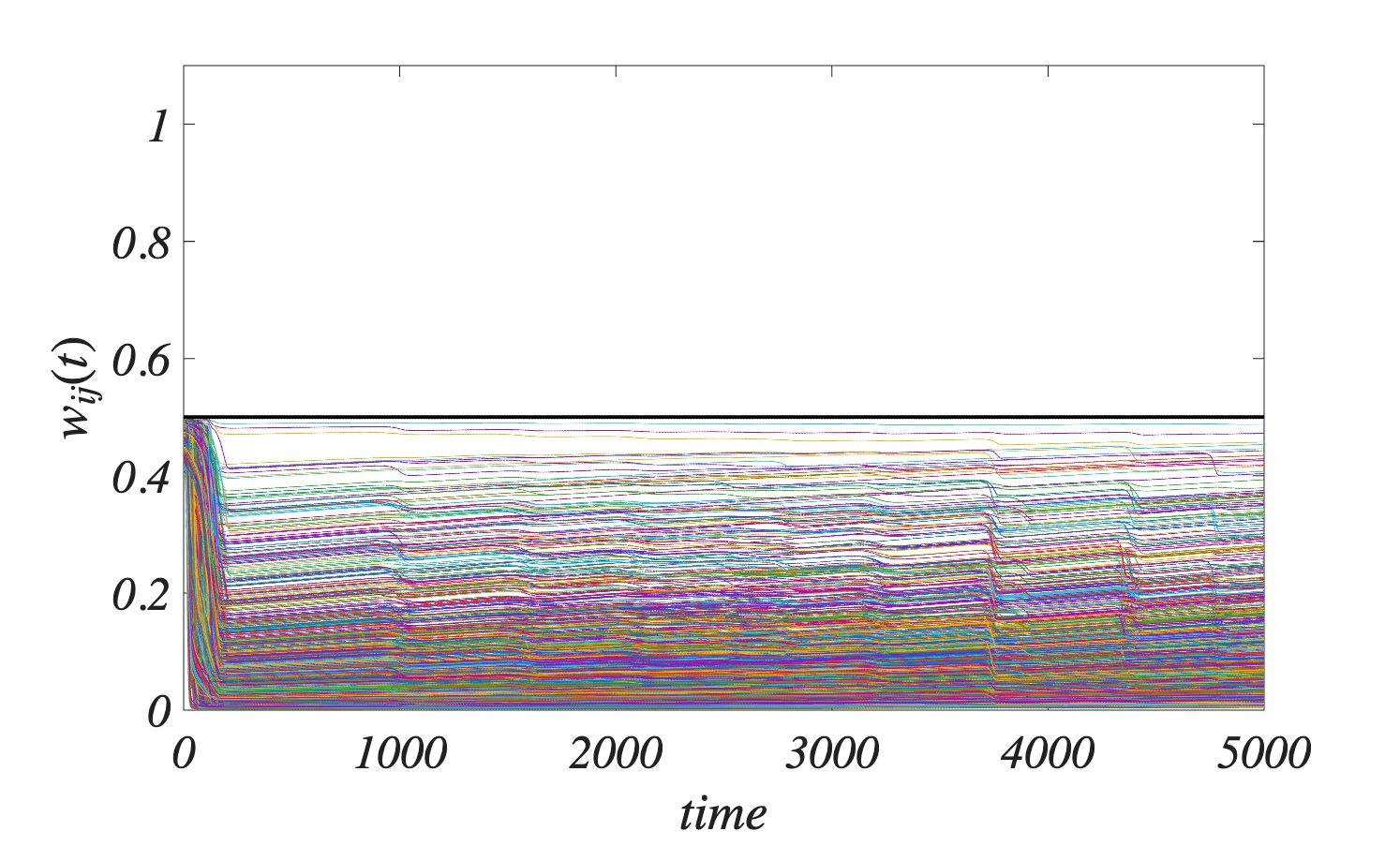}\quad
\includegraphics[scale=0.22]{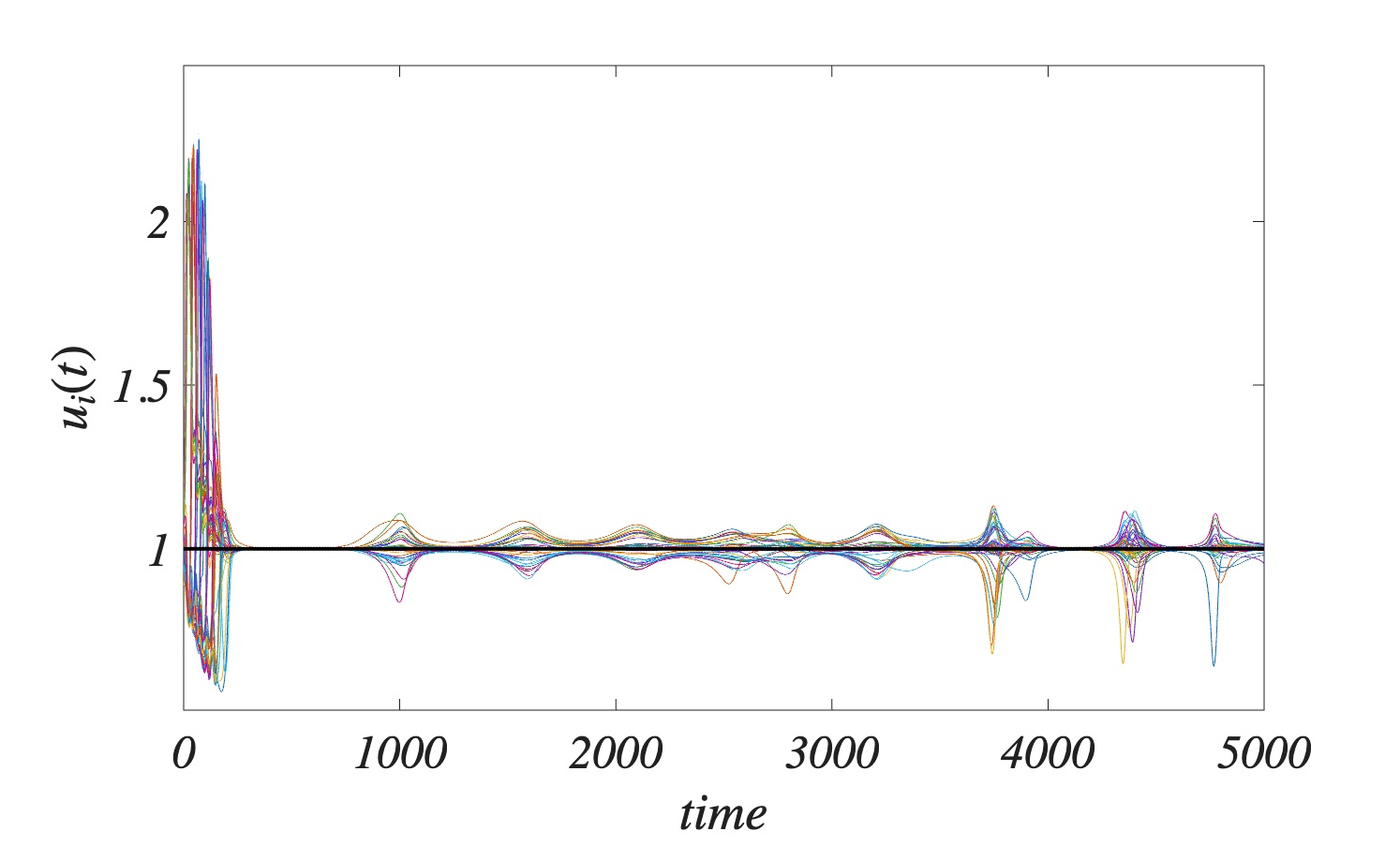}
\caption{Turing patterns induced by the quadratic adaptive response, $k(w)=w(1/2-w)$ and $A^*=5.0\, 10^{-4}$. Left panel:  dispersion relation for the initial static network, the red dots represent $\rho(\Lambda^{(\alpha)})$, where $\Lambda^{(\alpha)}$ denote the Laplacian eigenvalues, $\alpha=1,\dots,n$, while the blue curve stands for $\rho(x)$ and it has been drawn for ease of visualization; middle panel: time evolution of the weights $w_{ij}(t)$. Right panel: time evolution of species $u$ in each node, $u_i(t)$.}
\label{fig:AdaptTPparab}
\end{figure}

We thus assume to deal with a network composed by $n=52$ nodes, whose initial weights are uniformly drawn in the interval $(w_1^* - \delta , w_1^*)$, with $w_1^*=1/2$ and $\delta=0.1$, while initial densities of species $u$, resp. $v$, are drawn in the interval $(u^*-\delta,u^*+\delta)$, resp. $(v^*-\delta,v^*+\delta)$. The system evolution is described by Eq.~\eqref{eq:modeluvwBrus} with $k(w)=w(w_1^*-w)$, $w_1^*=1/2$ and $A^*=5.0\, 10^{-4}$; the remaining model parameters are set to $b = 4.0$, $c=7.8$, $D_u = 0.07$ and $D_v = 0.7$. In this way, patterns cannot emerge in the static network case, being the dispersion relation negative (see left panel of Fig.~\ref{fig:AdaptTPparab}). Because of adaptation, weights initially decrease (see middle panel of Fig.~\ref{fig:AdaptTPparab}) and thus the dispersion relation can now assume positive values, and patterns emerge as shown in the right panel of Fig.~\ref{fig:AdaptTPparab}.
We can observe that all the links are still active, i.e. $w_{ij}(t)>0$, for all $i\neq j$ and $t>0$, however, they evolve in time with sudden changes that induce a bursty behavior in $u_i(t)$, returning a sort of intermittent creation and disappearance of patterns. Let us observe that by using larger values of $A^*$ we can obtain patterns similar to the ones shown in the right column of Fig.~\ref{fig:Astarvarparab} (data not shown).

\section{Simulations with global amplitude in the parabola-like case}
\label{sec:globAmpli}

\begin{figure*}[ht]
\centering
\includegraphics[scale=0.25]{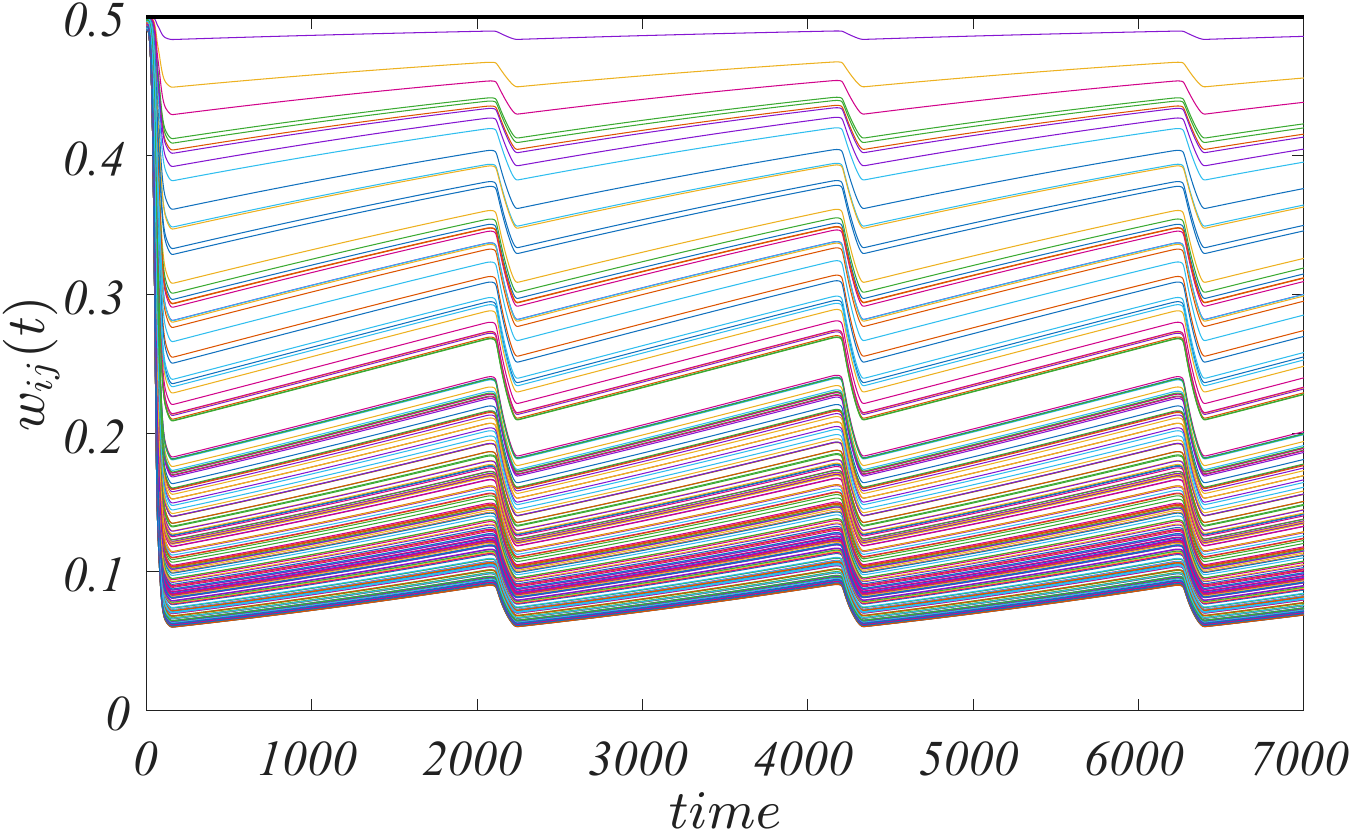}\includegraphics[scale=0.25]{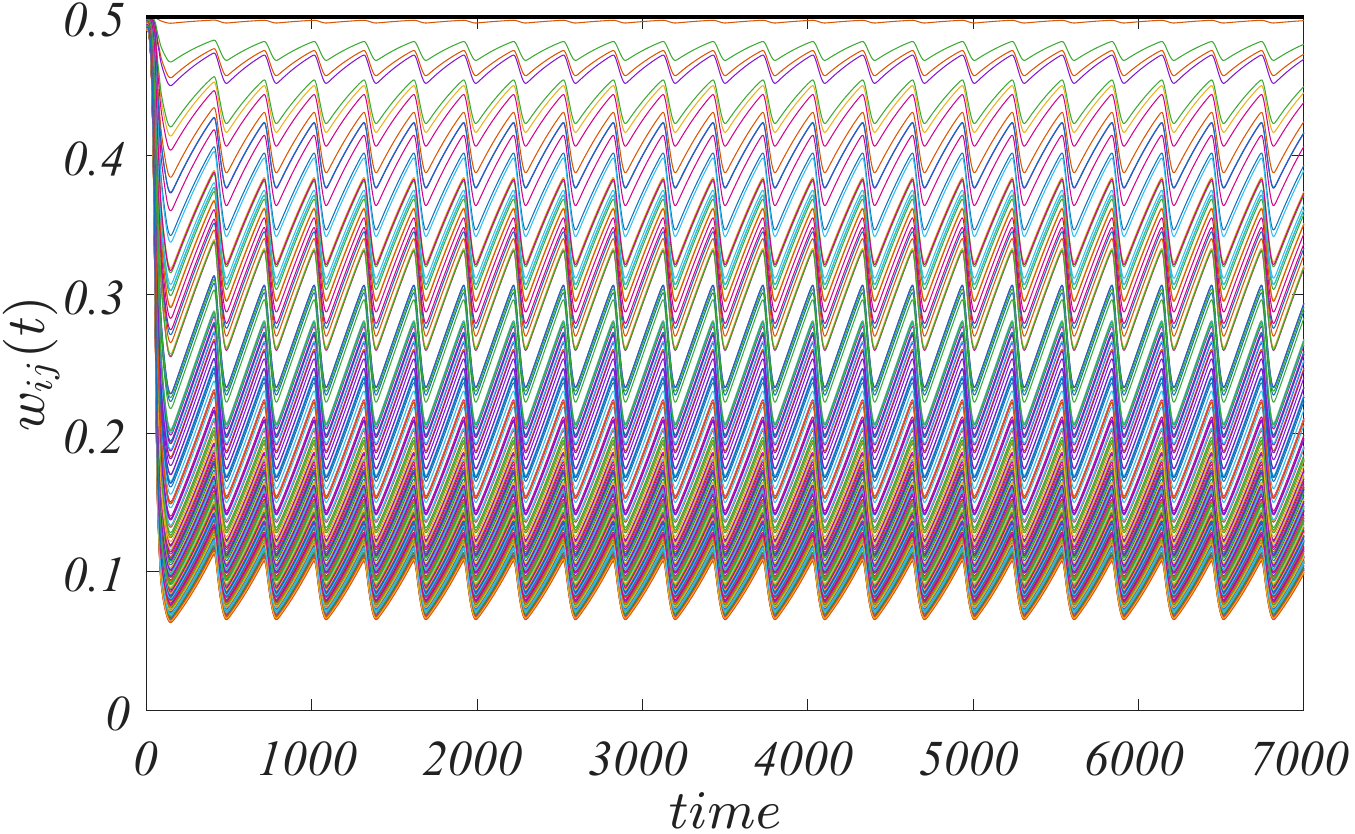}\includegraphics[scale=0.25]{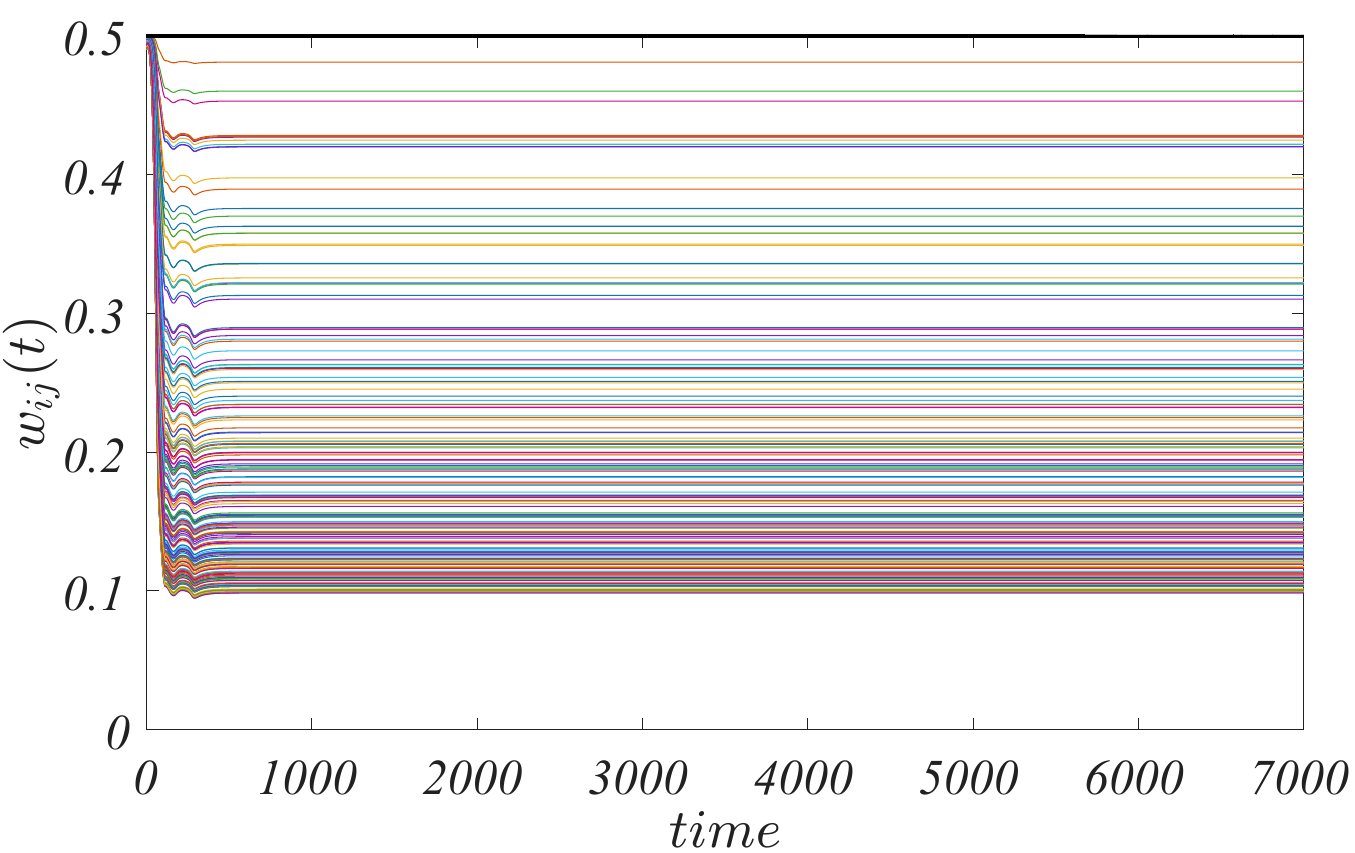}\\
\includegraphics[scale=0.25]{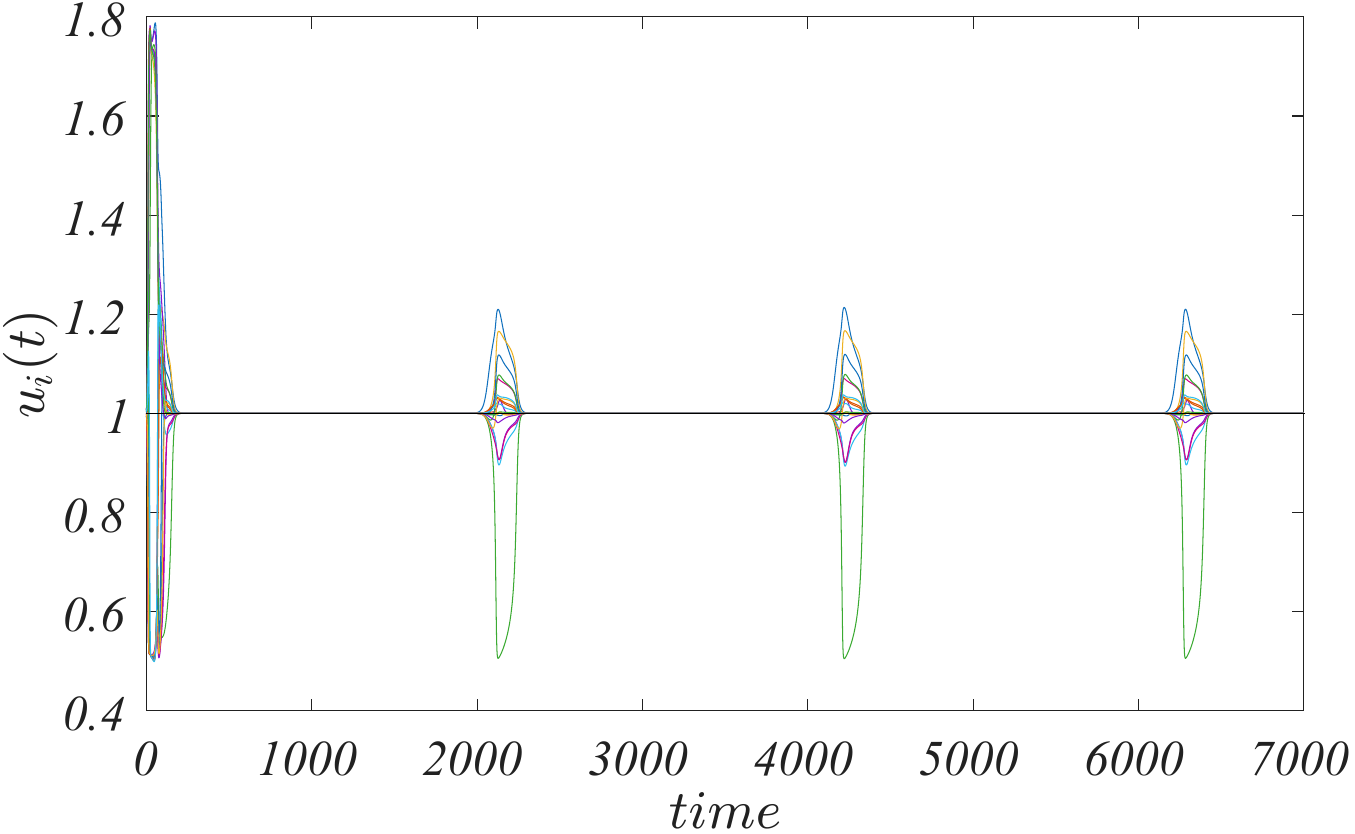}\includegraphics[scale=0.25]{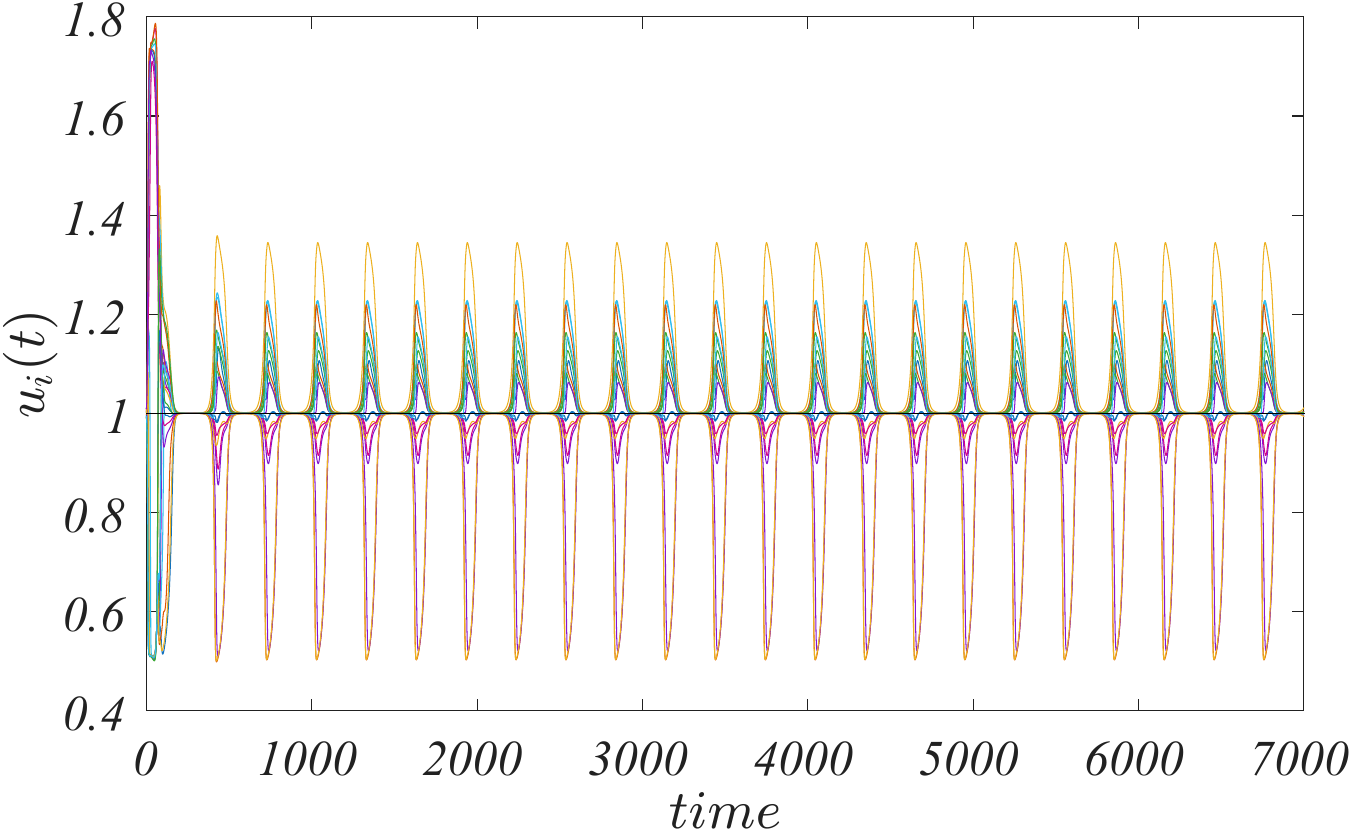}\includegraphics[scale=0.25]{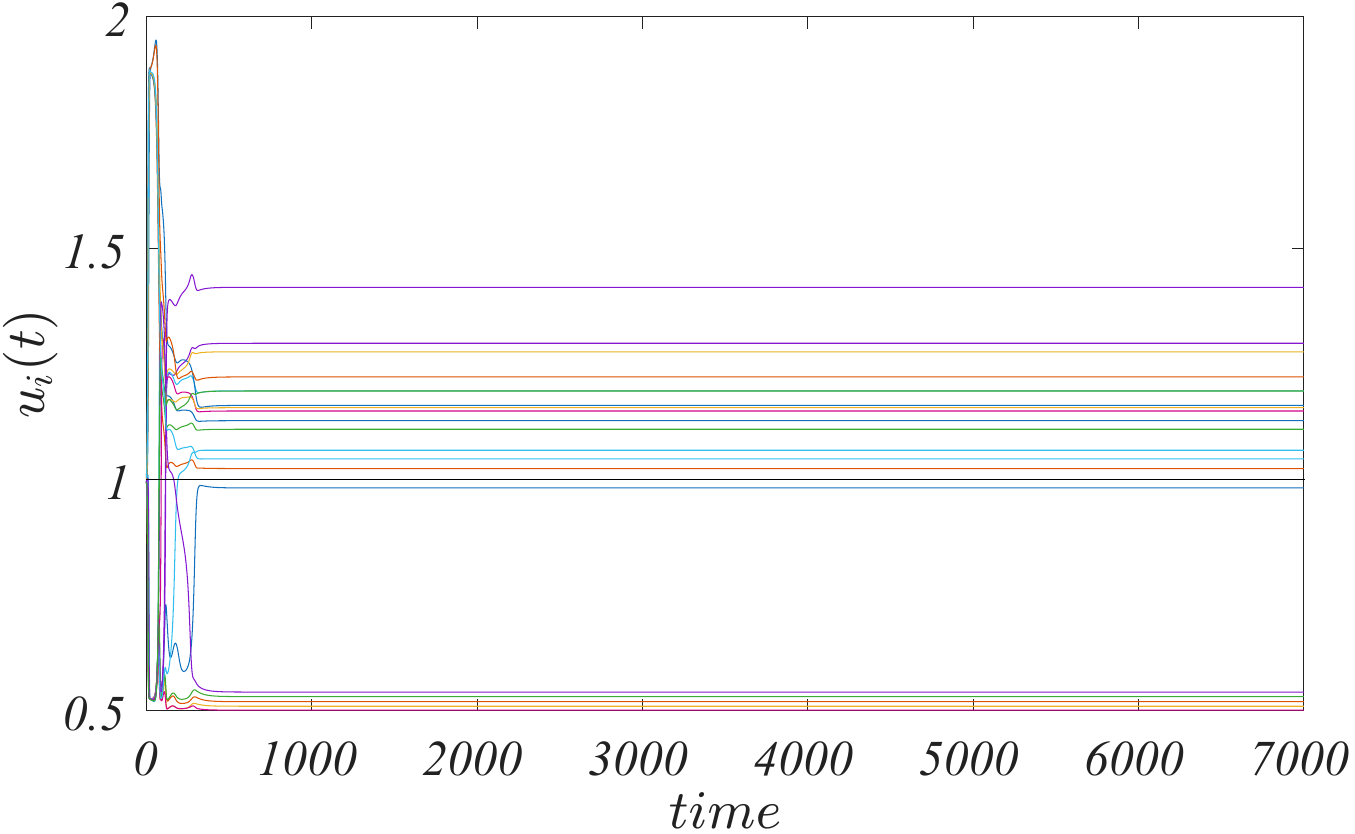}\\
\includegraphics[scale=0.28]{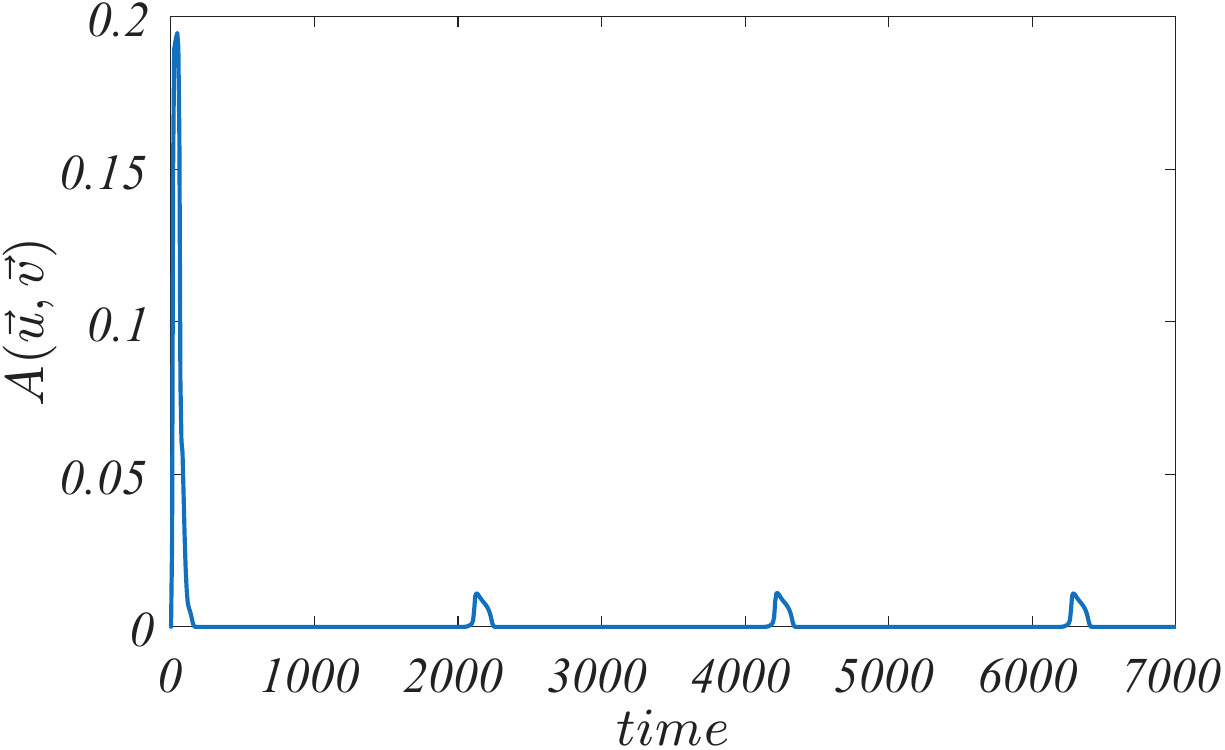}\includegraphics[scale=0.28]{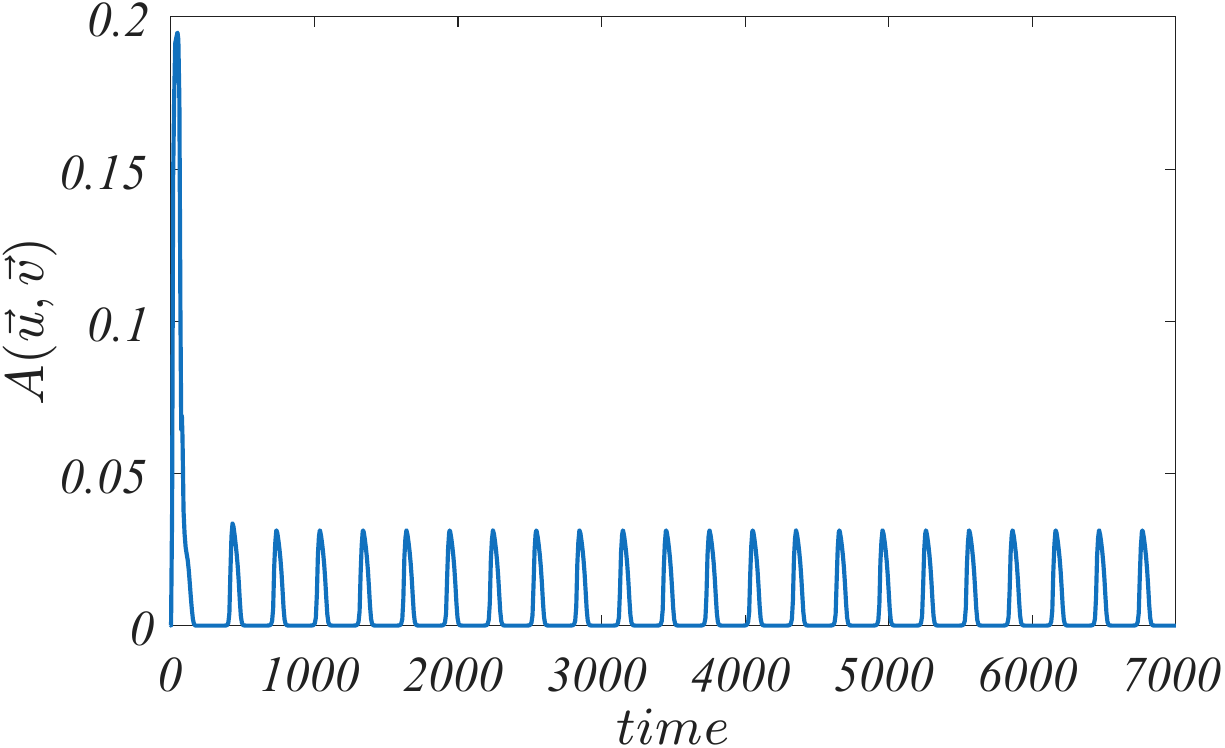}\includegraphics[scale=0.28]{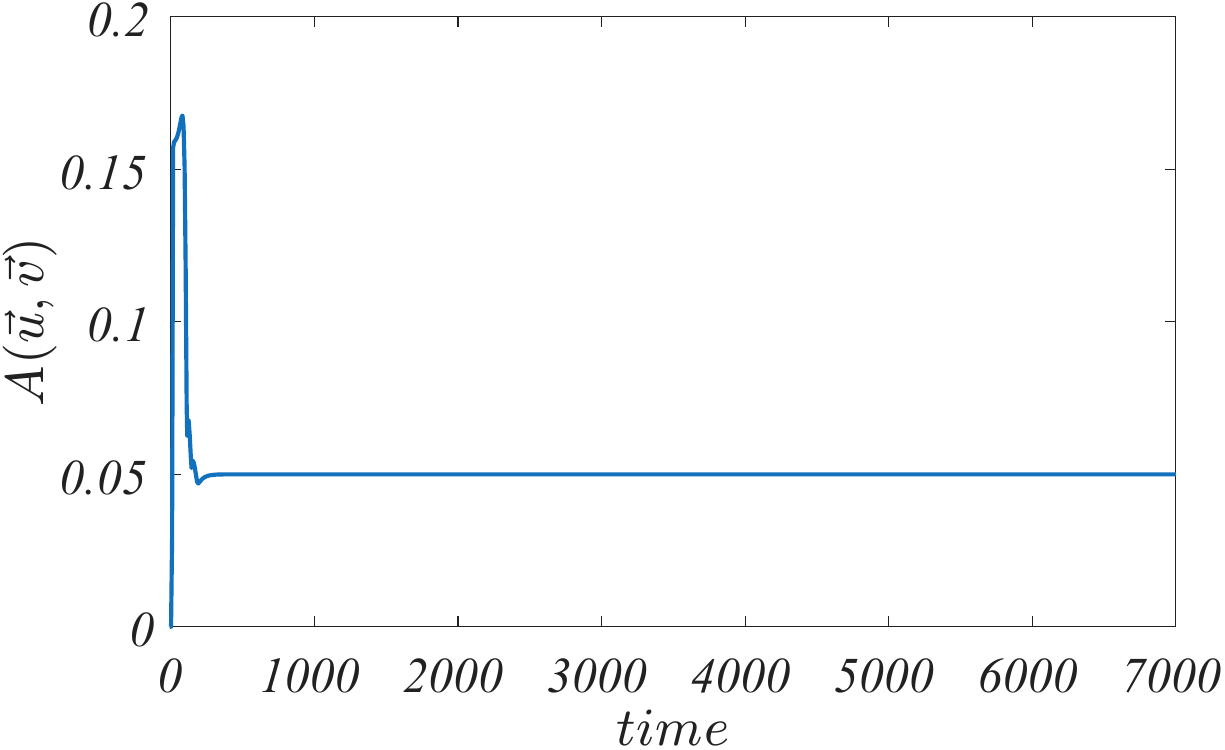}
\caption{The impact of $A^*$ on the dynamics defined via the global amplitude by using the parabola-like adaptive response. Top row: $w_{ij}(t)$, middle row: $u_i(t)$, lower row $A_{loc}(t)$. Left column $A^*=5.0\,10^{-4}$, middle column $A^*=5.5\,10^{-3}$, right column $A^*=5.0\,10^{-2}$.}
\label{fig:4withAuv}
\end{figure*}

For sake of completeness, we report in Fig.~\ref{fig:4withAuv} the results of the numerical simulations for the parabola-like case by using the global amplitude defined in the main text as
\begin{equation*}
A(\vec{u},\vec{v})=\frac{1}{2n}\sum \left[(u_i(t)-u^*)^2+(v_i(t)-v^*)^2\right]\, ,
\end{equation*}
where $\vec{u}=(u_1,\dots,u_n)^\top$, resp. $\vec{v}=(v_1,\dots,v_n)^\top$, denotes the state vector. In the figure, each column  corresponds to different values of $A^*$, while rows display, from top to bottom, the time evolution of the weights $w_{ij}(t)$, the dynamics of specie $u$ on each node, and the global amplitude $A(\vec{u},\vec{v})$ over time.

The behavior of $A(\vec{u},\vec{v})$ is qualitatively consistent with the one we observed for the local amplitude $A_{\mathit{loc}}$ discussed in the main text (see Fig.~\ref{fig:AdaptTPparab}), confirming that the global measure captures the same underlying dynamical mechanisms.

In the left column of Fig.~\ref{fig:4withAuv}, for small values of $A^*$, the system exhibits intermittent dynamics, with alternating phases where the global amplitude increases as the system goes away from the homogeneous equilibrium and phases where it decreases as the dynamics go back to it, leading to an irregular bursting behavior, also visible in the middle panel for $u_i(t)$.

For intermediate values of $A^*$, represented in the middle column of Fig.~\ref{fig:4withAuv}, a similar behavior is observed, but, as it is the case for the local amplitude, on a faster time scale. Finally, in the right column of Fig.~\ref{fig:4withAuv}, corresponding to sufficient large values of $A^*$, the intermittent regime can not be sustained and, after a transient, the amplitude converges to a steady value, consistent with the emergence of patterns.

\section{A more general Adaptive Mechanism. Application to networked FitzHugh-Nagumo oscillators}
\label{sec:anotheradapt}

In the main text, we have studied an adaptive mechanism~\eqref{eq:modelw3a} where the evolution of the weight $w_{ij}$ depends on the states of the nodes $i$ and $j$ via the local patterns amplitude~\eqref{eq:Aloc}, namely, a measure of the local distance for the homogeneous equilibrium. The aim of this section is to show that the proposed theory extends beyond this case, selected for pedagogical sake, and allows us to consider more general adaptive responses. Because of its relevance we decided to present a result in view of an application to patterns formation in a network of FitzHugh-Nagumo oscillators~\cite{FitzHugh1961,Nagumo1962,Rinzel1973,CarlettiNakao2020} (FHN), mimicking thus interacting neurons whose synapses adapt. To keep working in the same framework proposed in the main text and largely accepted in the literature, we assume again links weights evolve in time with an adaptive mechanism depending on the difference between state vectors of head and tail nodes of the link. More precisely, we considered the following response function
\begin{equation}
\label{eq:modelw}
 \frac{dw_{ij}}{dt} = w_{ij}\left[ a h(u_i-u_j,v_i-v_j)-b(w_{ij}+1)\right]\, ,
\end{equation}
with $a>b >0$, $u_i$ and $v_i$, denote the state of the $i$-th FHN neuron ruled by the following set of equations \teo{written by using non-dimensional variables and parameters}
\begin{equation}
\begin{dcases}
\frac{du_i}{dt} &=\mu u_i-u_i^3-v_i+D_u \sum_{j=1}^n \mathcal{L}_{ij}u_j\\
\frac{dv_i}{dt} &=\gamma (u_i-\alpha v_i )+D_v  \sum_{j=1}^n \mathcal{L}_{ij}v_j\, ,
\end{dcases}
\label{eq:FHNnet}
\end{equation}
where the Laplacian matrix, $\mathbf{\mathcal{L}}$, is again defined by
\begin{equation*}
 \mathcal{L}_{ij}=a_{ij}w_{ij}-\delta_{ij}\sum_{\ell} a_{i\ell}w_{i\ell}\, ,
\end{equation*}
being $a_{ij}$ the adjacency matrix of the fixed underlying network and $w_{ij}(t)$ the time-evolving weights. $\mu$, $\gamma$, $\alpha$, $D_u$ and $D_v$ are positive model parameters. \teo{Once again, the network topology, i.e., number of nodes, links and their position, is fixed and captured by the adjacency matrix elements $a_{ij}$; on the other hand weights evolve as a function of the variables $u$ and $v$ (see Eq.~\eqref{eq:modelw}) and the dynamics of the latter depend on the weights~\eqref{eq:FHNnet} via the Laplace matrix $\mathcal{L}$. We are thus facing once again to a system defined on an adaptive network}. 

The response function~\eqref{eq:modelw} contains the nonlinear function $h(x,y)$ that should satisfy the following assumptions:
\begin{equation}
\label{eq:assumpth}
h(x,y)\ll 1 \text{ if $|x|\gg 1$ and $|y|\gg 1$,} \, ah(0,0) > b \text{ and }h(-x,-y)=h(x,y)\, .
\end{equation}
The last condition ensures that, if weights are symmetric at time $t=0$, then they will remain symmetric for all $t>0$, indeed
\begin{eqnarray*}
 \frac{d(w_{ij}-w_{ji})}{dt}&=& aw_{ij} h(u_i-u_j,v_i-v_j)-bw_{ij}^2 - bw_{ij}-aw_{ji} h(u_j-u_i,v_j-v_i)+bw_{ji}^2 + bw_{ji}\\
 &=& (w_{ij}-w_{ji}) \left[a h(u_i-u_j,v_i-v_j)-b-b(w_{ij}+w_{ji})\right]\, ,
\end{eqnarray*}
from which the claim follows. Notice that $ah(0,0) > b$ implies that $h(0,0)>0$. Let us observe that the response function proposed in~\cite{Chakravartula2017} given by
\begin{equation}
\label{eq:hfunc}
h(x,y)=e^{-c(x^2+y^2)}\, ,
\end{equation}
$c >0$, satisfies the above conditions~\eqref{eq:assumpth}.

Let us briefly describe the rationale of the above model. If the states of neurons $i$ and $j$ are far each other, i.e., $|u_i-u_j|\gg 1$ and $|v_i-v_j|\gg 1$, namely they are desynchronized, because of the first condition~\eqref{eq:assumpth}, then Eq.~\eqref{eq:modelw} reduces to
\begin{equation*}
 \frac{dw_{ij}}{dt}\sim -bw_{ij}(w_{ij}+1)\, ,
\end{equation*}
namely the weight between $i$ and $j$ decreases, and thus those nodes will behave independently of each other, and they will converge to the homogeneous equilibrium, that is they will get synchronized. Let us observe that Eq.~\eqref{eq:modelw} always admits $w_{ij}=0$ as solution, however if $|u_i-u_j|\ll 1$ and $|v_i-v_j|\ll 1$, namely, the neurons are synchronized, the latter simplifies into
\begin{equation*}
 \frac{dw_{ij}}{dt}\sim w_{ij}(ah(0,0)-b)\, ;
\end{equation*}
because $ah(0,0)>b$, the weight $w_{ij}$ increases and thus neurons $i$ and $j$ will lose synchrony. Finally, let us observe that Eq.~\eqref{eq:modelw} admits a second solution in the case of (almost) synchronized neurons $i$ and $j$, namely 
\begin{equation}
\label{eq:wstarsol}
w_{ij}^*=a_{ij}\left(\frac{a}{b}h(0,0)-1\right)\, .
\end{equation}
 This is the equilibrium weight we will consider in the process of Turing instability.

To study the emergence of Turing patterns~\cite{CarlettiNakao2020} in the FHN adaptive model, let us consider the stationary homogeneous equilibrium for the state vectors and the link weights
\begin{equation}
\label{eq:equilibhomog}
u_i=u^*=0\, , v_i=v^*=0\text{ and }w_{ij}^*=a_{ij}\left(\frac{a}{b}h(0,0)-1\right)=:a_{ij}w^*\, .
\end{equation}
It is trivial to check that $(u^*,v^*)=(0,0)$ is a solution of system~\eqref{eq:FHNnet}. Assume moreover this solution to be stable for each isolated neuron, namely the Jacobian matrix $\mathbf{J}_0=\left(
\begin{smallmatrix}
\mu & -1\\ \gamma& -\alpha\gamma
\end{smallmatrix}
\right)$, has a negative spectrum. A straightforward computation allows to conclude that the latter condition is satisfied if
\begin{equation}
\label{eq:FHNstab}
\alpha \mu <1 \text{ and }\mu<\alpha \gamma\, .
\end{equation}

Our goal is to prove that under suitable conditions on the diffusion coefficients and the network topology, the null solution becomes unstable, namely Turing patterns arise. To prove our claim, we will perform a linear stability analysis by setting $u_i=\delta u_i+u^*$, $v_i=\delta v_i+v^*$ and $w_{ij}=\delta w_{ij}+w_{ij}^*$. Let us observe that
\begin{eqnarray*}
 \mathcal{L}_{ij}&=&a_{ij}(w_{ij}^*+\delta w_{ij})-\delta_{ij}\sum_\ell a_{i\ell}(w_{i\ell}^*+\delta w_{i\ell})=a_{ij}\left[a_{ij}\left(\frac{a}{b}h(0,0)-1\right)+\delta w_{ij}\right]-\delta_{ij}\sum_\ell a_{i\ell}\left[a_{i\ell}\left(\frac{a}{b}h(0,0)-1\right)+\delta w_{i\ell}\right]\\
 &=&\left(\frac{a}{b}h(0,0)-1\right) \left(a_{ij}-\delta_{ij}\sum_\ell a_{i\ell}\right)+\delta w_{ij} a_{ij}-\delta_{ij}\sum_\ell a_{i\ell}\delta w_{i\ell}\\
 &=&\left(\frac{a}{b}h(0,0)-1\right)L_{ij} +\delta w_{ij} a_{ij}-\delta_{ij}\sum_\ell a_{i\ell}\delta w_{i\ell}=w^*L_{ij} +\delta w_{ij} a_{ij}-\delta_{ij}\sum_\ell a_{i\ell}\delta w_{i\ell}\, ,
\end{eqnarray*}
where we introduced the Laplacian matrix of the underlying unweighted network~\eqref{eq:Lap}.

The linearization of Eqs.~\eqref{eq:modelw} and~\eqref{eq:FHNnet} gives
\begin{equation}
\label{eq:modeldeltauv}
\begin{cases}
\displaystyle \frac{d\delta u_i}{dt} = \mu\delta u_i- \delta v_i+D_uw^*\sum_j L_{ij}\delta u_j\\
\displaystyle \frac{d\delta v_i}{dt} = \gamma\delta u_i-\alpha\gamma \delta v_i+D_vw^*\sum_j L_{ij}\delta v_j\\
\displaystyle \frac{d\delta w_{ij}}{dt} = a_{ij}w^* a \left[\partial_u h(0,0) (\delta u_i-\delta u_j)+\partial_v h(0,0) (\delta v_i-\delta v_j)\right]-ba_{ij}w^* \delta w_{ij}\, ,
\end{cases}
\end{equation}

Let us introduce the $(2n)$-vector  $\delta \vec{x}=(\delta u_1,\delta v_1,\dots,\delta u_n,\delta v_n,)^\top$ and $\delta \vec{w}=(\delta w_{11},\dots,\delta w_{1n},\delta w_{21},\dots,\delta w_{2n},\dots,\delta w_{n1},\dots,\delta w_{nn})^\top$ that can be built by ``unrolling'' the matrix with elements $\delta w_{ij}$ row by row to form a $n^2$-vector. The system~\eqref{eq:modeldeltauv} can thus be rewritten in compact form as follows
\begin{equation}
\label{eq:modeldeltauvMat}
\frac{d}{dt}\left(
\begin{matrix}
 \delta \vec{x}\\\delta\vec{w}
\end{matrix}\right)=\left(
\begin{matrix}
 \mathbf{S} & \mathbf{O}\\\mathbf{C} & -bw^* \mathbf{A}'
\end{matrix}
\right)\left(
\begin{matrix}
 \delta \vec{x}\\\delta\vec{w}
\end{matrix}\right)\equiv \mathbf{M}\left(
\begin{matrix}
 \delta \vec{x}\\\delta\vec{w}
\end{matrix}\right)\,,
\end{equation}
With again $\mathbf{A}'=\mathrm{diag}(a_{11},a_{12},\dots,a_{1n},\dots, a_{nn})$, i.e., the $n^2$ diagonal matrix whose elements are the entries of the adjacency matrix $\mathbf{A}$ ``unrolled''. $\mathbf{M}$ is thus a $(2n+n^2)\times (2n+n^2)$ block matrix,  $\mathbf{O}$ is a rectangular matrix of size $2n\times n^2$ whose entries are all $0$. The matrix $\mathbf{S}$ is given by
\begin{equation*}
\mathbf{S}=\mathbf{I}_n\otimes \mathbf{J}_0+w^*L\otimes \left(
\begin{smallmatrix}
 D_u & 0\\0 & D_v
\end{smallmatrix}\right)\, , 
\end{equation*}
 and $\mathbf{C}$ is a $n^2\times 2n$ matrix, whose explicit form is not needed in the following analysis. $\mathbf{I}_{n}$ is the $n$-identity matrix,

The stability of system~\eqref{eq:modeldeltauvMat} is determined by the spectrum of the matrix $\mathbf{M}$ and because of its block shape the latter is the spectrum of $\mathbf{S}$ together with the eigenvalue $-bw^*$ that has multiplicity the number of links of $\mathbf{A}$, and the eigenvalue $0$ with multiplicity given by the ``number of empty entries'' in $\mathbf{A}$. 

Let us thus study the spectrum of $\mathbf{S}$. To make a step forward, let us introduce the eigenbasis of the Laplacian matrix ${L}$, namely ${L} \vec{\phi}^{(\alpha)} = \Lambda^{(\alpha)}\vec{\phi}^{(\alpha)}$, $\alpha=1,\dots, n$. From the properties of the Laplacian matrix we know that $\Lambda^{(1)}=0$, $\vec{\phi}^{(1)}\sim (1,\dots,1)^\top$, $\Lambda^{(\alpha)}<0$ for all $\alpha >1$ and $\vec{\phi}^{(\alpha)}\cdot \vec{\phi}^{(\beta)}=\delta_{\alpha\beta}$. Let us eventually define the matrix $\mathbf{V}$ given by
\begin{equation*}
 \mathbf{V}=\left(
\begin{matrix}
 \Phi\otimes \mathbf{I}_2 & \mathbf{O}\\\mathbf{O}^\top & \mathbf{I}_{n^2}
\end{matrix}\right)\, ,
\end{equation*}
where $\Phi$ is the $n\times n$ matrix whose columns are the eigenvectors $\vec{\phi}^{(\alpha)}$, namely $\Phi^\top\mathcal{L}\Phi=\mathrm{diag}(0,\Lambda^{(2)},\dots,\Lambda^{(n)})\equiv \mathbf{\Lambda}$. One can thus obtain
\begin{equation*}
 \mathbf{V}^\top \mathbf{M}\mathbf{V} = \left(
\begin{matrix}
 \mathbf{I}_n\otimes \mathbf{J}_0+w^*\mathbf{\Lambda}\otimes \left(
\begin{smallmatrix}
 D_u&0\\0& D_v
\end{smallmatrix}\right) & \mathbf{O}\\\mathbf{C}' & -bw^* \mathbf{A}'
\end{matrix}\right)\, ,
\end{equation*}
hence the spectrum of $\mathbf{S}$ is obtained by solving the $n$ characteristic problems
\begin{equation}
\label{eq:reldispeqApp}
\det \left[ \mathbf{J}_0+w^*\Lambda^{(\alpha)}\left(
\begin{smallmatrix}
 D_u&0\\0& D_v
\end{smallmatrix}\right)-\lambda \mathbf{I}_2\right]=0\quad \alpha=1,\dots,n\, ,
\end{equation}
from which one can determine the dispersion relation, $\rho(\Lambda^{(\alpha)})=\max \Re\lambda_{\pm}(\Lambda^{(\alpha)})$ where
\begin{equation*}
 \lambda_{\pm}(\Lambda^{(\alpha)}) = \frac{\mathrm{tr}\mathbf{J}_0+w^*\Lambda^{(\alpha)}(D_u+D_v)\pm\sqrt{\left(\mathrm{tr}\mathbf{J}_0+w^*\Lambda^{(\alpha)}(D_u+D_v)\right)^2-4 \Delta}}{2}\, ,
\end{equation*}
and
\begin{equation*}
\Delta = \left[\det \mathbf{J}_0+w^*\Lambda^{(\alpha)}\left(D_u \partial_v g+D_v\partial_u f\right)+(w^*)^2D_uD_v(\Lambda^{(\alpha)})^2\right]\, . 
\end{equation*}

Let us recall that the isolated equilibrium $(u^*,v^*)$ is stable, hence $\mathrm{tr}\mathbf{J}_0<0$. The sign of $\rho$ can thus be determined as follows. Because $\Lambda^{(\alpha)}\leq 0$ and $w^*>0$ we get $\mathrm{tr}\mathbf{J}_0+w^*\Lambda^{(\alpha)}(D_u+D_v)<0$. Then the condition to have $\rho(x) >0$ is $\Delta <0$, namely
\begin{equation*}
\begin{cases}
\displaystyle \det \mathbf{J}_0 - \frac{(D_u\partial_v g+D_v\partial_u f)^2}{4D_uD_v}<0\\
\displaystyle \frac{D_u\partial_v g+D_v\partial_u f}{2w^* D_uD_v}>0 \, ,
\end{cases}
\end{equation*}
and, being $w^*>0$ and $D_uD_v>0$, it is equivalent to
\begin{equation}
\label{eq:Turing2}
\begin{cases}
\displaystyle \det \mathbf{J}_0 - \frac{(D_u\partial_v g+D_v\partial_u f)^2}{4D_uD_v}<0\\
\displaystyle D_u\partial_v g+D_v\partial_u f>0\, .
\end{cases}
\end{equation}
By using the expression for $\mathbf{J}_0$ we get
\begin{equation}
\label{eq:Turing3}
\begin{cases}
\displaystyle (1-\alpha \mu)\gamma - \frac{(-\alpha \gamma D_u+\mu D_v)^2}{4D_uD_v}<0\\
\displaystyle -\alpha \gamma D_u+\mu D_v>0\, .
\end{cases}
\end{equation}

So, by taking into account~\eqref{eq:FHNstab}, the conditions for the emergence of Turing patterns are given by (see also~\cite{CarlettiNakao2020}) 
\begin{eqnarray}
\label{eq:TPcondFHN}
\alpha \mu <1 \\ 
\mu<\alpha \gamma \\
\mu D_v>\alpha \gamma D_u \\
(1-\alpha \mu)\gamma - \frac{(-\alpha \gamma D_u+\mu D_v)^2}{4D_uD_v}<0\, .
\end{eqnarray}
Because the dynamics evolve on a network, namely a discrete support, the above conditions are necessary but not sufficient, indeed one must require the existence of Laplacian eigenvalues, $\Lambda^{(\alpha)}$, falling in the interval $(\Lambda_{-},\Lambda_{+})$ where
\begin{eqnarray}
\Lambda_{-} = -\frac{1}{w^*}\left(\frac{\mu}{2D_u}-\frac{\alpha\gamma}{2D_v}\right)-\frac{1}{2w^*}\sqrt{\left(\frac{\mu}{D_u}-\frac{\alpha\gamma}{D_v}\right)^2-\frac{4\gamma(1-\mu\alpha)}{D_uD_v}},
\cr
\Lambda_{+} = -\frac{1}{w^*}\left(\frac{\mu}{2D_u}-\frac{\alpha\gamma}{2D_v}\right)+\frac{1}{2w^*}\sqrt{\left(\frac{\mu}{D_u}-\frac{\alpha\gamma}{D_v}\right)^2-\frac{4\gamma(1-\mu\alpha)}{D_uD_v}}\, .
\label{eq:rootreldispFHNmain}
\end{eqnarray}
The latter condition ensures a positive dispersion relation once evaluated on the Laplacian spectrum.

\begin{figure}[ht]
\centering
\includegraphics[scale=0.22]{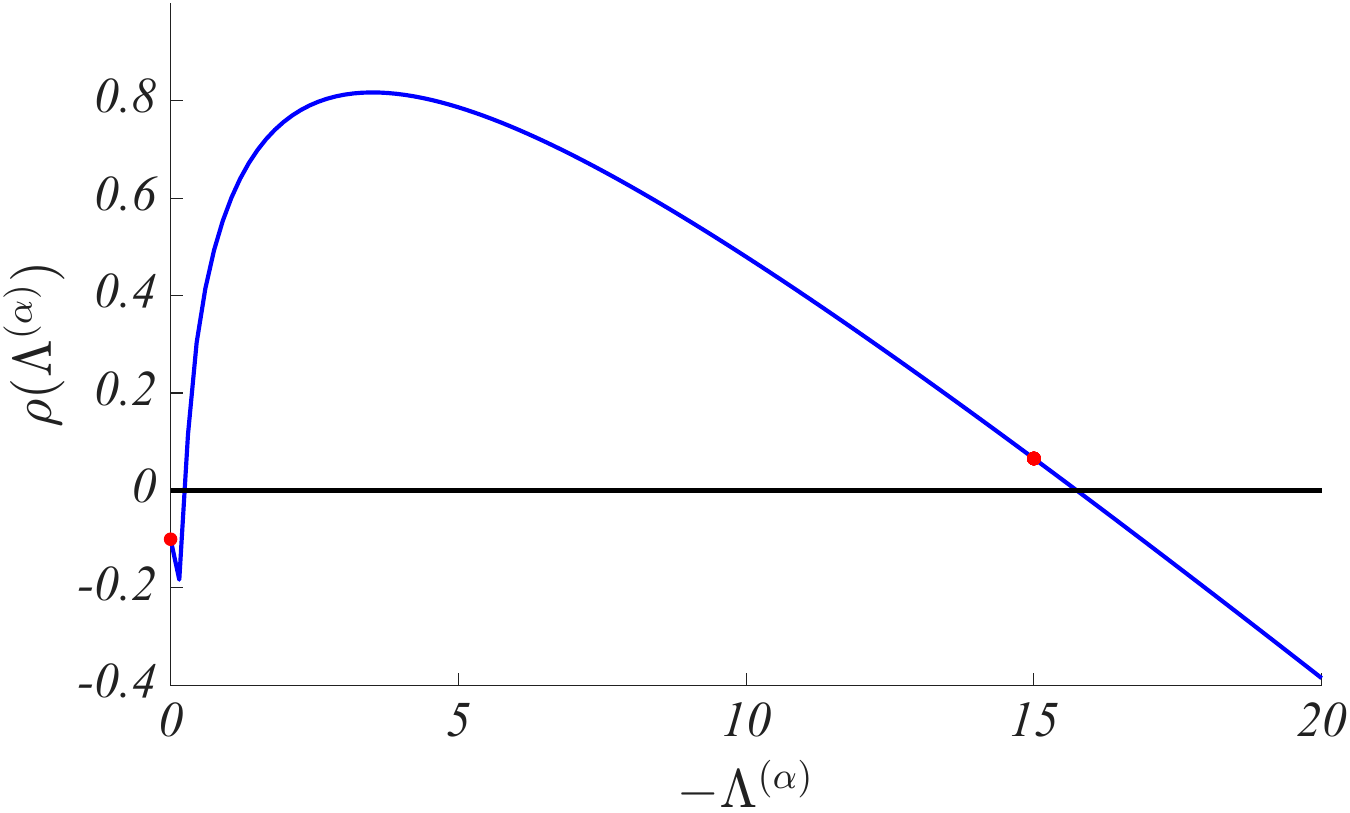}\quad
\includegraphics[scale=0.22]{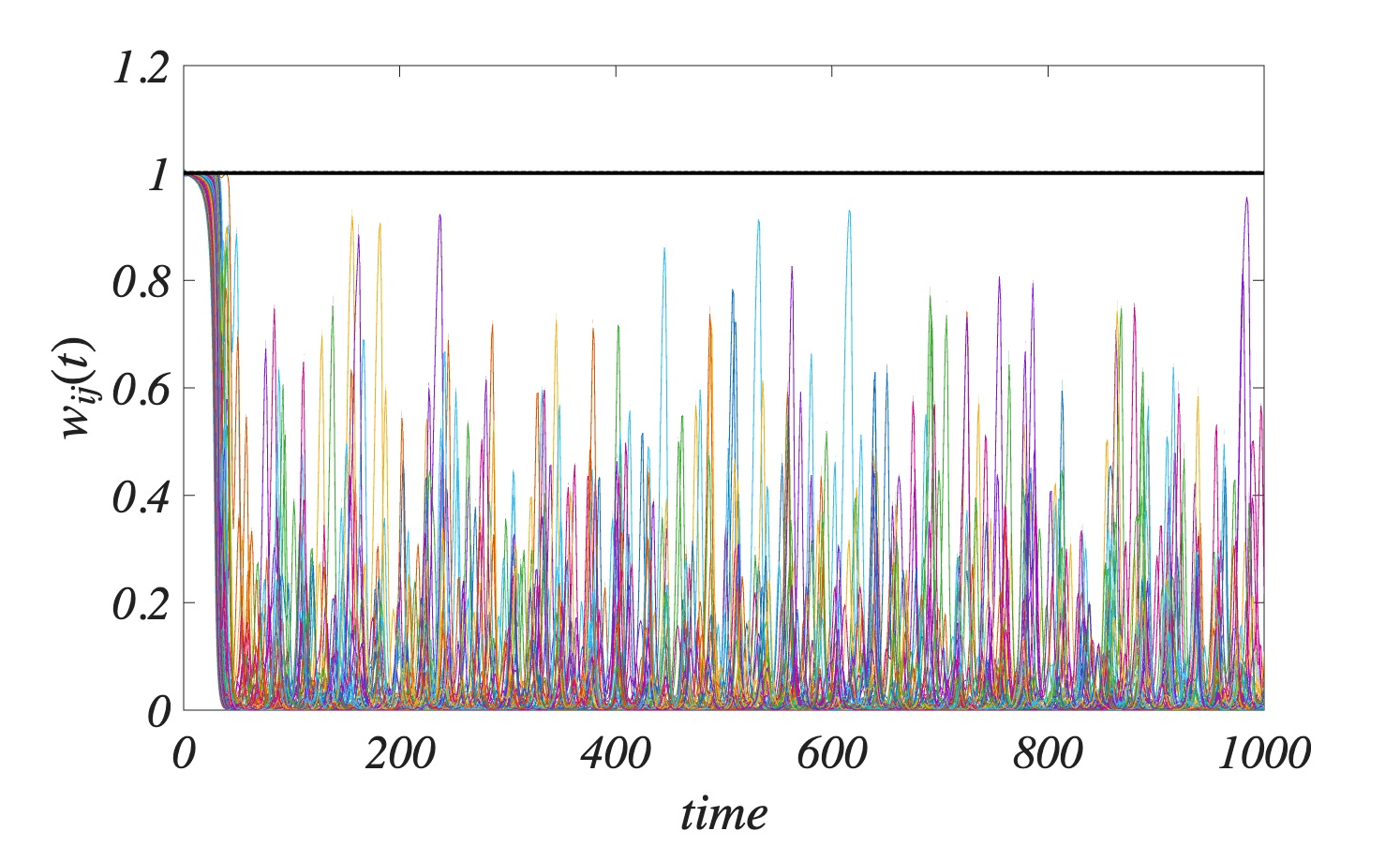}\quad
\includegraphics[scale=0.22]{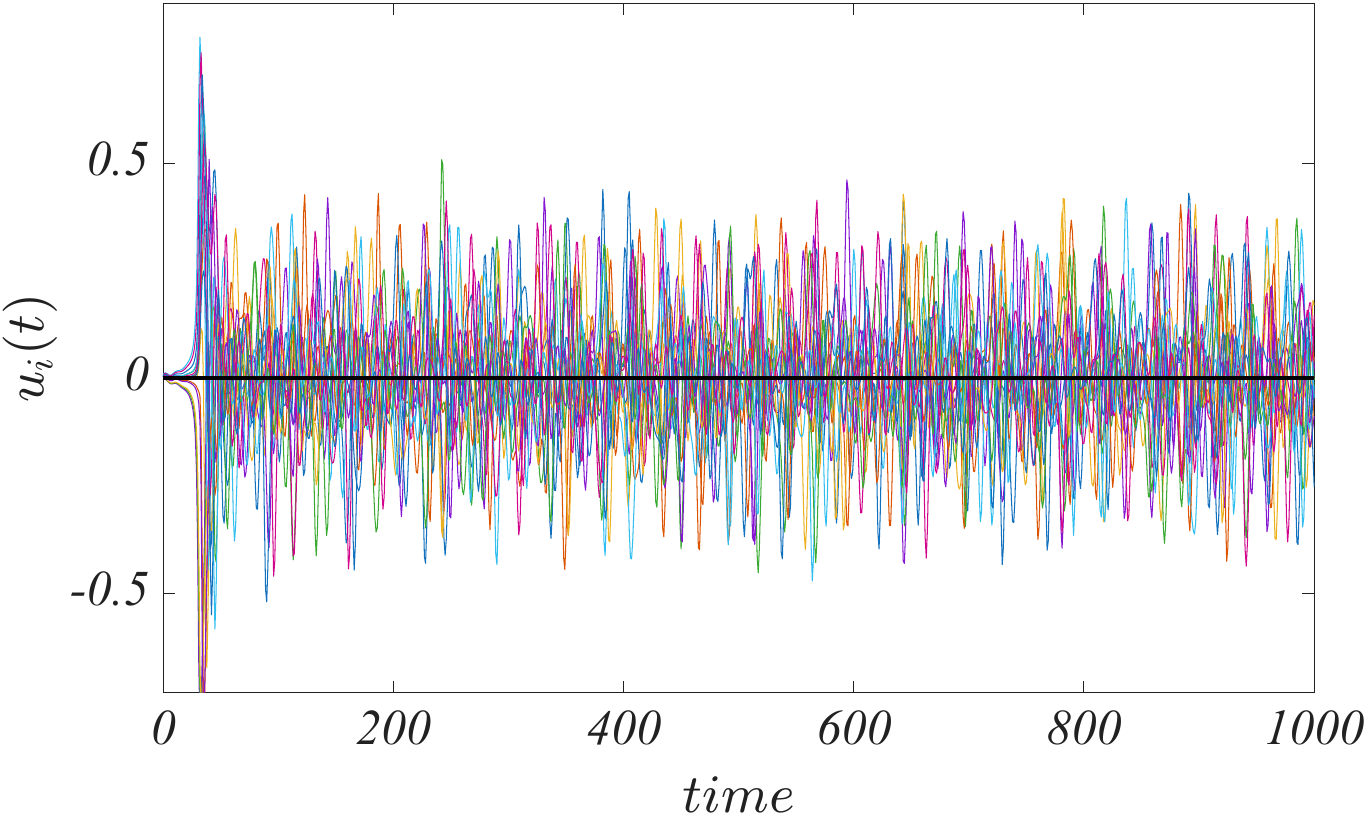}
\caption{Turing patterns in the FHN adaptive model with response function~\eqref{eq:modelw} and~\eqref{eq:hfunc}. Left panel: dispersion relation for the initial static network, the red dots represent $\rho(\Lambda^{(\alpha)})$, where $\Lambda^{(\alpha)}$, $\alpha=1,\dots,n$, denote the Laplacian eigenvalues, while the blue curve stands for $\rho(x)$ and it has been drawn for ease of visualization; middle panel: time evolution of the weights $w_{ij}(t)$. Right panel: time evolution of species $u$ in each node, $u_i(t)$.}
\label{fig:AdaptTPexp}
\end{figure}

\begin{figure}[H]
\centering
\includegraphics[scale=0.32]{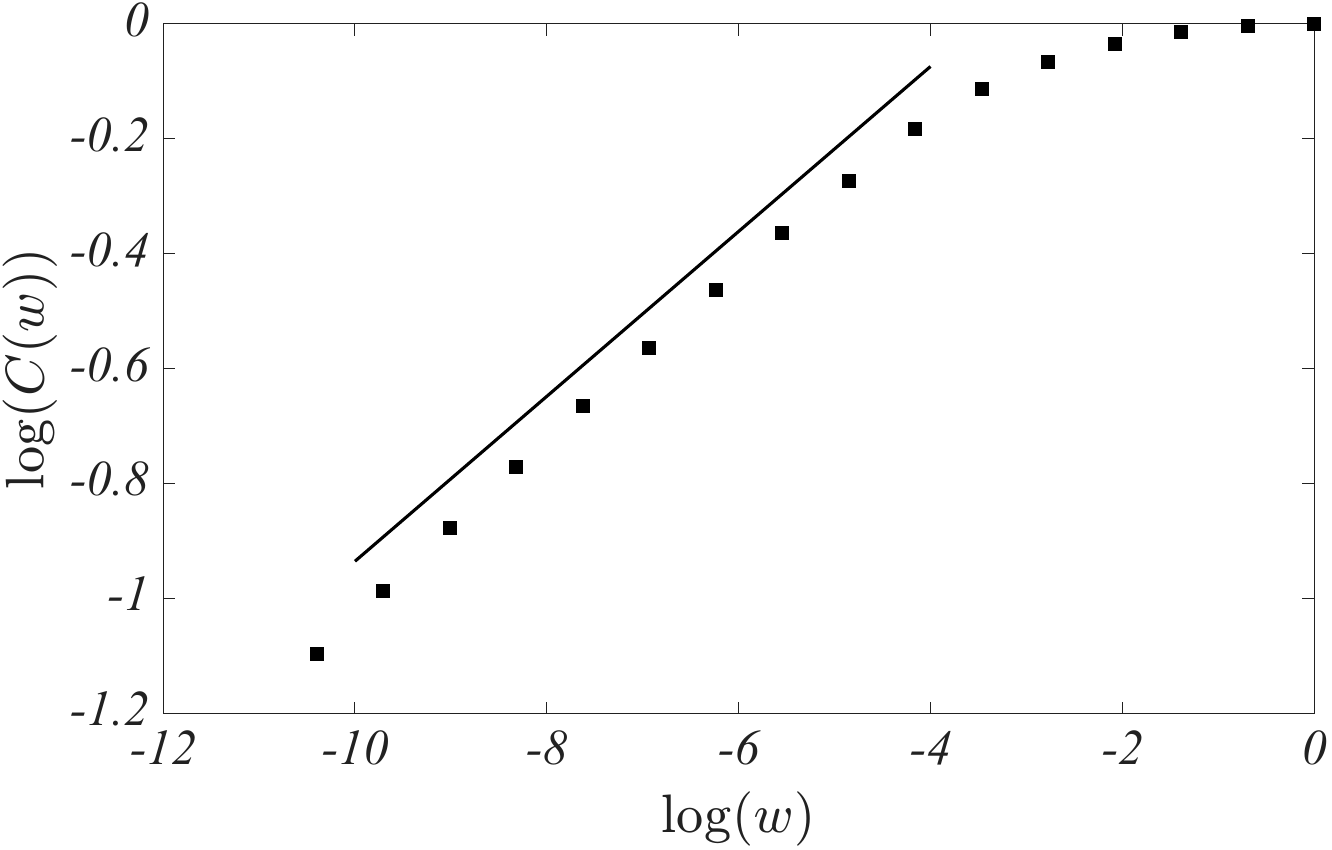}
\caption{The cumulative probability distribution, $C(w)$, of weights $w_{ij}(t)$, for $t\geq T_{\mathrm{trans}}=500$, in log-log scale. The straight line has slope $\sim 0.1435$.}
\label{fig:logCum}
\end{figure}

In Fig.~\ref{fig:AdaptTPexp} we report the results for the model
\begin{equation*}
\begin{dcases}
\frac{du_i}{dt} &=\mu u_i-u_i^3-v_i+D_u \sum_{j=1}^n \mathcal{L}_{ij}u_j\\
\frac{dv_i}{dt} &=\gamma (u_i-\alpha v_i )+D_v  \sum_{j=1}^n \mathcal{L}_{ij}v_j\\
 \frac{dw_{ij}}{dt} &= w_{ij}\left[ a e^{-c [(u_i-u_j)^2+(v_i-v_j)^2]}-b(w_{ij}+1)\right]
\end{dcases}
\end{equation*}
with parameters
\begin{equation*}
a = 1\, , b = 0.5\, , c = 12\, ,\alpha = 0.5\, , \mu = 1.8\, , \gamma = 4\, , D_u=0.1 \text{ and }D_v=1.0\, ,
\end{equation*}
the underlying support is a complete graph made of $n=15$ nodes, the weights are initialized uniformly at random in the interval $(w^* -\delta,w^*+\delta)$, where $w^* = \frac{a}{b}-1=1$. Both species $u$ and $v$ are drawn uniformly at random in the interval $(-\delta,\delta)$, with $\delta=0.01$. The dispersion relation (left panel) is positive, and thus patterns can emerge, indeed $u_i(t)$ move away from the equilibrium $u^*=0$ and oscillate around this value. This behavior causes the weights to oscillate between $0$ and the initial value $w^*=1$.
By an eyeball analysis of the evolution of weights and density of species $u$, it seems to emerge a bursty behavior. It is thus interesting to study the distribution of weights $w_{ij}(t)$. We thus gathered for all $i$ and $j$, the weights after a transient time $T_{\mathrm{trans}}=500$ and we computed the cumulative probability distribution, $C(w)$. The results reported in Fig.~\ref{fig:logCum} clearly show a power law decay for small $w$. Indeed the fit $\log C(w) = \kappa \log w$ returns for small $w$ an estimated value $\kappa \sim 0.1435$, that translated into a pdf of $w$ of the form $1/w^{1-\kappa}.$
In conclusion, we have shown the conditions for the emergence of Turing patterns for FHN models defined on an adaptive network, moreover the resulting patterns exhibit a bursty behavior with weights distributed according to a power law as already found in the literature~\cite{Chakravartula2017}.
\end{document}